\documentclass[12pt]{article}

\usepackage{amsmath}
\usepackage[utf8]{inputenc}
\usepackage{listings}
\usepackage{graphicx}
\usepackage{authblk}
\usepackage{amsfonts}
\usepackage{amsthm}
\usepackage{amsmath}
\usepackage{setspace}
\usepackage{amscd}
\usepackage[mathscr]{eucal}
\usepackage{indentfirst}
\usepackage{breakurl}
\usepackage{bbm}
\usepackage{graphicx}
\usepackage{graphics}
\usepackage[mode=buildnew]{standalone}
\usepackage{url}
\usepackage{comment}
\usepackage{natbib}
\usepackage{tikz}
\usetikzlibrary{intersections}
\usetikzlibrary{positioning}
\usetikzlibrary{shapes.multipart}
\usepackage{geometry}
\usepackage[colorlinks=true, linkcolor=red, citecolor=blue, urlcolor=blue]{hyperref}
\usepackage{xcolor}

\hypersetup{
    linktoc=all, % Links in both the section names and page numbers
    linkcolor=black, % Make links in TOC black
}
\usepackage[T1]{fontenc}
\usepackage{mathtools}

\let\oldFootnote\footnote
\newcommand\nextToken\relax

\renewcommand\footnote[1]{%
    \oldFootnote{#1}\futurelet\nextToken\isFootnote}

\newcommand\isFootnote{%
    \ifx\footnote\nextToken\textsuperscript{,}\fi}
    
\DeclareMathOperator{\E}{\mathbb{E}}

\DeclareMathOperator*{\argmax}{arg\,max}

\newcommand{\Var}{\mathrm{Var}}

\theoremstyle{plain}
\newtheorem{Th}{Theorem}
\newtheorem{Lemma}{Lemma}
\newtheorem{Cor}{Corollary}
\newtheorem{Prop}{Proposition}

 \theoremstyle{definition}
\newtheorem{Def}{Definition}

\newtheorem{Rem}{Remark}
\newtheorem{?}{Problem}
\newtheorem{Ex}{Example}
\newtheorem{Claim}{Claim}

\newtheorem{Assumption}{Assumption}

\newcommand{\R}{\mathbb{R}}

\newcommand{\N}{\mathbb{N}}

\newcommand{\cB}{\mathcal{B}}

\newcommand{\cS}{\mathcal{S}}
\newcommand{\cY}{\mathcal{Y}}

\DeclareMathOperator{\supp}{supp}

\usepackage{setspace}
\usepackage{etoolbox}

\usepackage{titlesec}

\titleformat{\section}
  {\normalfont\large\bfseries}{\thesection}{1em}{}

\titleformat{\subsection}
  {\normalfont\normalsize\bfseries}{\thesubsection}{1em}{}

\titleformat{\subsubsection}
  {\normalfont\normalsize\itshape}{\thesubsubsection}{1em}{}

\usepackage{footmisc}

\title{Data-Driven Mechanism Design: \\
Jointly Eliciting Preferences and Information
}

\author{Dirk Bergemann\thanks{Department of Economics, Yale University, dirk.bergemann@yale.edu} \quad Marek Bojko\thanks{Department of Economics, Yale University, marek.bojko@yale.edu} \quad Paul D\"utting\thanks{Google Research, duetting@google.com} \and Renato Paes Leme\thanks{Google Research, renatoppl@google.com} \quad Haifeng Xu\thanks{Department of Computer Science, University of Chicago and Google Research, haifengxu@uchicago.edu} \quad Song Zuo\thanks{Google Research, szuo@google.com}}

\date{\today}

\begin{document}

\maketitle

% Acknowledgment footnote on the title page
\renewcommand{\thefootnote}{\fnsymbol{footnote}} % Use symbols for footnotes
\footnotetext[1]{An earlier version of this paper appeared as an Extended Abstract in ACM EC'25. Dirk Bergemann gratefully acknowledges financial support from NSF SES 1948336 and SES 2049754 and ONR MURI. Haifeng Xu acknowledges financial support from   NSF Award
CCF-2303372 and AI2050   award G-24-66104 of Schmidt Sciences.  We thank Luis Hoderlein, Thibaut Horel, Elliot Lipnowski, Elchanan Mossel, Alessandro Pavan, Andrew Postlewaite and participants of seminars and conferences at ACM EC'25, Cornell, EEA Congress 2025, Georgetown, INFORMS 2025, TTIC Algorithmic Game Theory Workshop 2025, and Yale, as well as anonymous referees, for helpful comments. All errors are our own.} 
\renewcommand{\thefootnote}{\arabic{footnote}} % Reset footnote numbering style for the rest of the document

\linespread{1}
\begin{abstract}
    \noindent We study mechanism design in environments where agents have private preferences and private information about a common payoff-relevant state. In such settings with multi-dimensional types, standard mechanisms fail to implement efficient allocations. We address this limitation by proposing data-driven mechanisms that condition transfers on additional post-allocation information, modeled as an estimator of the payoff-relevant state. Our mechanisms extend the classic Vickrey–Clarke–Groves framework. We show they achieve exact implementation in posterior equilibrium when the state is fully revealed or utilities are affine in an unbiased estimator. With a consistent estimator, they achieve approximate implementation that converges to exact implementation as the estimator converges, and we provide bounds on the convergence rate. We demonstrate applications to digital advertising auctions and AI shopping assistants, where user engagement naturally reveals relevant information, and to procurement auctions with consumer spot markets, where additional information arises from a pricing game played by the same agents.
    \\
    \vspace{0in}
    \vspace{0in} 
\end{abstract}
\newpage

\linespread{1.25}

\newpage

%\tableofcontents

\newpage

%%%%%%%%%%%%%%%%%%%%%%%%%%%%%%%%%%%%%%%%%%%%%%%%%%%%
\section{Introduction}
%%%%%%%%%%%%%%%%%%%%%%%%%%%%%%%%%%%%%%%%%%%%%%%%%%%%

%%%%%%%%%%%%%%%%%%%%%%%%%%%%%%%%%%%%%%%%%%%%%%%%%%%%
\subsection{Motivation}
%%%%%%%%%%%%%%%%%%%%%%%%%%%%%%%%%%%%%%%%%%%%%%%%%%%%

The rise of data-rich digital environments has created new opportunities and challenges for mechanism design. In settings ranging from online advertising to artificial intelligence (AI)–generated content, participants often possess private information not only about their own preferences but also about underlying states that affect all agents’ payoffs. For example, in sponsored search auctions, advertisers may hold private information about their value per click as well as user behavior patterns that are informative about click-through and conversion rates. In emerging AI-driven environments, content providers increasingly rely on generative systems to produce output while possessing private information about both their intended content and the preferences of downstream consumers.\footnote{For instance, in new formats of digital advertising auctions, advertisers bid to appear in “sponsored’’ content generated by large language models (LLMs). At the time of writing, Perplexity AI, a widely used LLM-based chatbot with an integrated search engine, has begun incorporating such advertising auctions \citep{PerplexityFT2024}. Similar developments include AI shopping assistants recently introduced by Amazon \citep{Rufus1} and Walmart \citep{Sparky1}. Although these technologies remain nascent, the rising value of preference data for targeting suggests that aggregating information dispersed among advertisers will become increasingly critical for platform efficiency.} Efficiently aggregating and incentivizing the revelation of this multi-dimensional private information is essential for the optimal allocation. While the tension between disentangling preferences from information in mechanism design and related environments is well-documented \citep{jehiel2005allocative, lu2019bayesian}, these ``modern'' applications highlight the timeliness and relevance of revisiting these questions. At the same time, these applications generate abundant additional data through user interactions and platform feedback. In sponsored search auctions, such information includes clicks on displayed ads, while settings involving AI-generated content often yield richer data, such as direct user feedback and follow-up queries. As we will show, this additional data can be used to reconcile these forces.

In a static environment with interdependent values and quasi-linear preferences, we develop a mechanism-design framework that moves beyond traditional message-driven mechanisms by incorporating naturally available data. Rather than conditioning transfers solely on agents’ reports, our mechanisms condition transfers on additional information about the payoff-relevant state. This data-driven approach enables the implementation of efficient allocations—maximizing the sum of agents’ expected payoffs given their combined information—even in settings with multi-dimensional private types, where standard mechanisms provably fail \citep{maskin1992auctions,dasgupta2000efficient,jehiel2001efficient,jehiel2006limits}. We focus on implementation \emph{ex post} with respect to agents’ types, ensuring that no agent has an incentive to deviate from truthful reporting, given truthful reports by others, even after uncertainty about other agents’ types is resolved. This implementation notion, which we refer to as \emph{posterior equilibrium} implementation,\footnote{This notion is often referred to as \emph{ex-post} implementation. We use the term \emph{posterior equilibrium} to distinguish ex-post implementation with respect to agent types from ex-post implementation with respect to the state.} is robust to belief misspecification and eliminates incentives for espionage.

%%%%%%%%%%%%%%%%%%%%%%%%%%%%%%%%%%%%%%%%%%%%%%%%%%%%
\subsection{Framework and Results}
%%%%%%%%%%%%%%%%%%%%%%%%%%%%%%%%%%%%%%%%%%%%%%%%%%%%

In our model, agents hold private preferences over allocations and are endowed with private information about a common payoff-relevant state. We model each agent's information about the state as a signal and its realization. We expand the state space to the product probability space of the payoff-relevant state with a common prior and the unit interval with the Lebesgue measure representing residual randomness.\footnote{We extend the main model to heterogeneous priors in Section \ref{sec: Extensions and Discussion}.} A signal is a random variable on the expanded state space---a measurable function mapping the expanded state to a signal realization from a fixed space of feasible realizations. The formulation allows for arbitrary correlation of agents' signals, an important feature of our motivating examples, as agents might obtain data from similar sources with overlapping datasets. We do not assume signals are common knowledge. Instead, signals are private and form a component of agents' types. A mechanism elicits signals together with preferences and signal realizations. The formulation reflects practical considerations regarding data sharing. Proper interpretation of the data requires not only access to the dataset but also documentation of the data-generating process, both of which are proprietary.

To address the implementation impossibility with standard mechanisms \citep{jehiel2001efficient,jehiel2006limits}, we relax the restriction that transfers rely solely on the agents' messages. Instead, we allow transfers to be conditioned on additional information about the state, while maintaining that allocations are determined only by messages. We formalize the additional information as an estimator of the payoff-relevant state available to the designer after allocation but before finalizing transfers, with the estimator's data-generating process being common knowledge. The estimator is allowed to depend on the chosen allocation and is assumed independent of agents' signals conditional on the payoff-relevant state and the allocation. In our motivating examples, this approach captures scenarios where the designer gathers data on user interactions both within and beyond the auction environment, thereby gaining insights into user preferences and demand (i.e., the state).

% \footnote{We also analyze the scenario where it is the agents, rather than the designer, who acquire additional information about the state. In this setting, we propose a two-stage mechanism: in the first stage, agents report their types, and in the second stage, they report the additional information they have acquired. The designer then constructs a leave-one-out estimator for each agent using the second-stage reports, incorporating information from all agents except the one under consideration. By imposing the same conditions on these agent-specific estimators as those applied to the ``global'' estimator in our main analysis, the two-stage data-driven VCG mechanism achieves analogous results. See Section \ref{sec: Eliciting Additional Data from the Agents} for more details.}

We introduce a modified version of Vickrey–Clarke–Groves (VCG) mechanisms \citep{vickrey1961counterspeculation,clarke1971multipart,groves1973incentives}, which we call \textit{data-driven VCG mechanisms} (Definition \ref{def: data-driven VCG}).  In data-driven VCG transfers, the designer inserts the obtained estimate of the state into agents' payoffs. We then examine how various properties of the estimator affect the feasibility of implementation. We first show that in the benchmark case of full resolution of uncertainty about the payoff-relevant state, data-driven VCG mechanisms achieve implementation in posterior equilibrium (Proposition \ref{prop: ex-post}). Specifically, if the designer ultimately observes the true state, she can utilize ex-post utilities based on the reported types to calculate VCG payments, aligning agents' incentives. This alignment occurs because agents evaluate these transfers according to their true posterior beliefs, making truthful reporting optimal under efficient allocations.

Next, we turn to the more realistic case in which the designer observes only a noisy estimate of the state and analyze the implications of two key properties of estimators standard in statistics and econometrics. First, when the estimator is unbiased and agents’ utility functions are affine in the state, data-driven VCG mechanisms achieve implementation in posterior equilibrium (Theorem~\ref{prop: unbiased}). As anticipated by Jensen’s inequality, this result does not extend beyond affine utilities. Nevertheless, our applications to click-through auctions and AI-driven product recommendations illustrate the practical relevance of the result in environments with risk-neutral agents and appropriately defined states.

Second, we consider the property of consistency, where a sequence of estimators, indexed by the dataset size, converges in probability to the true state as the dataset size grows large. Subject to regularity conditions, Theorem \ref{prop: consistent estimator implementation} establishes that any corresponding sequence of data-driven VCG mechanisms achieves implementation in $\epsilon$-posterior equilibrium, where no agent has more than an $\epsilon$ additive utility loss of having reported truthfully after the resolution of uncertainty about others' types when other agents also report truthfully, with $\epsilon$ approaching zero as estimation becomes increasingly accurate.  That is, while reporting truthfully is an $\epsilon$-posterior equilibrium in any finite sample for large enough $\epsilon$ for any estimator, consistent estimators ensure the $\epsilon$ can be made arbitrarily small in the limit. Proposition \ref{prop: rate of convergence} in Appendix \ref{appendix: rate of convergence} further extends this result by linking the rate of convergence of the estimator to the rate at which $\epsilon$ approaches zero. Under suitable uniform integrability conditions, $\epsilon$ converges to zero at the same rate as the sequence of estimators converges to the true state.

%%%%%%%%%%%%%%%%%%%%%%%%%%%%%%%%%%%%%%%%%%%%%%%%%%%%
\subsection{Applications}
%%%%%%%%%%%%%%%%%%%%%%%%%%%%%%%%%%%%%%%%%%%%%%%%%%%%

We view one of the main contributions of the framework in its practical applicability. To illustrate this, we develop three applications.

We first apply our framework to click-through auctions. Following the canonical specification of agents' payoffs \citep{edelman2007internet,varian2007position}, we treat agents' values per click as their preference types and the click-through rates as the state. Within this framework, mechanisms based solely on per-impression payments conditioned only on agents' actions are subject to the same impossibility result identified earlier. In contrast, mechanisms utilizing click data fall within the class of data-driven mechanisms. 

Under a mild assumption on the richness of agents’ information, the data-driven pivot mechanism is the essentially unique data-driven mechanism that implements the efficient allocation in posterior equilibrium while satisfying individual rationality and no-subsidy conditions (Proposition~\ref{prop: Characterization of the VCG auction}). In the single-item case, this mechanism takes the form of a per-click second-price auction. These payments accord with our results, as the observed frequency of clicks is an unbiased estimator of the click-through rate when clicks are sampled independently and identically.\footnote{When click-through rates are agent-specific, data-driven VCG payments are no longer feasible using only click data from the current environment, and feasible per-click payment rules become vulnerable to manipulation.}

Our second application studies efficient product recommendations on platforms that interact with users sequentially, as in emerging AI shopping assistants such as Amazon’s Rufus \citep{Rufus1} and Walmart’s Sparky \citep{Sparky1}. Advertisers have private preferences over being recommended and private signals about the user’s product preferences, which constitute a common payoff-relevant state. The platform aggregates advertisers’ preferences with the user’s preferences. 

Relative to traditional advertising environments, this setting provides the platform with richer information through interactive conversations, including pre-allocation signals from the conversational context. We first show that such pre-allocation information cannot be used simultaneously for allocation and transfers when implementation is required ex post with respect to all information relevant for the allocation; post-allocation data therefore remains essential. We then develop a sequential extension in which advertisers report their types once at the outset while the platform issues recommendations over multiple rounds, progressively refining its information about user preferences. In this dynamic setting, we seek mechanisms that are robust to deviations: even if advertisers were allowed to revise their reports during the conversation, they would not benefit from doing so when others report truthfully. To this end, we construct data-driven dynamic team mechanisms, adapting the class of mechanisms introduced by \citet{athey2013efficient} to our environment with interdependent values.

In our third application, we illustrate that the additional data may also originate from a game played by the agents themselves. We introduce a model of procurement auctions in which firms’ costs depend on both common and idiosyncratic components. Implementing the corresponding data-driven pivot mechanism requires an estimate of the cost of the second-best product. In many settings, sellers not only participate in procurement auctions but also compete in consumer spot markets. Standard methods from the industrial organization literature \citep{berry1994estimating,berry1995automobile} can be used to combine consumer demand estimates with a pricing equilibrium inversion to recover firms’ costs. When demand estimates are consistent, this procedure yields consistent cost estimates, allowing us to apply our implementation continuity result.

%%%%%%%%%%%%%%%%%%%%%%%%%%%%%%%%%%%%%%%%%%%%%%%%%%%%
\subsection{Related Literature}
%%%%%%%%%%%%%%%%%%%%%%%%%%%%%%%%%%%%%%%%%%%%%%%%%%%%

This paper contributes to multiple strands of the mechanism design literature, including:
(i) efficient mechanism design in settings with interdependent values, (ii) sponsored search auctions, and (iii) mechanism design for AI-generated content.

The literature on mechanism design with interdependent values is extensive (see \cite{milgrom1982theory} and the follow-on literature).  In interdependent value settings with commonly known signals, \cite{maskin1992auctions}, \cite{dasgupta2000efficient}, \cite{jehiel2001efficient}, and \cite{jehiel2006limits} established an implementation impossibility for message-driven mechanisms when agents have multi-dimensional types. Given that agents in our model possess private information regarding both the state and their preferences, private types are inherently multi-dimensional. The generic impossibility results of \cite{jehiel2001efficient} and \cite{jehiel2006limits} apply to our framework.

Several papers propose ways to circumvent implementation impossibility. In message-driven settings, \citet{mclean2015implementation} show that posterior equilibrium implementation is feasible when agents have zero informational size and establish continuity as informational size vanishes. Relatedly, \citet{mclean2004informational} show that when informational size is small, efficient Bayesian implementation can be achieved via modified VCG mechanisms with side bets. \citet{mclean2017dynamic} reduce the amount of information transmitted by the agents to the mechanism using two-stage mechanisms. By contrast, we allow agents to have arbitrary informational size and focus on single-round mechanisms that implement efficient allocations in posterior equilibrium.

\citet{mezzetti2004mechanism} and \citet{riordan1988optimal} study mechanisms with contingent payments. In particular, \citet{mezzetti2004mechanism} assumes that agents observe their ex-post payoffs after the allocation is determined and proposes a two-stage mechanism: agents first report their types to determine the allocation, and subsequently report realized payoffs to compute transfers. This approach, however, has several drawbacks. The second reporting stage may be infeasible or costly, and, as emphasized by \citet{jehiel2005allocative}, the mechanism is fragile because agents are completely indifferent over reports in the second stage. Moreover, this class of mechanisms lacks an analogue of the pivotal mechanism suited to our applications.

In contrast, our approach eliminates the need for a second reporting stage or access to agents’ realized payoffs, thereby reducing communication requirements. By conditioning transfers on an estimate of the state, data-driven VCG mechanisms afford substantially greater flexibility. For example, in digital advertising environments, they encompass per-click second-price auctions and VCG auctions more broadly, whereas no corresponding mechanism exists within the class of \citet{mezzetti2004mechanism}. Outcome data of the type we rely on is already widely collected by digital platforms, and we show how such potentially noisy data can be leveraged for implementation.

In a principal–agent model, \citet{riordan1988optimal} demonstrate that an ex-post signal about the agent’s type allows the principal to implement the first-best allocation via a lottery mechanism akin to \citet{cremer1988full}. This insight extends to our framework: one could combine a lottery based on the estimator with standard VCG payments to elicit both components of agents’ types truthfully, provided a suitable stochastic relevance condition on agents’ beliefs holds. However, this condition fails in settings of interest. In particular, when the prior is itself a feasible posterior, it lies in the convex hull of the other posteriors. We elaborate on these issues in Section~\ref{sec: alternative approaches}.\footnote{These papers also assume a commonly known prior over the state and agents' signal realizations. Our model relaxes this assumption. Specifically, both the random variable defining an agent’s signal and its realization are private information that must be elicited by the designer, increasing the potential for strategic misreporting. This distinction is particularly relevant in practical settings, where assuming common knowledge of each agent’s data-generating process, especially in the presence of large and heterogeneous datasets, is unrealistic. Hence, our results accommodate more heterogeneity and offer a higher degree of robustness \citep{bergemann2005robust}.}

This paper is also related to the literature on auctions involving contingent payments and public ex-post information (\cite{hansen1985auctions} and the follow-on literature). Several papers examine how data on agent types can facilitate efficient outcomes in environments with adverse selection \citep{braverman2022data, she2022can, liang2024data} and dynamically evolving market participants \citep{jiang2015data}. %Our focus diverges by exploring how the designer can leverage data on a common, payoff-relevant stochastic factor to design transfers that encourage truthful reporting of agents' preferences and information.

Our first application contributes to the literature on sponsored search auctions. The canonical model due to \citet{edelman2007internet} and \citet{varian2007position} assumes that bidders have a value per click and ad slots have non-increasing click-through rates. Both studies propose essentially the same refinement of pure Nash equilibrium, showing that under this concept, the Generalized-Second Price (GSP) mechanism yields an efficient allocation.
The GSP mechanism can be implemented without knowledge of the click-through rates. This is also true for the VCG mechanism, but requires a more careful implementation in which bidders get charged and credited \citep{varian2009online,varian2014vcg}. 

\citet{milgrom2010simplified} argues that soliciting a single bid (value per click) can be beneficial but also warns that model mis-specification may cause advertisers’ values to misalign with observable clicks. Motivated by this, \citet{DuttingFP19} and \citet{DuttingFP24} examine standard position auctions with mis-specified bidding languages; and establish a ranking between standard position auction formats in regard to the format's ability to support an efficient equilibrium, when values follow different click-through rates than those used in the auction. Our work approaches this differently, by modeling the click-through rates as private information that the bidders can report to the auctioneer.

We also contribute to the emerging literature on mechanism design for AI-generated content \citep{duetting2024mechanism, soumalias2024truthful, dubey2024auctions, hajiaghayi2024ad}. \citet{banchio2024ads} study the optimal timing of ad placement for conversational AI assistants. We instead examine a sequential interaction model aimed at efficient product recommendations in an environment where advertisers possess private information about user preferences.

%%%%%%%%%%%%%%%%%%%%%%%%%%%%%%%%%%%%%%%%%%%%%%%%%%%%
\subsection{Outline}
%%%%%%%%%%%%%%%%%%%%%%%%%%%%%%%%%%%%%%%%%%%%%%%%%%%%

Section \ref{sec: model} introduces the formal model. In Section \ref{sec: message-driven mechanisms}, we analyze the implementation problem in the setting of standard message-driven mechanisms. In Section \ref{sec: data-driven mechanisms}, we define data-driven mechanisms, introduce the data-driven VCG mechanism, and present our implementation results within this framework. Section \ref{sec: applications} contains our applications. Finally, Section \ref{sec: discussion and conclusion} proposes directions for future research and concludes the paper.

%%%%%%%%%%%%%%%%%%%%%%%%%%%%%%%%%%%%%%%%%%%%%%%%%%%%
\section{Model}\label{sec: model}
%%%%%%%%%%%%%%%%%%%%%%%%%%%%%%%%%%%%%%%%%%%%%%%%%%%%

This section introduces the general model and defines efficient allocations as the central object of interest. We also present parametric examples to illustrate the framework.

%%%%%%%%%%%%%%%%%%%%%%%%%%%%%%%%%%%%%%%%%%%%%%%%%%%%
\subsection{Set-up}
%%%%%%%%%%%%%%%%%%%%%%%%%%%%%%%%%%%%%%%%%%%%%%%%%%%%

There is a set $N \equiv \{1,\dots,n\}$ of agents. Let $X$ be a compact space of feasible allocations. 

\paragraph{Payoffs.} Payoff uncertainty is represented by a set of possible states of the world $\Omega$, a compact metric space endowed with the corresponding Borel $\sigma$-algebra $\mathcal{B}(\Omega)$. We denote the metric on $\Omega$ by $d_{\Omega}$; we follow this notational convention for all metric spaces considered in this paper. Each agent is endowed with a private \textit{preference type} $\theta_i \in \Theta_i$, where $\Theta_i$ is a measurable space. Define $\Theta = \prod_{i \in N} \Theta_i$. Agent $i$'s ex-post utility is quasi-linear in the payoff and transfers: $U_i: X \times \Omega \times \Theta_i \times \mathbb{R} \rightarrow \mathbb{R}$ given by:
\begin{equation*}
    U_i(x,\omega,\theta_i,t_i) = u_i(x,\omega,\theta_i) + t_i,
\end{equation*}
where we assume $u_i$ is common knowledge. We assume $u_i$ is bounded, continuous in $x$ for each $\theta_i$ and $\omega$, and measurable in $\omega$ for each $x$ and $\theta_i$.

\paragraph{Information.}

There is a commonly known probability measure $\pi^0 \in \Delta(\Omega)$ with full support. Each agent is endowed with private information about the common payoff-relevant state $\omega \in \Omega$. Drawing on \cite{gentzkow2017bayesian} and \cite{green2022two}, the underlying stochastic structure is as follows. We expand the state space $(\Omega, \cB(\Omega), \pi^0)$, by taking a product with the uniform distribution on $[0,1]$, to the \textit{expanded state space} $(\Omega \times [0,1], \cB(\Omega) \otimes \cB([0,1]), \pi^0 \times \lambda)$, where $\lambda$ denotes the Lebesgue measure. The uniform distribution on $[0,1]$ captures residual randomness beyond the uncertainty about the state $\omega$ and serves as a correlation device. We denote a typical realization of such a \textit{residual state} by $r \in [0,1]$.

For each agent $i$, let $\mathcal{S}_i$ be a compact metric space of possible signal realizations endowed with the corresponding Borel $\sigma$-algebra $\mathcal{B}(\mathcal{S}_i)$. Define $\mathcal{S} = \prod_{i \in N} \mathcal{S}_i$, with the corresponding product $\sigma$-algebra. Each agent $i$'s information is embodied in a measurable function $\mathbf{S}_i: \Omega \times [0,1] \rightarrow \mathcal{S}_i$. The mapping $\mathbf{S}_i$ will be referred to as $i$'s \textit{signal} and is assumed to be agent $i$'s private information. Denote by $\Psi_i$ the set of all feasible signals for agent $i$ and by $\Psi = \prod_{i \in N} \Psi_i$ the product space. We assume $\Psi$ is endowed with a metric $d_{\Psi}$.\footnote{The choice of a metric on this space of measurable functions is immaterial for our results.} Each $\mathbf{S}_i \in \Psi_i$ induces a random variable with the corresponding law given by $P_{\mathbf{S}_i} \equiv (\pi^0 \times \lambda) \circ \mathbf{S}_i^{-1}$. Similarly, a profile of signals $\mathbf{S} \in \Psi$ induces a random variable with the corresponding joint distribution $P_{\mathbf{S}} \equiv (\pi^0 \times \lambda) \circ \mathbf{S}^{-1}$. Preference types are assumed to be independent of the payoff-relevant state. 

A signal realization $s = \mathbf{S}(\omega, r) \in \mathcal{S}$ leads agents to update their beliefs. The posterior is given by a regular conditional probability distribution given by a Markov kernel $\pi_{\mathbf{S}}: \cB(\Omega) \times \mathcal{S} \rightarrow [0,1]$:\footnote{The Markov kernel is $P_{\mathbf{S}}$-a.e. unique. To avoid issues with possible non-uniqueness on measure zero sets, we assume there is a fixed, commonly known, regular conditional probability distribution for each $\mathbf{S} \in \Psi$. Whenever such an ambiguity arises in later sections, we make an analogous assumption.}
\begin{equation*}
    \pi_{\mathbf{S}}(B | \omega) = \lambda( \{ r \in [0,1]: \mathbf{S}(\omega, r) \in B \}) \quad \forall B \in \cB(\cS), \omega \in \Omega.
\end{equation*}
We define $i$'s expected payoff $v_i: X \times \Theta_i \times \mathcal{S} \times \Psi \rightarrow \R$ as follows:
\begin{equation*}
    v_i(x,\theta_i,s,\mathbf{S}) = \int_{\Omega} u_i(x,\omega, \theta_i) d \pi_{\mathbf{S}}(\omega | s).
\end{equation*}

Define $\Xi_i \equiv \{(\theta_i, s_i, \mathbf{S}_i) \in \Theta_i \times \mathcal{S}_i \times \Psi_i: s_i \in \supp P_{\mathbf{S}_i}\}$ to be $i$'s \textit{type space}, with a typical element $\xi_i$. Define $\Xi = \prod_{i \in N} \Xi_i$ to be the product space. We note that the type space of agents is larger than that considered in the related literature. Specifically, an agent's type includes not only the agent's signal realization but also the random variable governing the information technology itself. Finally, an \textit{instance} is a tuple $\Gamma = (N, X, \Omega, \pi^0, \Xi, (U_i)_{i \in N})$.

%%%%%%%%%%%%%%%%%%%%%%%%%%%%%%%%%%%%%%%%%%%%%%%%%%%%
\paragraph{Efficiency.}
%%%%%%%%%%%%%%%%%%%%%%%%%%%%%%%%%%%%%%%%%%%%%%%%%%%%

This paper focuses on implementing the efficient allocation rule based on the combined information held by agents while allowing for residual uncertainty about the payoff-relevant.

\begin{Def}[Efficient allocation rule]\label{def: efficient allocation}
    The deterministic allocation rule $x: \Xi \rightarrow X$ is efficient if, for all $\xi = (\theta, s, \mathbf{S}) \in \Xi$, it satisfies:%\footnote{We could alternatively formulate the efficient allocation rule as a function of agents' preference types and posterior beliefs. The latter is determined by agents' signals $\mathbf{S}$ and realizations $s$. We use the current formulation for ease of exposition, as the allocation in subsequent sections will be determined based on agents' reports of their types.}
    \begin{equation}\label{eq: ex-post efficiency}
        x(\xi) \in \argmax_{x \in X} \sum_{i \in N} v_i(x,\theta_i,s, \mathbf{S}).
    \end{equation}
\end{Def}

The argmax correspondence has non-empty and compact values; an efficient allocation function is thus guaranteed to exist. We fix an arbitrary selection $x^{*}$ from this correspondence. 

%%%%%%%%%%%%%%%%%%%%%%%%%%%%%%%%%%%%%%%%%%%%%%%%%%%%
\subsection{Examples}
%%%%%%%%%%%%%%%%%%%%%%%%%%%%%%%%%%%%%%%%%%%%%%%%%%%%

We illustrate the model with the following parametric examples, which will serve as running examples later on. We apply our results to further settings in Section \ref{sec: applications}.

\begin{Ex}[Joint prediction with private biases]\label{ex: joint prediction with private biases}
    Agent $i$'s payoff is given by
    \begin{equation*}
        u_i(x,\omega,\theta_i) = -(x - \theta_i - \omega)^2,
    \end{equation*}
    for one-dimensional $x, \theta_i, \omega$, for any agent $i \in N$. The common prior on $\omega$ is given by $N(0,1)$. Each agent $i$ is endowed with a signal $\mathbf{S}_i = \omega + \epsilon_i$, where $\epsilon_i \sim N(0,\sigma_i^2)$ is distributed independently across agents. The variance parameter $\sigma_i^2$ is agent $i$'s private information, along with the realization of the signal $s_i$. Collecting such signals yields $\Psi_i$. 

    We interpret $\theta_i$ as the \textit{bias} agent $i$ seeks to introduce into the allocation. Moreover, each signal contributes to a more precise estimate of the state $\omega$. We assume $\theta$ is stochastically independent of the state $\omega$ and agents' signals. The efficient allocation is given by
    \begin{equation*}
        x^{*}(\theta, s, \mathbf{S}) = \frac{1}{n} \sum_{i \in N} \theta_i + \E[\omega | \mathbf{S} = s] = \frac{1}{n} \sum_{i \in N} \theta_i + \frac{\sum_{i \in N} \frac{s_i}{\sigma_i^2}}{1 + \sum_{i \in N} \frac{1}{\sigma_i^2}}.
    \end{equation*}
\end{Ex}

\begin{Ex}[Advertising auctions]\label{ex: ad auction}
    We consider a canonical model of click-through auctions \citep{edelman2007internet, varian2007position}. There are $K$ advertising slots to be allocated among $n \geq K$ agents. Each agent $i$ is risk-neutral and obtains a payoff $\theta_i \in [0,\overline{\theta}]$ each time her ad is clicked, representing her \emph{value per click}. Each slot $k$ has an associated \textit{click-through rate} (CTR) $\omega_k \in [0,1]$, the probability that an ad placed in that slot is clicked. We assume the ex-post ranking of slots is commonly known: $\omega_1 \geq \omega_2 \geq \dots \geq \omega_K$ almost surely. An allocation $x_{ik}$ denotes the probability that agent~$i$ is assigned to slot~$k$, yielding the bidding-stage payoff
    \begin{equation*}
    u_i(x,\omega,\theta_i) = \sum_{k=1}^K \theta_i \cdot \omega_k \cdot x_{ik}.
    \end{equation*}
    Agents’ values per click are independent of CTRs and of agents’ information about them.
    
    Since the ex-post ranking of slots is common knowledge, the efficient allocation is unique up to tie-breaking and corresponds to assortative matching between agents and slots by values per click and CTRs. Consequently, agents’ information about the state $(\mathbf{S},s)$ does not affect the socially efficient allocation, though it may still create incentives for strategic manipulation.
\end{Ex}

%%%%%%%%%%%%%%%%%%%%%%%%%%%%%%%%%%%%%%%%%%%%%%%%%%%%
\section{Mechanisms and Implementation}\label{sec: mechanisms and implementation}
%%%%%%%%%%%%%%%%%%%%%%%%%%%%%%%%%%%%%%%%%%%%%%%%%%%%

In this section, we introduce and contrast message- and data-driven mechanisms and define our implementation notion.

%%%%%%%%%%%%%%%%%%%%%%%%%%%%%%%%%%%%%%%%%%%%%%%%%%%%
\subsection{Message-Driven Mechanisms}\label{sec: message-driven mechanisms}
%%%%%%%%%%%%%%%%%%%%%%%%%%%%%%%%%%%%%%%%%%%%%%%%%%%%

In a message-driven mechanism, agents select from a menu of messages, and both the allocation and transfers are determined solely as functions of these messages. This is the standard notion of a mechanism. The timeline is illustrated in Figure \ref{fig: timeline standard revelation mechanisms}.

\begin{Def}[Message-driven direct mechanism]\label{def: standard direct mechanism}
    A \textit{message-driven direct mechanism} is a pair $(x,t)$ where $x: \Xi \rightarrow X$ is the outcome function and $t: \Xi \rightarrow \mathbb{R}^N$ is the transfer function. 
\end{Def}

Our environment admits a revelation principle by standard arguments. Hence, it is without loss of generality to focus on direct mechanisms and truth-telling.

%%%%%%%%%%%%%%%%%%%%%%%%%%%%%%%%%%%%%%%%%%%%%%%%%%%%%%%%%%%%%%%%%%%%%%%%%%%%%%%%%%%%
\begin{figure}[t]
\centering
\begin{tikzpicture}[
 node distance = 0mm and 0.03\linewidth,
    box/.style = {inner xsep=0pt, outer sep=0pt,
                  text width=0.25\linewidth,
                  align=center, font=\footnotesize}
                    ]
\node (n1) [box]
        {};
\node (n2) [box, below right=of n1.north east]
        {};
\node (n3) [box, below right=of n2.north east]
        {};
        
\draw[black, very thick, -latex, color = black]    (n1.north west) -- (n3.north east) -- + (3mm,0);

\draw (n1.north) -- + (0,3mm) node[above] {\textcolor{black}{\begin{tabular}{c} states and types \\ realized \end{tabular}}};
\draw (n2.north) -- + (0,3mm) node[above] {\textcolor{black}{\begin{tabular}{c} reports \end{tabular}}};
\draw (n3.north) -- + (0,3mm) node[above] {\textcolor{black}{\begin{tabular}{c} final allocation \\ and transfers  \end{tabular}}};
\end{tikzpicture}
\caption{Timing of a message-driven direct revelation mechanism (Definition \ref{def: standard direct mechanism}).}
\label{fig: timeline standard revelation mechanisms}
\end{figure}
%%%%%%%%%%%%%%%%%%%%%%%%%%%%%%%%%%%%%%%%%%%%%%%%%%%%%%%%%%%%%%%%%%%%%%%%%%%%%%%%%

%%%%%%%%%%%%%%%%%%%%%%%%%%%%%%%%%%%%%%%%%%%%%%%%%%%%
\subsection{Data-Driven Mechanisms}\label{sec: data-driven mechanisms}
%%%%%%%%%%%%%%%%%%%%%%%%%%%%%%%%%%%%%%%%%%%%%%%%%%%%

Data-driven mechanisms relax the restriction that allocations and transfers depend solely on agents’ reports. Instead, we allow transfers to condition on additional information about the payoff-relevant state $\omega \in \Omega$ available after the allocation is chosen.

Data-driven mechanisms condition the allocation only on the reported messages but can condition transfers on the payoff-relevant state or an estimate thereof. The resulting timeline is summarized in Figure~\ref{fig: timeline data-driven mechanisms}. 

\begin{Def}[Data-driven direct mechanism]\label{def: direct mechanism with data-driven transfers}
    A \textit{data-driven direct mechanism} is a pair $(x,t)$ where $x: \Xi \rightarrow X$ is the outcome function and $t: \Xi \times \Omega \rightarrow \mathbb{R}^N$ is the transfer function.
\end{Def}

In our motivating applications, $\omega$ may represent user preferences, demand, or click-through rates. The designer collects information from user interactions in current or past environments, ad clicks, and demand estimation. In settings involving AI-generated content, signals such as time spent on specific prompts or follow-up queries provide particularly rich sources of data.

This information is used to estimate the payoff-relevant state $\omega$.  We formalize the data-generating process as an \emph{estimator} of $\omega$, obtained after the allocation is chosen but before final transfers are computed. Importantly, the estimator may depend on the selected allocation $x \in X$, reflecting that in digital environments additional data typically arise from users’ interactions with the implemented outcome.

\begin{Assumption}[Estimator]\label{assumption: estimator}
    There is a commonly known measurable mapping
    \begin{equation*}
        \widehat{\omega}: \Omega \times [0,1] \times X \rightarrow \Omega,
    \end{equation*}
    which defines a random variable $\widehat{\omega}^x \equiv \widehat{\omega}(\cdot, \cdot, x)$ on the expanded state space $(\Omega \times [0,1], \cB(\Omega) \otimes \cB([0,1]), \pi^0 \times \lambda)$, an \textit{estimator} of the state, for any allocation $x \in X$. For any signal profile $\mathbf{S} \in \Psi$ and allocation $x \in X$, the random variable $\widehat{\omega}^x$ is conditionally independent of $\mathbf{S}$ given the payoff-relevant state. Given a final allocation $x \in X$, the designer receives a realization of $\widehat{\omega}^x$ before the final payments are determined. This information is verifiable by the agents.
\end{Assumption}

\begin{figure}[t]
\centering
\begin{tikzpicture}[
 node distance = 0mm and 0.01\linewidth,
    box/.style = {inner xsep=0pt, outer sep=0pt,
                  text width=0.18\linewidth,
                  align=center, font=\footnotesize}
                    ]
\node (n1) [box]
        {};
\node (n2) [box, below right=of n1.north east]
        {};
\node (n3) [box, below right=of n2.north east]
        {};
\node (n4) [box, below right=of n3.north east]
        {};
\node (n5) [box, below right=of n4.north east]
        {};
        
\draw[black, very thick, -latex, color = black]    (n1.north west) -- (n5.north east) -- + (3mm,0);

\draw (n1.north) -- + (0,3mm) node[above] {\textcolor{black}{\begin{tabular}{c} states and types \\ realized \end{tabular}}};
\draw (n2.north) -- + (0,3mm) node[above] {\textcolor{black}{\begin{tabular}{c} reports \end{tabular}}};
\draw (n3.north) -- + (0,3mm) node[above] {\textcolor{black}{\begin{tabular}{c} final \\ allocation \end{tabular}}};
\draw (n4.north) -- + (0,3mm) node[above] {\textcolor{black}{\begin{tabular}{c} estimate \\ obtained \end{tabular}}};
\draw (n5.north) -- + (0,3mm) node[above] {\textcolor{black}{\begin{tabular}{c} final \\ transfers \end{tabular}}};
\end{tikzpicture}
\caption{Timing of a data-driven direct revelation mechanism (Definition \ref{def: direct mechanism with data-driven transfers}).}
\label{fig: timeline data-driven mechanisms}
\end{figure}

The conditional independence assumption allows us to separate uncertainty about the state from the residual uncertainty in the estimator, conditional on the state. The assumption is trivially satisfied in the case of full revelation of the state.

Note that by the conditional independence assumption and the law of iterated expectations, the expectation of any integrable function $f: \Omega \to \mathbb{R}$ is given by
\begin{equation*}
\E[f(\widehat{\omega}^x) | \mathbf{S} = s] = \int_{\Omega} \int_{[0,1]} f \left( \widehat{\omega}(\omega, r, x) \right)\, d\lambda(r)  d\pi_{\mathbf{S}}(\omega | s),
\end{equation*}
for any allocation $x \in X$. This expression first averages the function $f$ over the randomness in the estimator conditional on the payoff-relevant state, and then takes the expectation with respect to the posterior distribution of the state. By construction of the expanded state space and the corresponding random variables, the residual randomness embedded in $r$ generates the distribution of the estimator conditional on $\omega$. We expand on the role of this property in the next section when presenting our main implementation results.

%%%%%%%%%%%%%%%%%%%%%%%%%%%%%%%%%%%%%%%%%%%%%%%%%%%%
\subsection{Implementation}
%%%%%%%%%%%%%%%%%%%%%%%%%%%%%%%%%%%%%%%%%%%%%%%%%%%%

We focus on efficient implementation \emph{ex post} with respect to agents’ types. That is, truthful reporting is optimal for each agent, given that others report truthfully, \emph{ex post} in types (though not necessarily in the state). Since full resolution of uncertainty about the state plays an important role in the next section, we save the terminology of \textit{ex post} to this section, and we refer to the resulting equilibrium concept as a \emph{posterior equilibrium}.\footnote{Posterior implementation introduced by \citet{green1987posterior} for a general mechanism requires agents’ strategies to be mutually optimal given the specific information revealed by the mechanism. In a direct revelation mechanism, this corresponds to conditioning on the realized message profile, that is, the reported types, which yields the notion adopted here. The idea that agents have no regret about truthful reporting after uncertainty about others’ types has been resolved is often referred to as an \emph{ex-post equilibrium} \citep{bergemann2002information}. While these two concepts coincide for direct mechanisms, they generally differ \citep{jehiel2007posterior}. Given the importance of resolving uncertainty about the state in our environment, we adopt the terminology of \emph{posterior equilibrium} in our discussion of implementation.} This equilibrium notion is appealing because it does not require agents to form beliefs or higher-order beliefs about others’ types and eliminates incentives for espionage.

In data-driven mechanisms, since agents do not observe the realization of the estimator at the reporting stage but know its data-generating process, they form beliefs about its realization. We denote the expected transfer in data-driven mechanism $(x,t)$ for a profile of signals $\mathbf{S} \in \Psi$, signal realizations $s \in \mathcal{S}$, and type reports $\xi' \in \Xi$ by
\begin{equation}\label{eq: t bar}
    \overline{t}_i(\xi',s,\mathbf{S}) \equiv \E[t_i(\xi', \widehat{\omega}^{x(\xi')}) | s, \mathbf{S}] = \int_{\Omega} \int_{[0,1]} t_i(\xi', \widehat{\omega}( \omega, r, x(\xi'))) d \lambda(r) d \pi_{\mathbf{S}}(\omega | s).    
\end{equation}
Message-driven mechanisms can be viewed as a special case of data-driven mechanisms with no additional data and hence a trivial estimator, for example given by the prior mean. With a slight abuse of notation, we write $\overline{t}_i(\xi',s,\mathbf{S}) = t_i(\xi')$ for the expected transfer in a message-driven mechanism. In the formal definition of posterior equilibrium implementation below, we refer to both message- and data-driven mechanisms simply as mechanisms.

\begin{Def}[Posterior equilibrium implementation]\label{def: posterior equilibrium with data-driven mechanisms}
    Mechanism $(x,t)$ \textit{permits implementation in posterior equilibrium} if for each agent $i \in N$ and types $\xi = (\theta, s, \mathbf{S}) \in \Xi$:
    \begin{equation*}
        v_i(x(\xi),\theta_i,s,\mathbf{S}) +  \overline{t}_i(\xi, s,\mathbf{S}) \geq v_i(x(\xi_i', \xi_{-i}),\theta_i,s,\mathbf{S}) +  \overline{t}_i(\xi_i', \xi_{-i}, s, \mathbf{S}) \quad \forall \xi_i' \in \Xi_i. 
    \end{equation*}
\end{Def}

It is well known that efficient allocation rules are not implementable by message-driven mechanisms in multi-dimensional environments with interdependent values \citep{maskin1992auctions, dasgupta2000efficient, jehiel2001efficient, jehiel2006limits}. The generic impossibility results of \citet{jehiel2001efficient} and \citet{jehiel2006limits} apply to those environments within our framework that fall into their scope: whenever agents have private preference types, overall types are multi-dimensional, even if signals about the state are one-dimensional. We review these results in Appendix~\ref{sec: appendix message-driven mechanisms}. However, not all environments we study fall within this subset, including our running examples. For these examples, we establish implementation impossibility directly. Example~\ref{ex: joint prediction with private biases} features a continuum allocation space, which is not explicitly covered by \citet{jehiel2001efficient} or \citet{jehiel2006limits}. Example~\ref{ex: ad auction} considers an environment that lies outside the generic class of \citet{jehiel2006limits} under certain conditions on the signal structure. In the next section, we introduce data-driven VCG mechanisms and show how they overcome these impossibility results.

%%%%%%%%%%%%%%%%%%%%%%%%%%%%%%%%%%%%%%%%%%%%%%%%%%%%
\section{Data-Driven VCG Mechanisms}\label{sec: data-driven VCG}
%%%%%%%%%%%%%%%%%%%%%%%%%%%%%%%%%%%%%%%%%%%%%%%%%%%%

We introduce the class of \emph{data-driven VCG} mechanisms, the central focus of our analysis, and present our main implementation results. Data-driven VCG mechanisms extend the classical VCG framework \citep{vickrey1961counterspeculation,clarke1971multipart,groves1973incentives} by incorporating an estimate of the state into agents’ \emph{ex-post} payoffs.

\begin{Def}[Data-driven VCG] \label{def: data-driven VCG}
    A data-driven direct mechanism $(x^{*},t)$ is a \textit{data-driven VCG mechanism} if $x^{*}$ is an efficient allocation rule and for each $i \in N$, the transfer, as a function of reports $\xi \in \Xi$ and estimate $\widehat{\omega}^{*}$ of $\omega$, takes the form
    \begin{equation}\label{eq: ddVCG}
    t_i(\xi,\widehat{\omega}^{*}) \equiv h_i(\xi_{-i},\widehat{\omega}^{*}) + \sum_{j \neq i} u_j(x^{*}(\xi),\widehat{\omega}^{*},\theta_j),
    \end{equation}
    for an arbitrary integrable function $h_i$.
\end{Def}

An important element of the class is the \textit{data-driven pivot mechanism}, where
\begin{equation}\label{eq: dd pivot}
    h_i(\xi_{-i}, \widehat{\omega}^{*}) \equiv - \sum_{j \neq i} u_j(x^{*}(\xi_{-i}), \widehat{\omega}^{*}, \theta_j).
\end{equation}

A central property of data-driven VCG transfers is that they depend on reported signals only through the resulting allocation. Once the allocation is determined, the designer can compute transfers without knowing agents’ beliefs. 

We next examine desirable properties of the estimator and their implications for implementation with data-driven VCG mechanisms. We then compare this class to alternative mechanisms with contingent transfers proposed in the literature and discuss extensions.

%%%%%%%%%%%%%%%%%%%%%%%%%%%%%%%%%%%%%%%%%%%%%%%%%%%%
\subsection{The Special Case of Ideal Data}\label{sec: noiseless data}
%%%%%%%%%%%%%%%%%%%%%%%%%%%%%%%%%%%%%%%%%%%%%%%%%%%%

As a natural benchmark, we first consider the case of ideal data: the designer observes the payoff-relevant state $\omega$ without any noise or imperfections after the allocation but before the transfers are finalized. In such a world, the estimator would fully reveal the state: $\widehat{\omega}(\omega, r, x) = \omega$ for every $\omega \in \Omega$, $r \in [0,1]$, and $x \in X$.

\begin{Prop}[Implementation with ideal data]\label{prop: ex-post}
    If $\widehat{\omega}(\omega, r, x) = \omega$ for every $\omega \in \Omega$, $r \in [0,1]$, and $x \in X$, every data-driven VCG mechanism permits implementation in posterior equilibrium.
\end{Prop}

Each agent evaluates data-driven VCG transfers and gross payoffs in expectation using her true beliefs. Thus, if all other agents report truthfully, agent $i$'s net expected payoff aligns with the social objective, ensuring truthfulness by a standard argument.

\begin{proof}
    Fix an arbitrary data-driven VCG mechanism and agent $i$. Assume the rest of the agents report truthfully. Then for any $\xi \in \Xi$ and reports $\xi_i' \in \Xi_i$, the expected transfer is given by
    \begin{equation*}
        \overline{t}_i(\xi_i', \xi_{-i}, s, \mathbf{S}) = h_i(\xi_{-i}, s, \mathbf{S}) + \sum_{j \neq i} v_j(x^{*}(\xi_i', \xi_{-i}), \theta_j, s, \mathbf{S}),
    \end{equation*}
    where, with some abuse of notation, we integrated out realizations of the estimator $\widehat{\omega}$ from $h_i$. The net utility is
    \begin{equation*}
        h_i(\xi_{-i}, s, \mathbf{S}) + \sum_{j \in N} v_j(x^{*}(\xi_i', \xi_{-i}), \theta_j, s, \mathbf{S}).
    \end{equation*}
    Since $x^{*}$ is efficient, this expression is maximized at $\xi_i$.
\end{proof}

We illustrate the transfer scheme under the construction of Example \ref{ex: joint prediction with private biases}.

\begin{Ex}[Joint prediction with private biases: data-driven VCG]\label{ex: joint prediction with private biases: data-driven VCG}
    Example \ref{ex: joint prediction with private biases} continued.  We illustrate that under the data-driven pivot mechanism, agents are rewarded for providing valuable information about $\omega$, with rewards increasing in the value of their contribution. However, agents must also pay for steering the social decision in favor of their preferences. The expected transfer in the truthful equilibrium under the pivot mechanism is given by
    \begin{align*}
        \overline{t}_i(\xi, s, \mathbf{S}) &= - \sum_{j \neq i} (\overline{\theta}_{-i} - \theta_j)^2 + \left( \E[\omega | \mathbf{S} = s] - \E[\omega | \mathbf{S}_{-i} = s_{-i}] \right)^2,
    \end{align*}
    where $\overline{\theta}_{-i} = \frac{1}{n-1} \sum_{j \neq i} \theta_j$.
    
    We highlight the following properties.  First, in expectation, the agent is \textit{paid} for increasing the prediction accuracy. This is most clearly seen if preferences are aligned. Then the transfer of agent $i$ is given by $( \E[\omega | \mathbf{S} = s] - \E[\omega | \mathbf{S}_{-i} = s_{-i}] )^2$. This transfer is always non-negative and strictly positive if agent $i$'s signal induces a different posterior mean when combined with other agents' signal than when based on $(\mathbf{S}_{-i}, s_{-i})$ alone. Second, the more accurate the information agent $i$ provides, as measured by a reduction in the expected residual variance $\E [ \Var[\omega | \mathbf{S} = s] | \mathbf{S}_{-i} = s_{-i}] - \Var[\omega | \mathbf{S}_{-i} = s_{-i}]$, the higher is the agent's expected compensation.\footnote{Formally, this follows from the law of total variance.
    %$\Var[ \E[\omega | \mathbf{S} = s] | \mathbf{S}_{-i} = s_{-i}] = \Var[\omega | \mathbf{S}_{-i} = s_{-i}] - \E[ \Var[\omega | \mathbf{S} = s] | \mathbf{S}_{-i} = s_{-i}]$, noting that by the law of iterated expectations, $\Var[ \E[\omega | \mathbf{S} = s] | \mathbf{S}_{-i} = s_{-i}] = \E [( \E[\omega | \mathbf{S} = s] - \E[\omega | \mathbf{S}_{-i} = s_{-i}])^2 | \mathbf{S}_{-i} = s_{-i}]$. Thus, a reduction in $\E[ \Var[\omega | \mathbf{S} = s] | \mathbf{S} = s_{-i}] - \Var[\omega | \mathbf{S}_{-i} = s_{-i}]$ increases agent $i$'s expected payment. 
    In the parametric case with normal distributions, the ex-ante expected transfer of agent $i$ in the truthful equilibrium is $\frac{1 / \sigma_i^2}{(1 + \sum_{j \neq i} 1 / \sigma_j^2) (1+ \sum_{j \in N} 1 / \sigma_j^2)}$,
    which is strictly decreasing in $\sigma_i^2$.} Third, the agent \textit{pays} for introducing bias into the social decision, with the expected payment for introducing bias given by $(\theta_i - \theta_j)^2 / 4$. 
\end{Ex}

We discuss the relation between Proposition~\ref{prop: ex-post} and the main result of \citet{mezzetti2004mechanism}, as well as to \citet{riordan1988optimal}-style mechanisms, in Section~\ref{sec: alternative approaches}. We next explore the more realistic case of noisy data.

%%%%%%%%%%%%%%%%%%%%%%%%%%%%%%%%%%%%%%%%%%%%%%%%%%%%
\subsection{The General Case: Noisy Data}\label{sec: noisy data}
%%%%%%%%%%%%%%%%%%%%%%%%%%%%%%%%%%%%%%%%%%%%%%%%%%%%

The case of noiseless data provides a useful benchmark, but in practice the designer typically observes only a noisy estimate of the state. We focus on estimators satisfying two properties central in statistics and econometrics: unbiasedness and consistency.

\paragraph{Unbiased Estimators.} With an unbiased estimator and utility functions affine in the state, data-driven VCG mechanisms achieve implementation in posterior equilibrium. 

\begin{Th}[Unbiased estimator]\label{prop: unbiased}
    Suppose $\widehat{\omega}$ is an unbiased estimator of $\omega$ conditional on $\omega$ pointwise:
    \begin{equation*}
        \E[\widehat{\omega}^x | \omega] \equiv \int_{[0,1]} \widehat{\omega}(\omega, r, x) d \lambda(r) = \omega \quad \forall \omega \in \Omega, x \in X.
    \end{equation*}   
    If utility functions are affine in $\omega$, every data-driven VCG mechanism permits implementation in posterior equilibrium.
\end{Th}

\begin{proof}
    By the law of iterated expectations and conditional independence, for any allocation $x \in X$:
    \begin{equation*}
        \E[ \widehat{\omega}^x | \mathbf{S} = s] =  \int_{\Omega} \int_{[0,1]} \widehat{\omega}(\omega, r, x) d \lambda(r) d \pi_{\mathbf{S}}(\omega | s) = \int_{[0,1]} \omega d \pi_{\mathbf{S}} (\omega | s) = \E[\omega | \mathbf{S} = s].
    \end{equation*}
    Since utility functions are affine, it follows that for each $i \in N, \xi = (\theta, s, \mathbf{S}) \in \Xi, \xi' = (\theta', s', \mathbf{S}') \in \Xi$:
    \begin{equation*}
        \overline{t}_i(\xi', s, \mathbf{S}) = h_i(\xi_{-i}, s, \mathbf{S}) + \sum_{j \neq i} v_j(x^{*}(\xi'), \theta_j', s, \mathbf{S}),
    \end{equation*}
    having integrated out the estimator $\widehat{\omega}$ from $h_i$. The result now follows by an argument analogous to the proof of Proposition \ref{prop: ex-post}.
\end{proof}

By Jensen’s inequality, this result does not extend beyond affine utilities. Nonetheless, Theorem~\ref{prop: unbiased} remains practically relevant, as illustrated by our applications.

\begin{Ex}[Advertising auctions: data-driven VCG]\label{ex: Advertising auctions: data-driven VCG}
    Consider Example~\ref{ex: ad auction} in the single-item case. We first illustrate why message-driven variants of the second-price auction fail to implement the efficient allocation. Suppose the winning agent~$i$ pays $-\max_{j\neq i}\theta_j\,\E[\omega]$ and zero otherwise. Upon winning, agent~$i$’s payoff is $\theta_i \E[\omega\mid s]-\max_{j\neq i}\theta_j\,\E[\omega]$. If $\theta_i>\max_{j\neq i}\theta_j$ but $\theta_i \E[\omega\mid s]<\max_{j\neq i}\theta_j\,\E[\omega]$, truthful reporting is not optimal. An analogous failure arises when the winner’s payment is $-\max_{j\neq i}\theta_j\,\E[\omega\mid s_{-i}]$.

    Now assume clicks on the displayed ad are i.i.d.\ Bernoulli$(\omega)$, so that the observed click frequency $\widehat{\omega}^{*}$ is an unbiased estimator of the click-through rate. In the data-driven pivot mechanism, the winner pays
    \begin{equation}\label{eq: unbiased per click spa}
        \max_{j \neq i} \theta_j \cdot \widehat{\omega}^{*}.
    \end{equation}
    This mechanism admits a natural interpretation as a per-click second-price auction. We analyze the general case of the auction environment in Section~\ref{sec: application to click-through auctions}, where we show that Theorem~\ref{prop: unbiased} yields a novel characterization of the VCG auction in online advertising.
\end{Ex}

\paragraph{Consistent Estimators.} Next, we prove that data-driven VCG mechanisms with a consistent estimator of $\omega$ achieve implementation in $\epsilon$-posterior equilibrium, where $\epsilon$ can be made arbitrarily small as the estimator converges in probability. 

\begin{Def}[$\epsilon$-posterior equilibrium]\label{def: epsilon posterior eq}
    Fix $\epsilon \geq 0$. A data-driven direct mechanism $(x,t)$ \textit{permits implementation in $\epsilon$-posterior equilibrium} if for each $i \in N$ and types $\xi = (\theta, s, \mathbf{S}) \in \Xi$:
    \begin{equation*}
        v_i(x(\xi),\theta_i,s,\mathbf{S}) +  \overline{t}_i(\xi, s,\mathbf{S}) + \epsilon \geq v_i(x(\xi_i', \xi_{-i}),\theta_i,s,\mathbf{S}) +  \overline{t}_i(\xi_i', \xi_{-i}, s, \mathbf{S}) \quad \forall \xi_i' \in \Xi_i. 
    \end{equation*}
\end{Def}

In words, once the uncertainty about others' types is resolved, no agent regrets reporting truthfully by more than $\epsilon$ units of the numeraire.

%Truthful reporting constitutes an $\epsilon$-posterior equilibrium in finite samples for sufficiently large $\epsilon$ regardless of the estimator used. However, with consistent estimators and under the regularity conditions stated below, $\epsilon$ can be made arbitrarily small as the size of the dataset used to construct the estimator grows large.

We define the following notions of consistency of estimators. Recall that agents face two sources of uncertainty: uncertainty about the payoff-relevant state $\omega$, and residual randomness in the estimator conditional on $\omega$. Fixing any $\omega \in \Omega$ and $x \in X$, we consider the residual randomness generated by $r$. A sequence of estimators $\{\widehat{\omega}^x_m\}_m$ is said to be \textit{consistent for $\omega$} if the sequence $\{\widehat{\omega}_m(\omega, \cdot, x)\}_m$ converges in probability to $\omega$ as a random variable on $([0,1], \mathcal{B}([0,1]), \lambda)$:
\begin{equation}\label{eq: consistency}
    \forall \epsilon > 0: \quad \lim_{m \rightarrow \infty} \lambda(\{r \in [0,1]: d_{\Omega}(\widehat{\omega}_m(\omega, r, x), \omega) > \epsilon\}) = 0.
\end{equation}
We define $\{\widehat{\omega}_m\}_m$ to be \textit{pointwise consistent} if it is consistent for every $\omega \in \Omega$ and $x \in X$:
\begin{equation}\label{eq: pointwise consistency}
    \forall \epsilon > 0: \quad \lim_{m \rightarrow \infty} \lambda(\{r \in [0,1]: d_{\Omega}(\widehat{\omega}_m(\omega, r, x), \omega) > \epsilon\}) = 0 \quad \forall \omega \in \Omega, x \in X.
\end{equation}
Further, $\{\widehat{\omega}_m\}_m$ is \textit{uniformly consistent} if the convergence is uniform across $\omega$ and $x$:
\begin{equation}\label{eq: uniform consistency}
    \forall \epsilon > 0: \quad \lim_{m \rightarrow \infty} \sup_{\omega \in \Omega, x \in X} \lambda(\{r \in [0,1]: d_{\Omega}(\widehat{\omega}_m(\omega, r, x), \omega) > \epsilon\}) = 0.
\end{equation}

We also introduce a continuity notion for the estimator with respect to the allocation and for posterior beliefs $\pi_{\mathbf{S}}(\cdot | s)$ with respect to signals $\mathbf{S} \in \Psi$ and realizations $s \in \cS$. Suppose $X$ is a metric space with a metrix $d_X$. We call an estimator $\widehat{\omega}$ \textit{Lipschitz continuous in the allocation} if the mapping $x \mapsto \widehat{\omega}(\omega, r, x)$ is Lipschitz continuous uniformly in $(\omega, r)$. A sequence of estimators $\{\widehat{\omega}_m\}_m$ is \textit{uniformly Lipschitz continuous in the allocation} if there is a constant $L_X > 0$ such that $\widehat{\omega}_m$ is Lipschitz continuous in the allocation with a Lipschitz constant of $L_X$ for every $m \in \mathbb{N}$.

Further, we call posterior beliefs \textit{Lipschitz continuous} if the mapping $(\mathbf{S}, s) \mapsto \pi_{\mathbf{S}}(\cdot | s)$ satisfies the following Lipschitz condition with respect to the Wasserstein 1-distance $W_1$:
\begin{equation}\label{eq: lipschitz posteriors}
    \exists L > 0 \text{ s.t. } \forall \mathbf{S}_1, \mathbf{S}_2 \in \Psi, s_1, s_2 \in \cS: W_1(\pi_{\mathbf{S}_1}(\cdot | s_1), \pi_{\mathbf{S}_2}(\cdot | s_2)) \leq L \left( d_{\cS}(s_1, s_2) + d_{\Psi}(\mathbf{S}_1, \mathbf{S}_2) \right).
\end{equation}

We obtain the following implementation continuity result.\footnote{When $\Omega$, $X$, $\mathcal{S}$, and $\Psi$ are finite, pointwise consistency of the estimator is sufficient to obtain the result, as all other regularity conditions hold automatically. In particular, the Lipschitz condition on utility functions follows directly from the finiteness of \(\Omega\) and $X$. Moreover, since \(\Omega\) and $X$ are finite, convergence in probability conditional on each \(\omega \in \Omega\) and $x \in X$ in \eqref{eq: pointwise consistency} implies uniform convergence in probability across all \(\omega\) and $x$ in \eqref{eq: uniform consistency}. Likewise, the Lipschitz condition on posteriors in \eqref{eq: lipschitz posteriors} is satisfied when \(\mathcal{S}\) and \(\Psi\) are finite.}

\begin{Th}[Consistent estimator]\label{prop: consistent estimator implementation}
    Suppose $u_i$ is Lipschitz in $\omega$ uniformly in $x$ and $\theta_i$ for each $i \in N$.\footnote{That is, $\forall i \in N, \exists L_i > 0$ such that $\forall x \in X, \theta_i \in \Theta_i$ and $\omega_1, \omega_2 \in \Omega$, $|u_i(x,\omega_1,\theta_i) - u_i(x,\omega_2,\theta_i)| \leq L_i d_{\Omega}(\omega_1, \omega_2)$. Note that a sufficient condition for this to hold is Lipschitz continuity of $u_i$.} Fix a sequence of estimators $\{\widehat{\omega}_m\}_m$ such that either:
    \begin{enumerate}
        \item $\{\widehat{\omega}_m\}_m$ is uniformly consistent; or 
        \item $\{\widehat{\omega}_m\}_m$ is pointwise consistent and uniformly Lipschitz continuous in the allocation, posterior beliefs are Lipschitz continuous, and $\Psi$ is compact.
    \end{enumerate}
    Then there is a non-negative sequence $\{\epsilon_m\}_m$, with $\epsilon_m \rightarrow 0$ as $m \rightarrow \infty$, such that every data-driven VCG mechanism for $\widehat{\omega}_m$ permits implementation in $\epsilon_m$-posterior equilibrium for every $m \in \mathbb{N}$.
\end{Th}

The proof, contained in Appendix \ref{sec: appendix proof consistent estimator implementation}, proceeds as follows. For any $m$ and types $\xi \in \Xi$, define $\epsilon_i^m(\xi)$ to be the payoff loss for agent $i$ from reporting truthfully compared to $i$'s desired allocation when other agents report truthfully. Also let $\epsilon_m(\xi)$ be the maximum across agents and $\epsilon_m \equiv \sup_{\xi \in \Xi} \epsilon_m(\xi)$. Reporting truthfully is an $\epsilon_m$-posterior equilibrium. Further, as shown in Lemma \ref{lemma: bound on epsilon}, $\epsilon_m(\xi)$ is upper-bounded by a constant multiple of the difference of expected transfers for the ex-post case and for the $m$-th estimator $\widehat{\omega}_m$. As shown in Lemma \ref{lemma: bound via expected error}, the assumed regularity conditions allow us to further upper-bound this by a suitable supremum over the conditional expectation of the difference between the value of the estimator and the payoff-relevant state. Finally, Lemma \ref{lemma: uniform convergence} shows this upper bound converges to zero.

In Appendix \ref{appendix: rate of convergence}, we show that under suitable uniform integrability conditions, the sequence $\{\epsilon_m\}_m$ can converge to zero at essentially the same rate as the sequence of estimators converges to the true state: if $q_m d_{\Omega}(\widehat{\omega}_m, \omega)$ converges to zero in probability for a non-negative sequence $\{q_m\}_m$, $q_m \epsilon_m$ also converges to zero as $m \rightarrow \infty$. Up to a constant factor, this result provides an upper bound on the utility loss agents experience from reporting truthfully when others do the same, given a convergence rate of the estimator.

%%%%%%%%%%%%%%%%%%%%%%%%%%%%%%%%%%%%%%%%%%%%%%%%%%%%
\subsection{Different Mechanisms with Contingent Transfers}\label{sec: alternative approaches}
%%%%%%%%%%%%%%%%%%%%%%%%%%%%%%%%%%%%%%%%%%%%%%%%%%%%

%[Even in the ideal case where several other approaches exist, our approach has several advantages.]

\paragraph{Mezzetti's Mechanism.} 
\citet{mezzetti2004mechanism} proposes a two-stage generalization of VCG mechanisms. In the first stage, agents report their types. After the allocation is chosen and agents observe their ex-post payoffs, they report realized utilities in the second stage. Given utility reports $u'$, the transfer to agent~$i$ is
\begin{equation}
t_i(\xi', u') = \sum_{j \neq i} u_j' + h_i(\xi_{-i}', u_{-i}').
\end{equation}
for an arbitrary function $h_i$ of other agents' reports. Because agent~$i$’s transfer does not depend on $u_i'$, she is indifferent over second-stage reports, and truthful reporting is therefore optimal. By backward induction, truthful reporting in the first stage is optimal.

We remark that Proposition \ref{prop: ex-post} can be derived from implementation results of \cite{mezzetti2004mechanism} and \cite{athey2013efficient}.\footnote{Suppose we run \cite{mezzetti2004mechanism}'s mechanism but the designer imputes what the agents would have reported in the second stage assuming that the type report was truthful. The mechanism sets transfers as in \cite{mezzetti2004mechanism} based on these imputed payoffs. Because truthful reporting of types in the first stage and utilities in the second stage was an equilibrium in \cite{mezzetti2004mechanism}'s mechanism, reporting types truthfully remains optimal. Proposition \ref{prop: ex-post} follows.} Nonetheless, data-driven VCG mechanisms circumvent several key limitations of \citet{mezzetti2004mechanism}’s class, underscoring the practical appeal of the data-driven VCG approach.

First, data-driven VCG mechanisms reduce communication requirements. From a practical standpoint, two-stage mechanisms may be infeasible or costly to implement. Digital platforms already collect abundant additional data, so avoiding extra reports from agents further lowers communication burdens. Moreover, agents’ true ex-post payoffs need not be reported for implementation: it suffices to elicit agents’ types, which are reported truthfully in equilibrium.

Second, the Groves component in data-driven VCG mechanisms offers greater flexibility than in the mechanism of \citet{mezzetti2004mechanism}. The construction in \eqref{eq: ddVCG} yields a natural notion of a pivot mechanism, defined in \eqref{eq: dd pivot} and illustrated in Examples~\ref{ex: joint prediction with private biases: data-driven VCG} and \ref{ex: Advertising auctions: data-driven VCG}. This flexibility has important implications for online advertising auctions: data-driven VCG mechanisms accommodate per-click second-price auctions and VCG auctions more broadly. No corresponding mechanism exists within the class of \citet{mezzetti2004mechanism}, as the framework does not allow for the construction of counterfactual outcomes in which a focal agent is excluded.\footnote{No mechanism in the class of \citet{mezzetti2004mechanism} can generate payments matching the ex-post version of \eqref{eq: unbiased per click spa} in Example \ref{ex: Advertising auctions: data-driven VCG}. When agent~$i$ wins the item, the realized payoffs of all other agents are identically zero, whereas in the counterfactual allocation in which agent~$i$ is excluded, this need not be so. Consequently, the winner’s transfer must take the form $h_i(\xi_{-i}, u_{-i}) = h_i(\xi_{-i}, 0)$, which cannot replicate the pivotal payment $\max_{j \neq i} \theta_j \cdot \omega$.} Under a richness condition on the information agents may possess, we show in Section~\ref{sec: application to click-through auctions} that the data-driven pivot mechanism is the essentially unique data-driven mechanism that implements the efficient allocation while satisfying individual rationality and no-subsidy conditions in advertising auctions.

Third, in \cite{mezzetti2004mechanism}'s mechanisms, agents are completely indifferent among all second-stage reports. As emphasized by \citet{jehiel2005allocative}, this contrasts sharply with standard private-value VCG mechanisms, in which agents typically have strict incentives to report truthfully. Data-driven VCG mechanisms retain this latter property.

\paragraph{Riordan and Sappington's Mechanism.}

In a principal–agent model, \citet{riordan1988optimal} show that observing a signal about the agent’s type allows the principal to extract the full surplus, and hence implement the first-best, under a stochastic relevance condition on the conditional distribution of the signal analogous to that of \citet{cremer1988full}. 

This insight extends to our setting as follows. Consider the case of full ex-post revelation of the state $\omega$. Fixing the reports of all other agents at truthful messages, the focal agent’s report determines the induced posterior belief about the state. Thus, conditional on others’ truthful reports, the mechanism effectively elicits the focal agent’s posterior belief about the state. Suppose $\Omega$, $\mathcal{S}$, and $\Psi$ are all finite. Fix arbitrary $(s_{-i}, \mathbf{S}_{-i}) \in \mathcal{S}_{-i} \times \Psi_{-i}$, and define
\begin{equation*}
    \Pi_i(s_{-i}, \mathbf{S}_{-i}) = \left\{ \pi_{\mathbf{S}_i, \mathbf{S}_{-i}}(\cdot | s_i, s_{-i}) \in \Delta(\Omega): (s_i, \mathbf{S}_i) \in \mathcal{S}_i \times \Psi_i \right\}
\end{equation*}
to be the set of feasible posterior beliefs. In Appendix \ref{appendix: surplus extraction mechanisms}, we show that if for any agent $i \in N$ and any $(s_{-i}, \mathbf{S}_{-i}) \in \mathcal{S}_{-i} \times \Psi_{-i}$,\footnote{co($Y$) denotes the convex hull of a set $Y$.}
    \begin{equation}\label{eq: CM condition}
        \forall \pi_i \in \Pi_i(s_{-i}, \mathbf{S}_{-i}): \quad \pi_i \notin \text{co}(\Pi_i(s_{-i}, \mathbf{S}_{-i}) \setminus \{\pi_i\}),
    \end{equation}
there exists $\phi_i: \mathcal{S} \times \Psi \times \Omega \to \mathbb{R}$ for each $i \in N$ such that the mechanism $(x^{*}, t)$ with transfers given by
    \begin{equation}\label{eq: transfer CM}
        t_i(\xi,\omega) = \sum_{j \neq i} v_j(x^{*}(\xi), \theta_j, s, \mathbf{S}) + \phi_i(s,\mathbf{S},\omega),
    \end{equation}
for any $\xi = (\theta, s, \mathbf{S}) \in \Xi$, implements $x^{*}$ in posterior equilibrium.\footnote{Moreover, if agents' types consist only of information about the state, there is a mechanism that awards agents no information rent.}

However, condition~\eqref{eq: CM condition} cannot be satisfied in several settings of interest. Consider, for example, the independent-signals model of \citet{bergemann2002information}. The state space factors as $\Omega = \prod_i \Omega_i$, with each $\Omega_i$ finite, and the prior decomposes as $\pi^0(\omega) = \prod_i \pi^0_i(\omega_i)$. Agent~$i$ receives a signal informative only about $\omega_i$, with signals independent across agents. It is therefore without loss to take agent~$i$’s type to be her posterior belief over $\omega_i$. Let $\Pi_i$ denote the finite set of feasible posterior beliefs for agent~$i$. In this setting, condition~\eqref{eq: CM condition} becomes
\begin{equation}\label{eq: CM condition indep}
    \forall \pi_i \in \Pi_i:\qquad 
    \pi_i \notin \operatorname{co}\!\big(\Pi_i \setminus \{\pi_i\}\big).
\end{equation}
However, if the prior $\pi_i^0$ is itself a feasible posterior, then by Bayes’ rule
\begin{equation*}
    \pi_i^0 \in \operatorname{co}\!\big(\Pi_i \setminus \{\pi_i^0\}\big),
\end{equation*}
and condition~\eqref{eq: CM condition indep} fails. Consequently, mechanisms of the form~\eqref{eq: transfer CM} cannot be applied ``off the shelf’’, whereas data-driven VCG mechanisms align incentives in \emph{all} environments.

While mechanisms of the form~\eqref{eq: transfer CM} are not applicable in many environments of interest, they may be useful in other settings; namely, when an analogue of condition~\eqref{eq: CM condition} holds for a noisy estimator.

%%%%%%%%%%%%%%%%%%%%%%%%%%%%%%%%%%%%%%%%%%%%%%%%%%%%
\subsection{Extensions and Discussion}\label{sec: Extensions and Discussion}
%%%%%%%%%%%%%%%%%%%%%%%%%%%%%%%%%%%%%%%%%%%%%%%%%%%%

\paragraph{Eliciting Additional Data from Agents.} We have assumed the designer obtains additional information about the state through user engagement, feedback, or third-party sources. Alternatively, suppose that after the allocation is determined, each agent collects additional data about the state, and a second reporting stage is introduced to elicit this information. Based on the reported data, we construct a \textit{leave-one-out} estimator $\widehat{\omega}_{-i}$ for each agent $i$, derived solely from the information reported by the other agents. The corresponding data-driven VCG transfer is, for any realization $\widehat{\omega}_{-i}^{*}$ of the estimator $\widehat{\omega}_{-i}$,
\begin{equation}\label{eq: leave-one-out transfers}
    t_i(\xi,\widehat{\omega}_{-i}^{*}) \equiv h_i(\xi_{-i}, \widehat{\omega}_{-i}^{*}) + \sum_{j \neq i} u_j(x^{*}(\xi), \widehat{\omega}_{-i}^{*}, \theta_j).
\end{equation}

Analogously to the two-stage mechanism of \citet{mezzetti2004mechanism}, in the case of full revelation of $\omega$, there is a perfect Bayesian equilibrium where all agents report truthfully at both stages. In the second stage, since an agent’s transfer is independent of the agent's report, agents are indifferent to the reporting choice.

\paragraph{Heterogeneous Priors.} While agents might be endowed with heterogeneous and private signals, we have maintained that there is a common prior $\pi^0$ over $\Omega$. We can adapt the model to allow for heterogeneous priors as follows. Each agent $i$ is endowed with a full-support prior $\pi_i^0 \in \Pi_i \subseteq \Delta(\Omega)$, with $\pi_i$ being $i$'s private information. We define an agent-specific expanded state space $(\Omega \times [0,1], \mathcal{B}(\Omega) \otimes \mathcal{B}([0,1]), \pi_i^0 \times \lambda)$, which forms agent $i$'s subjective probability space. The rest of the specification follows our baseline model. In particular, a signal of each agent $j$ is a measurable mapping $\mathbf{S}_j : \Omega \times [0,1] \rightarrow \mathcal{S}_j$, which is $j$'s private information. Each signal $\mathbf{S}_j$ defines a random variable on $i$'s expanded state space, with the law given by $P_{\mathbf{S}_j, \pi_i^0} = (\pi_i^0 \times \lambda) \circ \mathbf{S}_j^{-1}$. Observing a profile of signals $\mathbf{S}$ and a profile of signal realizations $s$ leads agent $i$ to update her beliefs to a posterior regular conditional distribution $\psi_{\mathbf{S}, \pi_i^0}(\cdot | s)$. We define $i$'s expected payoff as
\begin{equation*}
    v_i(x,\theta_i,s,\mathbf{S}, \pi_i^0) = \int_{\Omega} u_i(x,\omega, \theta_i) d \psi_{\mathbf{S}, \pi_i^0}(\omega | s).
\end{equation*}
Define $\Xi_i \equiv \{(\theta_i, s_i, \mathbf{S}_i, \pi_i^0) \in \Theta_i \times \mathcal{S}_i \times \Psi_i \times \Pi_i: s_i \in \supp P_{\mathbf{S}_i, \pi_i^0}\}$ to be $i$'s type space, and by $\Xi$ the product space across agents. We adjust the definition of efficiency as follows.

\begin{Def}[Efficient allocation rule with heterogeneous priors]\label{def: efficient allocation het priors}
    The deterministic allocation rule $x: \Xi \rightarrow X$ is efficient if, for all $\xi = (\theta, s, \mathbf{S}, \pi^0) \in \Xi$, it satisfies:
    \begin{equation}\label{eq: ex-post efficiency}
        x(\xi) \in \argmax_{x \in X} \sum_{i \in N} v_i(x,\theta_i,s, \mathbf{S}, \pi_i^0).
    \end{equation}
\end{Def}

Our implementation results readily extend to this framework and Definition \ref{def: efficient allocation het priors}. Specifically, we assume there is a commonly known measurable mapping $\widehat{\omega}: \Omega \times [0,1] \times X \to \Omega$, which remains independent of agents' signals conditional on the payoff-relevant state under the obtained outcome, for any agent $i$, signal profile $\mathbf{S} \in \Psi$, and prior $\pi_i^0 \in \Pi_i$. Under this formulation, all properties of estimators discussed in this section remain well-defined across agent types, as we always condition on the payoff-relevant state and treat the estimator as a random variable on the residual probability space $([0,1], \mathcal{B}([0,1]), \lambda)$.

\paragraph{Bayesian Interpretation of Additional Information about the State.} We defined and analyzed properties of data-driven VCG mechanisms under a primarily frequentist perspective on additional information. A Bayesian approach can also be considered. In Appendix \ref{appendix: bayesian interpretation}, we formalize additional information about the payoff-relevant state as a further signal independent of agents' signals conditionally on the payoff-relevant state. We identify two ways to define data-driven VCG mechanisms under this framework and analyze their implications for implementation. Here, we provide a brief discussion.

First, we use the posterior mean of the payoff-relevant state given the additional signal as an estimator in data-driven VCG transfers. We obtain analogous results. In particular, we establish that under a suitable version of \textit{posterior consistency} of additional signals, the corresponding sequence of posterior means forms a consistent sequence (Lemma \ref{lemma: consistency of the posterior mean}). This, in turn, yields an analogous continuity result to Theorem \ref{prop: consistent estimator implementation} (Corollary \ref{cor: posterior consistent posterior mean implementation}). 

Second, rather than directly substituting an estimate of the payoff-relevant state into agents' payoffs, we can use the posterior distribution implied by the additional signal to compute expected payoffs and define \textit{Bayesian data-driven VCG} transfers based on them (Definition \ref{def: Bayesian data-driven VCG}). We obtain analogous implementation results. In particular, posterior consistency—either uniform or pointwise with a Lipschitz condition on posteriors as in \eqref{eq: lipschitz posteriors}—suffices to establish a continuity result analogous to Theorem \ref{prop: consistent estimator implementation} (Proposition \ref{prop: posterior consistency in Bayesian data-driven VCG}).

\paragraph{State-Revealing Signals.} The results in this and the previous section apply to environments in which agents’ signals, when combined, fully reveal the state. In such settings, the designer cannot rely solely on reported signals and preference types to determine transfers when agents have strictly positive informational size, as this would again permit manipulation.

\paragraph{Interdependent Preferences.}
The arguments no longer hold if we introduce additional interdependence into agents' payoffs. Specifically, if the payoff function $u_i$ of each agent $i$ depends on the entire type profile $\theta \in \Theta$, such that $u_i: X \times \Theta \times \Omega \rightarrow \mathbb{R}$,
our results no longer go through. The failure is illustrated in Example~\ref{ex: rem Interdependent preferences} in Appendix~\ref{sec: appendix further results and derivations}.

%%%%%%%%%%%%%%%%%%%%%%%%%%%%%%%%%%%%%%%%%%%%%%%%%%%%%%%%%%%%
\section{Applications}\label{sec: applications}
%%%%%%%%%%%%%%%%%%%%%%%%%%%%%%%%%%%%%%%%%%%%%%%%%%%%%%%%%%%%

%%%%%%%%%%%%%%%%%%%%%%%%%%%%%%%%%%%%%%%%%%%%%%%%%%%%%%%%%%%%
\subsection{Advertising Auctions}\label{sec: application to click-through auctions}
%%%%%%%%%%%%%%%%%%%%%%%%%%%%%%%%%%%%%%%%%%%%%%%%%%%%%%%%%%%%

In this section, we apply our framework to a canonical model of click-through auctions \citep{edelman2007internet, varian2007position} introduced in Example \ref{ex: ad auction}. We further micro-found the model as follows. We assume there is a fixed number of impressions $M > 0$ per advertising slot.\footnote{The assumption of a fixed number of impressions is not essential. For example, instead of fixing the number of impressions, the ad display could be auctioned for a specific duration, with impressions arriving stochastically. All objects below can be defined on a per-impression basis, and all results remain valid under this interpretation.} We assume a stationary environment and model clicks as binary random variables $\mathbf{Z}_{imk} \in \{0,1\}$ drawn i.i.d. according to the Bernoulli distribution given by the click-through rate (CTR) $\omega_k \in [0,1]$ for each advertiser $i$ and slot $k$.\footnote{This assumption can also be relaxed and alternative models of user search behavior may be considered. As will be clear below, the only essential aspect of the construction is that the proportion of impressions resulting in a click when the ad of agent $i$ is displayed in position $k$ is an unbiased estimator of the click-through rate $\omega_k$.} Denote a realization of $\mathbf{Z}_{imk}$ by $z_{imk}$. Agent $i$'s ex-post payoff conditional on being allocated slot $k$ is given by $\theta_i \cdot \sum_{m=1}^M \mathbf{Z}_{imk}$. At the bidding stage, only the expected payoff matters for agents' incentives. Hence, we adopt the following payoff function for each agent $i$:
\begin{equation*}
    u_i(x, \omega, \theta_i) = M \cdot \theta_i \cdot \sum_{k=1}^K x_{ik} \cdot \omega_k,
\end{equation*}
where $M \cdot \omega_k$ represents the expected number of clicks conditional on being allocated slot $k$.  As explained in Example \ref{ex: ad auction}, the efficient outcome is given by an assortative matching of agents to slots based on their values per click and the CTRs, respectively.

Recall that in message-driven direct mechanisms, allocations and transfers depend only on reported types. This class includes per-impression payments but excludes per-click payments, and the corresponding implementation impossibility applies. Payments based on click data, by contrast, fall within the broader class of data-driven mechanisms. Under our maintained assumptions, the fraction of impressions resulting in a click is an unbiased estimator of the slot’s true CTR.

Fix a profile of values per click $\theta$ and, without loss of generality, assume it is ordered in descending order. The efficient allocation $x^{*}(\theta)$ assigns agent~$k$ to slot~$k$. Define
\begin{equation*}
    \widehat{\omega}_k^{x^{*}(\theta)}
    = \frac{1}{M} \sum_{m=1}^M \mathbf{Z}_{kmk}.
\end{equation*}
This is an unbiased estimator of $\omega_k$ and satisfies the conditional independence assumption of Assumption~\ref{assumption: estimator}.

The corresponding \emph{data-driven pivot mechanism} is $(x^{*},t)$, with transfers defined as follows. For any realization $\widehat{\omega}^{*}$ of the estimator $\widehat{\omega}^{x^{*}(\theta)}$,
\begin{equation}\label{eq: quasi ex-post payoffs ad auction}
    t_i(\theta,\widehat{\omega}^{*})
    = - M \sum_{k=i}^{K}
    \big( \widehat{\omega}^{*}_{k}-\widehat{\omega}^{*}_{k+1} \big)\theta_{k+1},
\end{equation}
for $i \le K$, and $t_i(\theta,\widehat{\omega}^{*})=0$ otherwise, where we set $\widehat{\omega}^{*}_{k}=0$ for all $k>K$.\footnote{This mechanism is commonly referred to as the VCG auction in the advertising literature; see, for example, \citep{varian2014vcg}.}

Observe that neither the allocation nor the transfers depend on agents’ reports about the state. Accordingly, the mechanism need not elicit such information, reducing communication requirements. In environments with a single advertising slot, the mechanism admits an interpretation as a per-click second-price auction.

For purposes of the induced reporting game, the transfer, taken in expectation conditional on the state $\omega$, can be treated as an \emph{ex-post} transfer. The expected transfer is given by
\begin{equation}\label{eq: expectation VCG ad auctions}
    \E \left[ t_i(\theta, \widehat{\omega}^{x^{*}(\theta)}) \mid \omega \right]
    = - M \cdot \sum_{k = i}^K \left( \omega_k - \omega_{k+1} \right) \theta_{k+1} \quad , \forall \omega \in \Omega,
\end{equation}
whenever $i \leq K$, and is zero otherwise.

We show the data-driven pivot mechanism implements the efficient decision in posterior equilibrium and satisfies individual rationality and no subsidy conditions formally introduced below. Moreover, any data-driven mechanism that satisfies these conditions must have transfers of the form in \eqref{eq: expectation VCG ad auctions} under an additional assumption.

\begin{Def}[No subsidy] Data-driven mechanism $(x,t)$ with estimator $\widetilde{\omega}$ provides no subsidy if for any agent $i \in N$,
    \begin{equation*}
        \E \left[ t_i(\theta, \widetilde{\omega}^{x(\theta)}) \mid \omega \right] \leq 0,
    \end{equation*}
    for any profile $\theta \in \theta$ and state $\omega \in \Omega$.
\end{Def}

That is, no agent is paid by the mechanism. As a consequence, the auctioneer’s revenue is non-negative. 

\begin{Def}[Individual rationality]
    Data-driven mechanism $(x,t)$ with estimator $\widetilde{\omega}$ is individually rational if for any agent $i \in N$,
    \begin{equation*}
        M \cdot \theta_i \cdot \sum_{k=1}^K \sum_{m=1}^M x_{ik}(\theta) \cdot \omega_k + \E \left[ t_i(\theta, \widetilde{\omega}^{x(\theta)}) \mid \omega \right] \geq 0,
    \end{equation*}
    for any profile $\theta \in \Theta$ and state $\omega \in \Omega$.
\end{Def}

In the result below, we assume that each agent may hold information that makes her more or less optimistic about the value of the slots, as formalized below. This condition ensures that information about click-through rates provides sufficient scope for profitable deviations in any proposed mechanism.

\begin{Assumption}[Informational richness]\label{assumption: information CTR}
    For every agent $i \in N$, there is a signal profile $\mathbf{S} \in \Psi$ and signal realizations $s_i', s_i \in \mathcal{S}_i$ and $s_{-i} \in \mathcal{S}_{-i}$ such that
    \begin{equation*}
        \E[\omega_k | \mathbf{S} = s] \geq  \E[\omega_k | \mathbf{S}_i = s_i', \mathbf{S}_{-i} = s_{-i}] \quad \forall k \in \{1,\dots,K\}
    \end{equation*}
    with a strict inequality for the $K$-th slot.
\end{Assumption}

Under Assumption~\ref{assumption: information CTR}, the pivot mechanism is the essentially unique mechanism, evaluating payoffs in expectation with respect to clicks, that satisfies our desiderata.

\begin{Prop}[Partial characterization of the pivot mechanism]\label{prop: Characterization of the VCG auction}
    The data-driven pivot mechanism implements the efficient decision $x^{*}$ in posterior equilibrium, is individually rational, and provides no subsidies. Conversely, suppose Assumption \ref{assumption: information CTR} holds and data-driven mechanism $(x^{*},t)$ with an estimator $\widetilde{\omega}$ satisfies these conditions. Then
    \begin{equation}\label{eq: expectation VCG ad auctions prop}
    \E \left[ t_i(\theta, \widetilde{\omega}^{x^{*}(\theta)}) \mid \omega \right] = - M \cdot \sum_{k = i}^K \left( \omega_k - \omega_{k+1} \right) \theta_{k+1} \quad , \omega-a.e.
\end{equation}
whenever $i \leq K$ and zero otherwise.
\end{Prop}

\begin{Rem}
    The remaining multiplicity concerns only the specification of the estimator.\footnote{For example, one may disregard a random subset of click data and still obtain an unbiased estimator.} Indeed, the transfer in \eqref{eq: expectation VCG ad auctions} can be interpreted as the data-driven pivot mechanism under full revelation of the state $\omega$, when agents’ ex-post payoffs are specified as in \eqref{eq: quasi ex-post payoffs ad auction}. In this case, the data-driven pivot mechanism is unique up to a measure-zero set of states.
\end{Rem}

The estimator $\widehat{\omega}^{x^{*}(\theta)}$ is unbiased. Since agents' payoffs are affine in $\omega$ at the reporting stage, the data-driven pivot mechanism implements $x^{*}$ by Theorem \ref{prop: unbiased}. It is easy to check the individual rationality and no subsidy conditions are also satisfied.

We prove the converse in Appendix~\ref{sec: appendix proof characterization of the VCG auction}. The argument parallels \citet{moulin1986characterizations}'s characterization of the pivot mechanism in private-value settings. We first assume that information is reported truthfully and show that, under the no-subsidy and individual-rationality conditions, any transfer scheme that elicits agents’ values per click must be a pivot mechanism. We then endogenize information reporting: implementation in posterior equilibrium implies that transfers cannot depend on agents’ reports of information. Finally, imposing individual rationality yields the expression in \eqref{eq: expectation VCG ad auctions prop}.

\begin{Rem}
In the single-item case, the data-driven pivot mechanism satisfies individual rationality and no-subsidy \emph{ex post for every realization of the click data}. The estimator of the single-slot CTR is simply the number of clicks received by the winner, so the winner~$i$’s ex-post net payoff is
\[
\theta_i \cdot \sum_{m=1}^M z_{im}
    - \max_{j \neq i} \theta_j \cdot \sum_{m=1}^M z_{im}
    = (\theta_i - \max_{j \neq i} \theta_j) \cdot \sum_{m=1}^M z_{im}
    \ge 0.
\]
By Proposition~\ref{prop: Characterization of the VCG auction}, and under Assumption~\ref{assumption: information CTR}, this is the unique data-driven mechanism based on click data that satisfies these properties.
\end{Rem}

Thus far, we have assumed that click-through rates do not depend on the identity of the assigned agent. We now relax this assumption and, for simplicity, focus on the single-item case. Suppose each agent~$i$ has an agent-specific click-through rate $\omega_i \in [0,1]$. The efficient allocation assigns the item to the agent with the highest expected payoff, that is, to agent~$i$ such that $\theta_i \, \E[\omega_i \mid \mathbf{S}=s]
    \ge \max_{j \neq i} \left\{ \theta_j \, \E[\omega_j \mid \mathbf{S}=s] \right\}$.
In this environment, information about the state is socially valuable. We continue to assume that clicks on the displayed ad are observed. 

The corresponding per-click second-price auction no longer aligns incentives. Fix an arbitrary agent~$i$ and suppose all other agents report truthfully. Conditional on others’ truthful reports, agent~$i$’s expected transfer is
\begin{equation*}
    - \max_{j \neq i} \theta_j \cdot x_i^{*}(\xi_i', \xi_{-i}) \cdot \frac{\E[\omega_j | \mathbf{S}_i' = s_i', \mathbf{S}_{-i} = s_{-i}]}{\E[\omega_i | \mathbf{S}_i' = s_i', \mathbf{S}_{-i} = s_{-i}]} \cdot M \cdot \E[\omega_i \mid \mathbf{S}=s],
\end{equation*}
where $\xi = (\theta, s, \mathbf{S}) \in \Xi$ is the true type profile and $\xi_i' = (\theta_i', s_i', \mathbf{S}_i') \in \Xi_i$ is the report of agent $i$.
Agent~$i$ now may have an incentive to misreport her information so as to inflate her own expected CTR while depressing those of other agents. By doing so, she can secure the item by promising a higher expected payment frequency, even though such payments are unlikely to materialize given her posterior beliefs.\footnote{To illustrate, consider two agents and $M=1$. Suppose that each agent~$i$ perfectly observes her own CTR $\omega_i$, while the prior over $\omega_i$ has positive variance: $\Psi_i$ is a singleton with $\mathcal{S}_i=\Omega_i$ and $\mathbf{S}_i(\omega,r)=\omega_i$ for all $\omega \in \Omega$ and $r \in [0,1]$. Assume further that $\theta_1 s_1 < \theta_2 s_2$ but $\theta_1 > \theta_2$. If agent~2 reports truthfully and agent~1 reports $\xi_1'=(\theta_1', s_1')$, her expected net payoff under per-click second-price auction is
\begin{equation*}
    \left( \theta_1 - \theta_2 \cdot \frac{s_2}{s_1'} \right) \cdot x_1^{*}(\xi_1', \xi_2) \cdot s_1
    = \left( \theta_1 - \theta_2 \cdot \frac{s_2}{s_1'} \right) \cdot \mathbbm{1}_{\{\theta_1' s_1' \geq \theta_2 s_2\}} \cdot s_1.
\end{equation*}
Truthful reporting is not a posterior equilibrium: reporting $s_1'=\theta_1'=1$ yields a strictly higher payoff.}

The data-driven pivot mechanism instead requires an estimate of the CTR of other bidders: conditional on others' truthful reports, agent~$i$’s expected transfer is
\begin{equation}\label{eq: data-driven pivot individual CTR}
    M \cdot \sum_{j \neq i} \theta_j \cdot \E[\omega_j \mid \mathbf{S} = s] (x_j^{*}(\xi_i', \xi_{-i}) - x_{j}^{*}(\xi_{-i})).
\end{equation}
In particular, whenever agent $i$ obtains the slot, the expected payment is $M \cdot \theta_j \cdot \E[\omega_j \mid \mathbf{S} = s]$, where $j$ maximizes $\theta_l \cdot \E[\omega_l \mid \mathbf{S}_i' = s_i', \mathbf{S}_{-i} = s_{-i}]$ among agents $l \neq i$. Such a transfer scheme is infeasible without additional data.\footnote{\label{footnote: sacrifice efficiency}We could construct payments based on clicks in the current auction environment at the expense of  efficiency. For instance, consider two agents and a lottery that selects the efficient allocation with probability $p \in (0,1)$, assigning the slot randomly otherwise. Suppose the efficient allocation gives the slot to agent $i$. Under the lottery, agent $i$ receives the slot with probability $(1 + p)/2$, and agent $j$ with probability $(1 - p)/2$. Setting $i$'s payment as $2 \theta_j/(1 - p)$, whenever agent $j$'s ad is clicked when assigned the slot (occurring with probability $\omega_j$), makes $i$'s expected payment match an analogue of \eqref{eq: data-driven pivot individual CTR}. This method implements the inefficient allocation with probability $(1 - p) / 2$, which can be made arbitrarily close to 0 as $p$ approaches 1. However, as $p$ approaches 1, the payment  approaches infinity.}%

%%%%%%%%%%%%%%%%%%%%%%%%%%%%%%%%%%%%%%%%%%%%%%%%%%%%%%%%%%%%
\subsection{Product Recommendations by AI Shopping Assistants}\label{sec: efficient product recommendations}
%%%%%%%%%%%%%%%%%%%%%%%%%%%%%%%%%%%%%%%%%%%%%%%%%%%%%%%%%%%%

Generative AI environments facilitate richer information flows than traditional online advertising. Consider, for example, a user interacting with an AI shopping assistant.\footnote{For example, Amazon has recently introduced AI tools such as \emph{Help Me Decide} and \emph{Rufus AI} that provide product recommendations \citep{Rufus1,Rufus2}.
%see \href{https://www.aboutamazon.com/news/retail/amazon-things-to-buy-help-me-decide-gen-ai}{https://www.aboutamazon.com/news/retail/amazon-things-to-buy-help-me-decide-gen-ai} and \href{https://www.aboutamazon.com/news/retail/how-to-use-amazon-rufus}{https://www.aboutamazon.com/news/retail/how-to-use-amazon-rufus}. 
Other retailers, including Walmart, have also deployed AI shopping assistants \citep{Sparky1}.} %; see \href{https://www.walmart.com/cp/sparky/5291783}{https://www.walmart.com/cp/sparky/5291783}. 
Through iterative queries, the assistant elicits preferences before making a recommendation. This learning process continues post-recommendation: the platform updates its beliefs based on follow-up questions, direct feedback, or transaction data. We consider a platform that seeks to maximize the joint welfare of advertisers and the user.

%A natural question is whether such pre-allocation information acquired through the interaction with the user can be used to align agents’ incentives without any additional data. We show that the answer is negative unless the information obtained by the platform effectively renders the agents’ private information about the state obsolete, thereby reducing the environment to one with private values. Thus, additional information that the platform incorporates into only the transfer rules is essential for restoring incentive compatibility in these environments.

The mechanism design environment begins with a user query. Analogous to our click-through auction setting, the set of agents $N$ consists of advertisers. Each advertiser $i \in N$ obtains a payoff $\theta_i \in \Theta_i \equiv \mathbb{R}_{+}$ when her product is recommended. The user derives payoff $\omega_i \in \mathbb{R}$ from product~$i$ and $\omega_0 \in \mathbb{R}$ from an outside option, which we normalize to zero: $\omega_0 = 0$. Consequently, recommending product $i$ generates a total surplus of $\theta_i + \omega_i$. We denote the user's preference profile over the products by $\omega = (\omega_1, \dots, \omega_n) \in \Omega \equiv \mathbb{R}^{n}$.

The agents and the platform share a common prior over user preferences, $\omega \sim N(0, \Sigma)$ for a positive definite covariance matrix $\Sigma$. Each agent $i$ is endowed with a private signal $\mathbf{S}_i = \omega + \Sigma_i \mathbf{Z}_i$, where $\mathbf{Z}_i \sim_{i.i.d.} N(0, I)$ is independent of $\omega$. The matrix $\Sigma_i$ is the Cholesky factor of the signal's positive definite covariance matrix.

Before issuing a recommendation, the platform engages the user in a \emph{conversation}. This interaction generates an additional signal $\mathbf{C}_0 = \omega + \Sigma^{0}\mathbf{Z}^{0}$,
where $\mathbf{Z}^{0} \sim N(0,I)$ is independent of $\{\mathbf{Z}_i\}_{i=1}^n$ and of $\omega$. As before, we denote a realization by $c_0 \in \mathcal{C} \equiv \mathbb{R}^{n}$. This signal summarizes the context elicited by the assistant and provides information regarding $\omega$. For simplicity of exposition, we assume the random variable $\mathbf{C}_0$ is commonly known, while its realization is proprietary information of the platform.

We initially assume the assistant recommends a single product. The set of feasible allocations is the simplex $X \equiv \Delta_{n} \subset \mathbb{R}^{n+1}$, where $x_i$ denotes the probability of recommending product $i$, and $x_0$ denotes the probability of recommending the outside option. We extend the framework to multi-round interactions afterwards.

The platform determines the allocation based on advertiser values $\theta$ and the signals $(s, c_0, \mathbf{S}, \mathbf{C}_0)$ regarding user preferences. Let $\xi = (\theta, s, \mathbf{S})$ denote the agents' private information. The efficient allocation rule $x^{*} : \Xi \times \mathcal{C} \to X$ maximizes total expected surplus conditional on the available information. Thus, for any $(\xi, c_0)$, the rule solves:
\begin{equation*}
    x^{*}(\xi, c_0)
    \in \operatorname*{argmax}_{x \in X} \left\{
    \sum_{i \in N} \left(
        \theta_i +
        \mathbb{E}[\omega_i \mid \mathbf{S} = s,\, \mathbf{C}_0 = c_0]
    \right) x_i
    \right\}.
\end{equation*}
    
We first ask whether the allocation rule $x^{*}$ is implementable using only agents' reports and the context signal $c_0$. To this end, we define a \emph{message-driven mechanism with initial context} as a mechanism where both the allocation and transfers are conditioned on the realized signal $c_0$. This environment departs from our main model: the platform's signal constitutes a direct input to the social choice function.

Our goal is to provide agents with incentives to report truthfully that are robust to the details of the user’s interaction with the shopping assistant. We therefore require implementation \emph{ex post} with respect to all information used in the social decision,\footnote{Formally, a message-driven direct mechanism with initial context $(x,t)$ \emph{permits implementation in posterior equilibrium} if, for each agent $i \in N$, each type profile $\xi=(\theta,s,\mathbf{S})\in\Xi$, and each realization $c_0$ of the additional signal $\mathbf{C}_0$,
\begin{equation*}
    v_i\!\left(x(\xi,c_0),\theta_i,s,c_0,\mathbf{S},\mathbf{C}_0\right)+t_i(\xi,c_0)
    \;\geq\;
    v_i\!\left(x(\xi_i',\xi_{-i},c_0),\theta_i,s,c_0,\mathbf{S},\mathbf{C}_0 \right) + t_i(\xi_i',\xi_{-i}, c_0),
    \quad \forall\,\xi_i'\in\Xi_i.
\end{equation*}
} including the additional signal~$\mathbf{C}_0$.\footnote{Alternative objectives are possible. For instance, one could ask whether the efficient allocation can be implemented when agents evaluate payoffs in expectation with respect to $\mathbf{C}_0$. If the joint signal $(\mathbf{S},\mathbf{C}_0)$ were common knowledge and the stochastic relevance condition of \citet{cremer1988full} were satisfied, then lottery mechanisms in the spirit of \citet{cremer1988full} could implement the efficient allocation without additional information about the state. Such approaches, however, rely on strong informational assumptions: in practice, the platform’s procedure for generating $\mathbf{C}_0$ through its interaction with the user may be proprietary.} We show that this form of implementation is impossible to achieve with message-driven mechanisms.

\begin{Prop}[Impossibility without additional data]\label{prop: impossibility without additional data product recs}
    There is no message-driven mechanism with initial context that implements $x^{*}$ in posterior equilibrium.
\end{Prop}

\begin{proof}
    The environment can be recast by augmenting the set of agents to include the platform, with payoff $u(x,\omega) = \sum_{i \in N} x_i \omega_i$. The platform is endowed with information $(\mathbf{C}_0,c_0)$ about the state and has no private preference type. This setting fits within our main model and constitutes a special case of the environment studied by \citet{jehiel2001efficient} and \cite{jehiel2006limits}. The result then follows directly from their impossibility theorems.
\end{proof}

In contrast to classical advertising auctions, the platform observes substantially richer post-allocation information. Following an initial recommendation, the assistant elicits additional signals regarding user preferences, such as direct feedback, follow-up queries, or requests for alternatives. Crucially, these interactions reveal information about the user's underlying preference profile $\omega$ that extends beyond the value of the recommended item. Moreover, the interaction is inherently dynamic: the assistant engages the user over multiple rounds, updating recommendations in response to feedback. This sequential structure generates a stream of preference-relevant information unavailable in standard one-shot allocation environments.

\begin{figure}[t]
\centering
\begin{tikzpicture}[
 node distance = 0mm and 0.01\linewidth,
    box/.style = {inner xsep=0pt, outer sep=0pt,
                  text width=0.18\linewidth,
                  align=center, font=\footnotesize}
                    ]
\node (n1) [box] {};
\node (n2) [box, below right=of n1.north east] {};
\node (n3) [box, below right=of n2.north east] {};
\node (n4) [box, below right=of n3.north east] {}; % spacing only
\node (n5) [box, below right=of n4.north east] {};

% --- baseline: solid (up to Rec 2) ---
\draw[black, very thick]
    (n1.north west) -- (n2.north east) -- (n3.north east);

% --- dots in place of line between Rec 2 and Rec T ---
\draw[black, dotted, very thick]
    (n3.north east) -- (n5.north east);

% --- baseline: solid with arrow under Rec T only ---
\draw[black, very thick, -latex]
    (n5.north west) -- (n5.north east) -- +(3mm,0);

% --- labels (no label/tick for n4) ---
\draw (n1.north) -- +(0,3mm)
      node[above] {\begin{tabular}{c} Type reports \\ Context $\mathbf{C}_0$ \end{tabular}};
\draw (n2.north) -- +(0,3mm)
      node[above] {\begin{tabular}{c} Rec.\ 1 \\ Response $\mathbf{C}_1$ \end{tabular}};
\draw (n3.north) -- +(0,3mm)
      node[above] {\begin{tabular}{c} Rec.\ 2 \\ Response $\mathbf{C}_2$ \end{tabular}};
% no tick/label over the dots
\draw (n5.north) -- +(0,3mm)
      node[above] {\begin{tabular}{c} Rec.\ T \\ Final response $\mathbf{C}_T$ \end{tabular}};

\end{tikzpicture}
\caption{Timeline of a conversation and product recommendations of an AI shopping assistant.}
\label{fig: timeline conversation AI assistant}
\end{figure}

We extend the framework to a finite horizon $T \geq 1$. In each period $\tau \in \{1, \dots, T\}$, the assistant selects a deterministic allocation $x_{\tau} \in X$.\footnote{In the efficient allocation, ties occur on a set of Lebesgue measure zero.} Following the recommendation, the user's interaction generates a new signal $\mathbf{C}_{\tau} = \omega + \Sigma^{\tau}(h_{\tau}, x_{\tau}) \mathbf{Z}^{\tau}$. Here, $\mathbf{Z}^{\tau} \sim N(0, I)$ is independent of all other random variables. We define the public history at the beginning of period $\tau$ as:
\[
    h_{\tau}
    \equiv
    (c_0, x_1, c_1, \dots, x_{\tau-1}, c_{\tau-1}) \in H_{\tau},
\]
where $H_{\tau}$ denotes the set of feasible histories. $\Sigma^{\tau}(h_{\tau}, x_{\tau})$ is a deterministic matrix scaling the noise based on the history up to period $\tau$ and the allocation in period $\tau$. This timeline is illustrated in Figure \ref{fig: timeline conversation AI assistant}.

We assume that all participants accrue flow payoffs in each round and do not discount future payoffs. For advertisers, value is generated whenever their product is recommended; even absent an immediate transaction, an impression captures user attention, facilitating branding and information transmission. Similarly, in AI-assisted shopping, the user derives utility from each interaction as recommendations clarify attributes, trade-offs, or alternatives. Consequently, we model payoffs as accumulating over the horizon rather than realizing solely upon a final purchase. Finally, we fix the interaction length $T$ to reflect design constraints such as user attention spans and computational budgets. While stylized, this formulation captures the essential dynamics of the problem; we leave endogenous stopping times and richer search models for future research.

For any type profile $\xi = (\theta, s, \mathbf{S}) \in \Xi$, the social planner solves the following dynamic optimization problem:
\begin{equation*}
    \max_{\{x_{\tau}: H_{\tau} \to X\}_{\tau=1}^T}
    \mathbb{E} \left[
        \sum_{\tau = 1}^T
            \sum_{i \in N} (\theta_i + \omega_i) x_{i\tau}(h_{\tau})
        \,\middle|\,
        \mathbf{S} = s,\, \mathbf{C}_0 = c_0
    \right].
\end{equation*}
Let $x^{*}$ denote an optimal policy solving this problem.

Our objective is to implement the allocation rule $x^{*}$. We restrict attention to mechanisms that elicit advertisers’ preferences and information only once at the outset, as illustrated in Figure~\ref{fig: timeline conversation AI assistant}. Nevertheless, we seek to implement the optimal policy robustly: at any history $h_{\tau}$ of the sequential interaction, no advertiser would profitably deviate from truthful reporting if permitted to revise her report, with that report then remaining fixed for the remainder of the interaction with the user.

To formalize this notion, we consider a dynamic extension of data-driven mechanisms equipped with history-dependent state estimators $\widehat{\omega}_{\tau}^{h_{\tau}}$, where $\widehat{\omega}_{\tau}^{h_{\tau}}$ is used to compute transfers in period~$\tau$ following history~$h_{\tau}$. In each round~$\tau$, the mechanism specifies an allocation $x_{\tau}$ and transfers $t_{\tau}$ as functions of agents’ initial reports, the realized history, and the corresponding state estimate.

\begin{Def}[Posterior equilibrium implementation at a history]
Dynamic data-driven direct mechanism $(x, t)$ \textit{permits implementation in posterior
equilibrium at round $\tau$ and history $h_{\tau} \in H_{\tau}$} if for every
advertiser $i \in N$ and every type profile $\xi = (\theta, s, \mathbf{S}) \in
\Xi$,\footnote{Conditioning on $h_{\tau}$ refers to the belief corresponding to the history. We omit conditioning on the allocation for notational brevity.}
\begin{multline*}
    \sum_{\tau' \geq \tau}
    \mathbb{E}\!\left[
        x_{i\tau'}^{*}(\xi, h_{\tau'}) \, \theta_i
        +
        t_{i\tau'}(\xi, h_{\tau'}, \widehat{\omega}_{\tau'}^{h_{\tau'}})
        \,\middle|\,
        \mathbf{S} = s,\,
        h_{\tau}
    \right] \\
    \;\geq\;
    \sum_{\tau' \geq \tau}
    \mathbb{E}\!\left[
        x_{i\tau'}^{*}(\xi_i', \xi_{-i}, h_{\tau'}) \, \theta_i
        +
        t_{i\tau'}(\xi_i', \xi_{-i}, h_{\tau'}, \widehat{\omega}_{\tau'}^{h_{\tau'}})
        \,\middle|\,
        \mathbf{S} = s,\,
        h_{\tau}
    \right]
    \quad
    \forall \xi_i' \in \Xi_i.
\end{multline*}
\end{Def}

The insight of Proposition~\ref{prop: impossibility without additional data product recs} continues to apply: the stream of signals used in the recommendation process alone is insufficient to align incentives. The user’s final response is therefore indispensable. However, as we show below, at any prior history the subsequent conversation can be used to form more precise estimates of the state entering the transfers, thereby concentrating data-driven transfers around their ex-post version.

To this end, define the sequence of state estimates $\widehat{\omega}_{\tau}^{h_{\tau}*}$ by
\begin{equation*}
    \widehat{\omega}_{\tau}^{h_{\tau}*}
    = \frac{1}{T-\tau+1}\sum_{\tau'=\tau}^{T} c_{\tau'},
\end{equation*}
where $c_{\tau'}$ denotes the realized signal in period~$\tau'$. With a slight abuse of notation, the corresponding estimator is
\begin{equation*}
    \widehat{\omega}_{\tau}^{h_{\tau}}
    = \frac{1}{T-\tau+1}\sum_{\tau'=\tau}^{T} \mathbf{C}_{\tau'},
\end{equation*}
where the data-generating process for $\mathbf{C}_{\tau'}$ is evaluated conditional on the history~$h_{\tau}$ and allocation $x_{\tau}$. The estimator is conditionally independent of agents’ signals $\mathbf{S}$ given the state~$\omega$, history~$h_{\tau}$, and allocation $x_{\tau}$, and is pointwise unbiased: $\E\!\left[\widehat{\omega}_{\tau}^{h_{\tau}} \mid \omega, h_{\tau}, x_{\tau} \right] = \omega$ for any $\omega\in\Omega, h_{\tau}\in H_{\tau}, x_{\tau} \in X$.

The \emph{data-driven dynamic team mechanism} $(x^{*},t)$ defines period-$\tau$ transfers analogously to the dynamic team mechanism of \citet{athey2013efficient} for private-value environments:
\begin{equation*}
    t_{i\tau}(\xi,h_{\tau},\widehat{\omega}^{*})
    =
    \widehat{\omega}_{i\tau}^{h_{\tau}*} x^{*}_{i\tau}(\xi,h_{\tau})
    +
    \sum_{j \neq i}
    \big(\theta_j + \widehat{\omega}_{j\tau}^{h_{\tau}*}\big)
    x^{*}_{j\tau}(\xi,h_{\tau})
    + h_i(\xi_{-i},h_{\tau},\widehat{\omega}^{*}),
\end{equation*}
The term $h_i(\xi_{-i},h_{\tau},\widehat{\omega}^{*})$ is an arbitrary integrable function of other agents’ reports, the current and past histories, and the state estimates. Under this transfer rule, each agent is a residual claimant of the period-$\tau$ social surplus.

In Appendix~\ref{appendix: revenue AI shopping}, we propose a version of the data-driven dynamic team mechanism, akin to a pivot adjustment, that ensures the platform does not, in expectation, pay any advertiser more than her marginal contribution to the user’s surplus. We also discuss when individual rationality constraints are satisfied under this mechanism.

We obtain the following result. The proof is provided in Appendix~\ref{sec: appendix proof dynamic implementation product recs}.

\begin{Prop}[Dynamic implementation]\label{prop: dynamic implementation product recs}
    The data-driven dynamic team mechanism permits implementation in posterior equilibrium at each period $\tau \geq 1$ and history $h_{\tau} \in H_{\tau}$.
\end{Prop}

%%%%%%%%%%%%%%%%%%%%%%%%%%%%%%%%%%%%%%%%%%%%%%%%%%%%%%%%%%%%
\subsection{Procurement Auctions}\label{sec: procurement auctions}
%%%%%%%%%%%%%%%%%%%%%%%%%%%%%%%%%%%%%%%%%%%%%%%%%%%%%%%%%%%%

We show that the additional data used in our contingent transfers need not arise solely from interactions between non-strategic users and the mechanism’s output. In particular, in a procurement setting with a spot market, such data can instead be generated through an auxiliary game played by the agents themselves.

We consider a setting with a single buyer and a set of sellers $N$ participating in a procurement auction for the delivery of one unit of a good in a future period. Each seller $i \in N$ offers a potentially differentiated version of the good and can supply it at constant marginal cost $c_i(\theta_i,\omega)$, where $\theta_i \in \Theta_i$ is a privately known cost component and $\omega \in \Omega$ is a common cost shock realized only at the time of delivery. The buyer derives a commonly known gross payoff $b_i \in \mathbb{R}_{+}$ from procuring one unit from seller~$i$.\footnote{For example, a government agency may run a procurement auction for vehicles, with each seller offering a specific brand–model configuration.} Each seller~$i$ is endowed with a private signal $\mathbf{S}_i \in \Psi_i$ informative about the common shock~$\omega$. For example, sellers may possess private information about future input prices or labor market conditions that affect the costs of all firms.

An allocation consists of assignment probabilities $x \in X=\Delta_{n-1}$ and transfers $t \in \mathbb{R}^n$. Seller~$i$’s ex-post utility is
\begin{equation*}
    U_i(x,\omega,\theta_i,t_i)
    = t_i - x_i \cdot c_i(\theta_i,\omega),
\end{equation*}
while the buyer’s ex-post payoff is
\begin{equation*}
    U_0(x,\omega,t)
    = \sum_{i \in N} x_i \cdot b_i - \sum_{i \in N} t_i.
\end{equation*}

For any type profile $\xi=(\theta,s,\mathbf{S})\in\Xi$, an efficient allocation $x^{*}(\xi)$ solves
\begin{equation*}
    \max_{x \in X}
    \; \E \!\left[
        \sum_{i \in N} x_i \big( b_i - c_i(\theta_i,\omega) \big)
        \,\middle|\,
        \mathbf{S}=s
    \right].
\end{equation*}
Equivalently, up to tie-breaking, the contract is awarded to the seller~$i$ maximizing $\E \!\left[ b_i - c_i(\theta_i,\omega) \,\middle|\, \mathbf{S}=s \right]$.

We implement the efficient allocation using the data-driven pivot mechanism. Suppose the buyer obtains an estimate $\hat{c}_j$ of seller~$j$’s cost. When seller~$i$ wins the auction, her transfer is
\begin{equation*}
    t_i(\theta,s,\mathbf{S};\hat{c})
    = b_i - b_j + \hat{c}_j,
\end{equation*}
where $j$ denotes the seller maximizing $\E \!\left[
        b_k - c_k(\theta_k,\omega)
        \,\middle|\,
        \mathbf{S}_{-i}=s_{-i}
    \right]$ over $k \in N \setminus \{i\}$. The estimate $\hat{c}_j$ may be obtained from several sources. If uncertainty arises from input prices, the buyer may estimate these prices and substitute them into the cost functions. When available, the buyer may also rely on accounting-based cost estimates.\footnote{The role of accounting cost estimates in procurement and regulation has been emphasized by \citet{laffont1986using,laffont1987auctioning} and subsequent work; these papers, however, do not study their use for implementing efficient allocations under interdependent values.} Finally, in many environments sellers also compete in a consumer spot market. Observing prices and quantities in that market, together with a maintained model of conduct, allows the buyer to estimate marginal costs using standard methods from empirical industrial organization \citep{berry1994estimating,berry1995automobile}.

Suppose firms compete in prices in a differentiated-products consumer market under complete information, as commonly assumed in this literature \citep{berry1995automobile}. Fix agent~$i$ as the winner of the procurement auction. For any other firm $j \neq i$, profits are given by
\begin{equation*}
    \pi_j(p) = D_j(p)\,(p_j - c_j),
\end{equation*}
where $p_j$ denotes firm~$j$’s price and $D_j(p)$ its demand as a function of the full price vector~$p$. Assuming an interior optimum, the first-order condition for firm~$j$ is
\begin{equation*}
    \frac{\partial D_j(p)}{\partial p_j}\,(p_j - c_j) + D_j(p) = 0.
\end{equation*}
It follows that at any equilibrium price vector $p^{*}$,
\begin{equation*}
    p_j^{*}
    = c_j + \frac{D_j(p^{*})}{\left| \tfrac{\partial D_j(p^{*})}{\partial p_j} \right|}.
\end{equation*}

Suppose now that the buyer has an estimate of the demand function $\hat{D}_j(\cdot)$ and observes equilibrium prices $p^{*}$ from the consumer market. Assuming these prices arise from a Nash equilibrium of the pricing game, firm~$j$’s marginal cost can be estimated as:\footnote{For notational convenience, we suppress the dependence on the auction outcome.}
\begin{equation*}
    \hat{c}_j
    = p^{*}_j - \frac{\hat{D}_j(p^{*})}
    {\big| \tfrac{\partial \hat{D}_j(p^{*})}{\partial p_j} \big|}.
\end{equation*}

To illustrate this construction, consider a simple logit demand specification,
\begin{equation*}
    D_j(p) = M \cdot s_j(p;\alpha,\beta)
    = M \cdot \frac{\exp(x_j' \beta - \alpha p_j)}
    {1 + \sum_{k} \exp(x_k' \beta - \alpha p_k)},
\end{equation*}
where $M$ denotes market size, $s_j(p;\alpha,\beta)$ is firm~$j$’s market share, $x_j$ is a vector of observable product characteristics, and $\alpha$ and $\beta$ are demand parameters. This specification can be micro-founded using a random utility model \citep{berry1994estimating}. Under this functional form, given estimated demand parameters $(\hat{\alpha},\hat{\beta})$, the marginal cost estimate is
\begin{equation}\label{eq: MC estimate logit}
    \hat{c}_j
    = p^{*}_j - \frac{1}{\hat{\alpha}
    \big(1 - s_j(p^{*};\hat{\alpha},\hat{\beta})\big)}.
\end{equation}

If the demand parameter estimates $(\hat{\alpha},\hat{\beta})$ are consistent for $(\alpha,\beta)$, then the implied marginal-cost estimator $\hat{c}_j$ is also consistent. Importantly, that a firm sets prices strategically, both in the consumer spot market and in anticipation of its transfer in the procurement auction, does not affect this conclusion, provided the observed prices arise from an equilibrium of the auxiliary pricing game.\footnote{One way to obtain consistent demand estimates is to estimate demand in markets for the same class of products in which the firms under consideration do not operate. For example, demand for vehicles may be estimated using data from geographic regions without participation by the relevant firms, under the assumption that consumer preferences are invariant across regions. In such settings, consistency follows from standard sampling assumptions in those markets. Further, even when demand is estimated in markets where the same firms operate, the resulting estimates remain valid under standard instrumental-variables assumptions, independently of any equilibrium model of price-setting, provided supply-side moments are not used in estimation \citep{berry2021foundations}. In this case, however, the sampling assumptions required for law-of-large-numbers arguments are less likely to be satisfied.}

We obtain the following immediate corollary of our implementation continuity result.

\begin{Cor}[Procurement auction with a consumer spot market]
    Suppose consumer demand follows the logit specification and $\hat{\alpha}$ and  $\hat{\beta}$ are consistent estimates of $\alpha$ and $\beta$, respectively, regardless of which seller obtains the contract. Then the data-driven pivot mechanism with cost estimates given by \eqref{eq: MC estimate logit} implements the efficient allocation in $\epsilon$-posterior equilibrium, where $\epsilon \geq 0$ can be made arbitrarily small as $\hat{\alpha} \rightarrow_p \alpha$ and $\hat{\beta} \rightarrow_p \beta$.
\end{Cor}

%%%%%%%%%%%%%%%%%%%%%%%%%%%%%%%%%%%%%%%%%%%%%%%%%%%%%%%%%%%%
\section{Discussion and Conclusion}\label{sec: discussion and conclusion}
%%%%%%%%%%%%%%%%%%%%%%%%%%%%%%%%%%%%%%%%%%%%%%%%%%%%%%%%%%%%

We offered an approach to mechanism design that harnesses the natural flow of information in digital environments and beyond. By conditioning transfers on post-allocation data, we showed how to achieve implementation even in challenging multi-dimensional settings. Our framework provides a foundation for designing practical mechanisms in modern applications where rich feedback data is readily available. 

Several questions remain open for future research. One natural direction is to study implementation via message-driven mechanisms in Bayesian equilibrium. With correlated types, efficient implementation may be possible under stochastic relevance conditions \citep{cremer1988full, mcafee1992correlated, mclean2004informational}. When these conditions fail, however, it remains unclear whether message-driven mechanisms can align incentives. An alternative approach is to explore the use of linking mechanisms \citep{jackson2007overcoming, matsushima2010role}.

Another promising direction concerns the construction of the estimator. In this paper, we allow the estimator to depend on the realized allocation but restrict agents’ ability to influence it through strategic misreporting. An important avenue for future research is to further endogenize the estimator and relax the conditional-independence assumption. In some environments, agents may have incentives to misreport in order to induce allocations that generate less informative additional data, thereby weakening incentive constraints. This creates a trade-off between efficiency and incentive alignment: maintaining truthful revelation may require sacrificing efficiency through randomized allocations. Formalizing and analyzing this trade-off remains an interesting direction for future work.

Building on the preceding discussion and our application to recommendation systems, a promising direction for future research is the study of dynamic mechanisms in which estimators evolve over time, either within a single environment or across multiple environments. When current allocations affect future estimators, the challenges identified above persist. At the same time, as our application highlights, dynamic settings may provide the designer with richer information about the state and a history of agents’ reports, which can facilitate incentive alignment. Developing fully dynamic data-driven VCG mechanisms, building on the dynamic pivot mechanisms of \citet{bergemann2010dynamic} and the dynamic team mechanism of \citet{athey2013efficient}, is therefore an important avenue for future research.

%%%%%%%%%%%%%%%%%%%%%%%%%%%%%%%%%%%%%%%%%%%%%%%%%%%%%%%%%%%%%%%%%%%%%%%%%%%%%%
\begingroup
\setstretch{1}
\bibliographystyle{apalike}
\bibliography{MD_LLM_bib}
\endgroup
%%%%%%%%%%%%%%%%%%%%%%%%%%%%%%%%%%%%%%%%%%%%%%%%%%%%%%%%%%%%%%%%%%%%%%%%%%%%%%

%%%%%%%%%%%%%%%%%%%%%%%%%%%%%%%%%%%%%%%%%%%%%%%%%%%%%%%%%%%%
\appendix
%%%%%%%%%%%%%%%%%%%%%%%%%%%%%%%%%%%%%%%%%%%%%%%%%%%%%%%%%%%%

%%%%%%%%%%%%%%%%%%%%%%%%%%%%%%%%%%%%%%%%%%%%%%%%%%%%%%%%%%%%
\section{Proofs Appendix}\label{sec: appendix proofs}
%%%%%%%%%%%%%%%%%%%%%%%%%%%%%%%%%%%%%%%%%%%%%%%%%%%%%%%%%%%%

%%%%%%%%%%%%%%%%%%%%%%%%%%%%%%%%%%%%%%%%%%%%%%%%%%%%%%%%%%%%
\subsection{Theorem \ref{prop: consistent estimator implementation}}\label{sec: appendix proof consistent estimator implementation}
%%%%%%%%%%%%%%%%%%%%%%%%%%%%%%%%%%%%%%%%%%%%%%%%%%%%%%%%%%%%

Fixing an estimator $\widehat{\omega}$, for each agent $i \in N$, with a slight abuse of notation, we define $\overline{t}_i(x, \theta_{-i}, s, \mathbf{S})$ to be the expected transfer under the data-driven VCG mechanism for a profile of signals $\mathbf{S} \in \Psi$, signal realizations $s \in \mathcal{S}$, allocation $x \in X$, and preference types $\theta_{-i} \in \Theta_{-i}$, omitting the $h_i$ component of data-driven VCG payments:
\begin{equation*}
    \overline{t}_i(x, \theta_{-i}, s, \mathbf{S}) \equiv \sum_{j \neq i} \E[u_j(x, \widehat{\omega}^x, \theta_j) | \mathbf{S} = s] = \sum_{j \neq i} \int_{\Omega} \int_{[0,1]} u_j(x, \widehat{\omega}(\omega, r, x), \theta_j) d \lambda(r) d \pi_{\mathbf{S}}(\omega | s).
\end{equation*}

The first lemma establishes that for any data-driven VCG mechanism for the estimator $\widehat{\omega}$, implementation in an $\epsilon$-posterior equilibrium is feasible for $\epsilon$ no larger than a constant multiple of the distance between the expected VCG transfers in the ex-post case and those obtained under $\widehat{\omega}$.

\begin{Lemma}[Bound on $\epsilon$]\label{lemma: bound on epsilon}
    Fix an estimator $\widehat{\omega}$. Then there is an $\epsilon >0$ with
    \begin{equation*}
        \epsilon \leq 2 \max_{i \in N} \sup_{\xi=(\theta, s, \mathbf{S}) \in \Xi, x \in X} \left| \sum_{j \neq i} v_j(x, \theta_j, s, \mathbf{S}) - \overline{t}_i(x, \theta_{-i}, s, \mathbf{S}) \right|,
    \end{equation*}
    such that every data-driven VCG mechanism for $\widehat{\omega}$ permits implementation in $\epsilon$-posterior equilibrium. 
\end{Lemma}

\begin{proof}
    Fix an estimator $\widehat{\omega}$, a data-driven VCG mechanism for $\widehat{\omega}$, and types $\xi \in \Xi$. Let $i \in N$ be arbitrary. Suppose agents other than $i$ report truthfully. Define $\epsilon_i(\xi)$ as the utility loss from reporting truthfully:
    \begin{equation*}
        \epsilon_i(\xi) \equiv \sup_{x \in X} v_i(x, \theta_i, s, \mathbf{S}) + \overline{t}_i(x, \theta_{-i}, s, \mathbf{S}) - v_i(x^{*}(\xi), \theta_i, s, \mathbf{S}) - \overline{t}_i(x^{*}(\xi), \theta_{-i}, s, \mathbf{S}) \geq 0.
    \end{equation*}
    Let $\epsilon_i \equiv \sup_{\xi \in \Xi} \epsilon_i(\xi)$ and $\epsilon \equiv \max_{i \in N} \epsilon_i$. By construction, truthful reporting constitutes an $\epsilon$-posterior equilibrium. Further, observe that
    \begin{align*}
        \epsilon_i(\xi) &= \sup_{x \in X} \sum_{j \in N} v_j(x, \theta_j, s, \mathbf{S}) + \overline{t}_i(x, \theta_{-i}, s, \mathbf{S})
        - \sum_{j \neq i} v_j(x, \theta_j, s, \mathbf{S}) \\
        &\quad - \sum_{j \in N} v_j(x^{*}(\xi), \theta_j, s) - \overline{t}_i(x^{*}(\xi), \theta_{-i}, s, \mathbf{S})
        + \sum_{j \neq i} v_j(x^{*}(\xi), \theta_j, s, \mathbf{S}).
    \end{align*}
    Moreover, since $x^{*}$ is efficient, $\sum_{j \in N} v_j(x, \theta_j, s, \mathbf{S}) \leq \sum_{j \in N} v_j(x^{*}(\xi), \theta_j, s, \mathbf{S})$ for any $x \in X$. We obtain an upper bound:
    \begin{align*}
        \epsilon_i^m &= \sup_{\xi \in \Xi} \epsilon_i(\xi) \leq 2 \sup_{\xi=(\theta, s, \mathbf{S}) \in \Xi, x \in X} \left| \sum_{j \neq i} v_j(x, \theta_j, s, \mathbf{S}) - \overline{t}_i(x, \theta_{-i}, s, \mathbf{S}) \right|.
    \end{align*}
    Taking a maximum over $i \in N$ on both sides, we obtain the claim.
\end{proof}

Next, we show that the obtained bound can be further bounded from above by the expected distance between the estimator's value and the payoff-relevant state $\omega$ while taking the appropriate suprema.

\begin{Lemma}[Bound via the expected error]\label{lemma: bound via expected error}
    Suppose $u_j$ is Lipschitz in $\omega$ uniformly in $x$ and $\theta_j$ with a Lipschitz constant $L_j$ for each agent $j \in N$. Then, for each agent $i \in N$,
    \begin{align}
    &\sup_{\xi \in \Xi, x \in X} \left| \sum_{j \neq i} v_j(x, \theta_j, s, \mathbf{S}) - \overline{t}_i(x, \theta_{-i}, s, \mathbf{S}) \right| \\
    &\leq \sum_{j \neq i} L_j \sup_{\mathbf{S} \in \Psi, s \in \mathcal{S}, x \in X}  \int_{\Omega} \int_{[0,1]} d_{\Omega}(\widehat{\omega}(\omega, r, x), \omega) \, d \lambda(r) \, d \pi_{\mathbf{S}}(\omega | s) \label{eq:epsilon_bound_1} \\
    &\leq \sum_{j \neq i} L_j \sup_{\omega \in \Omega, x \in X} \int_{[0,1]} d_{\Omega}(\widehat{\omega}(\omega, r, x), \omega) \, d \lambda(r). \label{eq:epsilon_bound_2}
    \end{align}
\end{Lemma}

\begin{proof}
    Fix an arbitrary agent $i$. To prove the claim, note that using the Lipschitz property, the law of iterated expectations, and Jensen's inequality, we have $\forall \xi = (\theta, s, \mathbf{S}) \in \Xi$ and $x \in X$:
    \begin{align*}
        \left| \sum_{j \neq i} v_j(x, \theta_j, s, \mathbf{S}) - \overline{t}_i(x, \theta_{-i}, s, \mathbf{S}) \right| %\Bigg| \sum_{j \neq i} \int_{\Omega} u_j(x, \omega, \theta_j) d \pi_{\mathbf{S}}(\omega | s) \\
        %&\quad - \int_{\Omega} \int_{[0,1]} u_j(x, \widehat{\omega}(\omega, r, x), \theta_{j}) d \lambda(r) d \pi_{\mathbf{S}}(\omega | s) \Bigg| \\
        &\leq \sum_{j \neq i} \int_{\Omega} \int_{[0,1]} \left | u_j(x, \omega, \theta_j) - u_j(x, \widehat{\omega}(\omega, r, x), \theta_j) \right | d \lambda(r) d \pi_{\mathbf{S}}(\omega | s) \\
        &\leq  \sum_{j \neq i} L_j \int_{\Omega} \int_{[0,1]} d_{\Omega}(\widehat{\omega}(\omega, r, x), \omega) d \lambda(r) d \pi_{\mathbf{S}}(\omega | s),
    \end{align*}
    where $L_j$ is the Lipschitz constant on the utility function of agent $j$.
    
    Furthermore, observe that for any $\mathbf{S} \in \Psi$ and $s \in \mathcal{S}$, we have
    \begin{equation*}
        \int_{\Omega} \int_{[0,1]} d_{\Omega}(\widehat{\omega}(\omega, r, x), \omega) d \lambda(r) d \pi_{\mathbf{S}}(\omega | s) \leq \sup_{\omega \in \Omega, x \in X} \int_{[0,1]} d_{\Omega}(\widehat{\omega}(\omega, r, x), \omega) d \lambda(r).
    \end{equation*}
    Hence,
    \begin{equation*}
        \sup_{\mathbf{S} \in \Psi, s \in \mathcal{S}, x \in X}  \int_{\Omega} \int_{[0,1]} d_{\Omega}(\widehat{\omega}(\omega, r, x), \omega) d \lambda(r) d \pi_{\mathbf{S}}(\omega | s) \leq \sup_{\omega \in \Omega, x \in X} \int_{[0,1]} d_{\Omega}(\widehat{\omega}(\omega, r, x), \omega) d \lambda(r),
    \end{equation*}
    from which the result follows.
\end{proof}

\begin{Lemma}[Uniform convergence]\label{lemma: uniform convergence}
    The following statements hold:
    \begin{enumerate}
        \item Suppose $\{\widehat{\omega}_m\}_m$ is a uniformly consistent sequence of estimators. Then,
        \begin{equation*}
            \lim_{m \rightarrow \infty} \sup_{\omega \in \Omega, x \in X} \int_{[0,1]} d_{\Omega}(\widehat{\omega}_m(\omega, r, x), \omega) \, d \lambda(r) = 0.
        \end{equation*}
        \item Suppose $\{\widehat{\omega}_m\}_m$ is pointwise consistent and uniformly Lipschitz in the allocation, posterior beliefs are Lipschitz continuous, and $\Psi$ is compact. Then,
        \begin{equation*}
            \lim_{m \rightarrow \infty} \sup_{\mathbf{S} \in \Psi, s \in \mathcal{S}, x \in X}  \int_{\Omega} \int_{[0,1]} d_{\Omega}(\widehat{\omega}_m(\omega, r, x), \omega) \, d \lambda(r) \, d \pi_{\mathbf{S}}(\omega | s) = 0.
        \end{equation*}
    \end{enumerate}
\end{Lemma}

The first part follows by adapting a standard argument for obtaining convergence in expectation using convergence in probability and uniform integrability, which holds in our case by the compactness of $\Omega$, to uniform convergence. The second part is established by showing that the sequence of functions $\{(\mathbf{S},s, x) \mapsto \int_{\Omega} \int_{[0,1]} d_{\Omega}(\widehat{\omega}_m(\omega, r, x), \omega) \, d \lambda(r) \, d \pi_{\mathbf{S}}(\omega | s)\}_m$ is uniformly equicontinuous and applying the Arzelà–Ascoli Theorem. 

\begin{proof}[Proof of Part 1.]
   Let $\epsilon > 0$ be arbitrary. Denoting by $\text{diam}(\Omega) = \sup_{\omega, \omega' \in \Omega} d_{\Omega}(\omega, \omega')$ the diameter of $\Omega$ according to the metric $d_{\Omega}$, we have
    \begin{multline*}
        \int_{[0,1]} d_{\Omega}(\widehat{\omega}_m(\omega, r, x), \omega) d \lambda(r) = \int_{[0,1]} d_{\Omega}(\widehat{\omega}_m(\omega, r, x), \omega) \mathbbm{1}_{d_{\Omega}(\widehat{\omega}_m, \omega) \leq \frac{\epsilon}{2}} d \lambda(r) + \\
        \int_{[0,1]} d_{\Omega}(\widehat{\omega}_m(\omega, r, x), \omega) \mathbbm{1}_{d_{\Omega}(\widehat{\omega}_m, \omega) > \frac{\epsilon}{2}} d \lambda(r) \leq \frac{\epsilon}{2} + \text{diam}(\Omega) \lambda \left( \left\{r \in [0,1]: d_{\Omega}(\widehat{\omega}_m(\omega, r, x), \omega) > \frac{\epsilon}{2} \right\} \right)
    \end{multline*}
    Hence, taking the supremum over $\omega \in \Omega, x \in X$, the left-hand side is upper-bounded by
    \begin{equation*}
        \frac{\epsilon}{2} + \text{diam}(\Omega) \sup_{\omega \in \Omega, x \in X} \lambda \left( \left\{r \in [0,1]: d_{\Omega}(\widehat{\omega}_m(\omega, r, x), \omega) > \frac{\epsilon}{2} \right\} \right)
    \end{equation*}
    By the uniform consistency assumption, it follows that there is $M \in \mathbb{N}$ such that, for all $m \geq M$,
    \begin{equation*}
        \sup_{\omega \in \Omega, x \in X} \lambda \left( \left\{r \in [0,1]: d_{\Omega}(\widehat{\omega}_m(\omega, r, x), \omega) > \frac{\epsilon}{2} \right\} \right) < \frac{\epsilon}{2}
    \end{equation*}
    Hence, for all $m \geq M$:
    \begin{equation*}
        \sup_{\omega \in \Omega, x \in X} \int_{[0,1]} d_{\Omega}(\widehat{\omega}_m(\omega, r, x), \omega) d \lambda(r) \leq \epsilon.
    \end{equation*}
    Since $\epsilon > 0$ was arbitrary, the result follows.
\end{proof}

\begin{proof}[Proof of Part 2.]
   Define $\Psi_m: \Psi \times \cS \times X \rightarrow \mathbb{R}$ by
   \begin{equation*}
       \Psi_m(\mathbf{S}, s, x) \equiv \int_{\Omega} \int_{[0,1]} d_{\Omega}(\widehat{\omega}_m(\omega, r, x), \omega) d \lambda(r) d \pi_{\mathbf{S}}(\omega | s).
   \end{equation*}
    Observe that $\Psi_m \rightarrow 0$ as $m \rightarrow 0$ pointwise. We show this convergence is uniform. To this end, we show the sequence of functions $\{\Psi_m\}_m$ is uniformly equicontinuous and apply the Arzelà–Ascoli Theorem (\citet{folland1999real}, Theorem 4.44).

    Let $\epsilon > 0$ be arbitrary and fix $\delta = \frac{\epsilon}{\text{diam}(\Omega) L + L_X}$. Let $(\mathbf{S}_1, s_1, x_1)$ and $(\mathbf{S}_2, s_2, x_2)$ be arbitrary elements of $\Psi \times \mathcal{S} \times X$ such that $d_{\Psi}(\mathbf{S}_1, \mathbf{S}_2) + d_{\mathcal{S}}(s_1, s_2) + d_X(x_1,x_2) < \delta$. Observe that $|\Psi_m(\mathbf{S}_1, s_1, x_1) - \Psi_m(\mathbf{S}_2, s_2, x_2)|$ can be written as
    \begin{multline*}
         \left | \int_{\Omega} \int_{[0,1]} d_{\Omega}(\widehat{\omega}_m(\omega, r, x_1), \omega) d \lambda(r) d [ \pi_{\mathbf{S}_1}(\omega | s_1) - \pi_{\mathbf{S}_2}(\omega | s_2)] \right | \\
        + \left | \int_{\Omega} \int_{[0,1]} d_{\Omega}(\widehat{\omega}_m(\omega, r, x_1), \omega) -  d_{\Omega}(\widehat{\omega}_m(\omega, r, x_2), \omega) \lambda(r) d \pi_{\mathbf{S}_2}(\omega | s_2) \right |,
    \end{multline*}
    where the first summand is upper bounded by
    \begin{align*}
        \text{diam}(\Omega) \left | \int_{\Omega} d [ \pi_{\mathbf{S}_1}(\omega | s_1) - \pi_{\mathbf{S}_2}(\omega | s_2)] \right | \leq \text{diam}(\Omega) W_1(\pi_{\mathbf{S}_1}(\cdot | s_1), \pi_{\mathbf{S}_2}(\cdot | s_2)),
    \end{align*}
    where the last inequality follows by the dual representation of $W_1$ stemming from the Kantorovich-Rubinstein Theorem (\citet{villani2021topics}, Theorem 1.14). Using the Lipschitz property \eqref{eq: lipschitz posteriors}, we obtain $|\Psi_m(\mathbf{S}_1, s_1, x_1) - \Psi_m(\mathbf{S}_2, s_2, x_2)|$ is upper-bounded by
    \begin{align*}
        \text{diam}(\Omega) L (d_{\Psi}(\mathbf{S}_1, \mathbf{S}_2) + d_{\mathcal{S}}(s_1, s_2)) + L_X d_X(x_1, x_2) < (\text{diam}(\Omega) L + L_X) \delta = \epsilon.
    \end{align*}
    Therefore, $\{\Psi_m\}$ is uniformly equicontinuous. Moreover, the sequence is uniformly bounded by the compactness of $\Omega$, converges pointwise to a continuous function, and the sequence of functions as well as the pointwise limit is defined on a compact domain by the compactness of $\mathcal{S}$ and $\Psi$. Hence, by the Arzelà–Ascoli Theorem, the convergence is uniform.
\end{proof}

\begin{proof}[Proof of Theorem \ref{prop: consistent estimator implementation}]
    Fix an arbitrary $m \in \mathbb{N}$ and a data-driven VCG mechanism for $\widehat{\omega}_m$. We obtain an upper bound on feasible $\epsilon_m$ such that the data-driven VCG mechanism permits implementation in $\epsilon_m$-posterior equilibrium by applying Lemma \ref{lemma: bound on epsilon} and Lemma \ref{lemma: bound via expected error}. This upper bound converges to zero by Lemma \ref{lemma: uniform convergence}.
\end{proof}

%%%%%%%%%%%%%%%%%%%%%%%%%%%%%%%%%%%%%%%%%%%%%%%%%%%%%%%%%%%%
\subsection{Proposition \ref{prop: Characterization of the VCG auction}}\label{sec: appendix proof characterization of the VCG auction}
%%%%%%%%%%%%%%%%%%%%%%%%%%%%%%%%%%%%%%%%%%%%%%%%%%%%%%%%%%%%

\begin{Lemma}[Necessary conditions for implementation]\label{lemma: necessary conditions ex-post implementation}
    Suppose $(x^{*},t)$ is a data-driven mechanism that implements the efficient decision in posterior equilibrium. Suppose we fix all agents reporting their information about the state truthfully. Then the expected transfer of agent $i$ takes the form
    \begin{equation*}
        M \sum_{j \neq i} \sum_{k=1}^K \theta_j \cdot x_{jk}^{*}(\theta) \cdot \E[\omega_{k} | \mathbf{S} = s] + h_i(\theta_{-i}, s, \mathbf{S}),
    \end{equation*}
    where $\theta$ are agents' reports of their values per click, for any $\mathbf{S} \in \Psi$ and $s \in \mathcal{S}$.
\end{Lemma}

The result is a standard application of \cite{holmstrom1979groves}'s Theorem 2 and the fact that posterior implementation and dominant strategy implementation are equivalent in private value settings \citep{bergemann2005robust}.

\begin{proof}[Proof of Proposition \ref{prop: Characterization of the VCG auction}]
    Let $(x^{*},t)$ be a data-driven mechanism with an estimator given by $\widehat{\omega}^{x^{*}(\theta)}$ that implements $x^{*}$ in posterior equilibrium. Suppose agents report their information about the state truthfully. By Lemma \ref{lemma: necessary conditions ex-post implementation}, the expected transfer for an arbitrary agent $i$ must have the form
    \begin{equation*}
        M \sum_{j \neq i} \sum_{k=1}^K \theta_j \cdot x_{jk}^{*}(\theta) \cdot \E[\omega_{k} | \mathbf{S} = s] + h_i(\theta_{-i}, s, \mathbf{S}).
    \end{equation*}
    Now consider agent $i$ with $\theta_i = 0$. Note that $x^{*}(0,\theta_{-i}) = x^{*}(\theta_{-i})$. Further, combining the individual rationality and no subsidy condition, it follows that, for any $\theta \in \Theta$, the expected transfer of agent $i$ takes the form
    \begin{equation*}
        M \sum_{j \neq i} \sum_{k=1}^K \theta_j \cdot x_{jk}^{*}(\theta) \cdot \E[\omega_{k} | \mathbf{S} = s] - M \sum_{j \neq i} \sum_{k=1}^K \theta_j \cdot x_{jk}^{*}(\theta_{-i}) \cdot \E[\omega_{k} | \mathbf{S} = s].
    \end{equation*}
    Next, we endogenize reports of agents' information about the state. We fix the reports of agents other than $i$ to be truthful and consider the reporting incentives of agent $i$. We consider the following mutually exclusive and exhaustive cases, obtaining that the expected transfer must have the following form
    \begin{equation*}
        \overline{t}_i(\theta, s_i', s_{-i}, \mathbf{S}_i', \mathbf{S}_{-i}, s, \mathbf{S}) = 
            - M \cdot \sum_{k=i}^K (\E[\omega_k | \mathbf{S} = s] - \E[\omega_{k+1} | \mathbf{S} = s]) \cdot\theta_{k+1}
    \end{equation*}
    whenever $i \leq K$ and zero otherwise.
    
    \medskip

    \noindent \textbf{Case 1}: The expected transfer when the other agents report truthfully is
    \begin{equation*}
        M \sum_{j \neq i} \sum_{k=1}^K \theta_j \cdot x_{jk}^{*}(\theta) \cdot \E[\omega_{k} | \mathbf{S}_i' = s_i', \mathbf{S}_{-i} = s_{-i}] - M \sum_{j \neq i} \sum_{k=1}^K \theta_j \cdot x_{jk}^{*}(\theta_{-i}) \cdot \E[\omega_{k} | \mathbf{S} = s].
    \end{equation*}
    Consider agent $i$ that is not allocated a slot and suppose $\mathbf{S}$ satisfies the conditions of Assumption \ref{assumption: information CTR}. Then there are signal realizations under which the agent has a strictly profitable deviations: the agent reports signal realizations to maximize the expected transfer; a contradiction.

    \medskip
    \noindent \textbf{Case 2}: The expected transfer when the other agents report truthfully is
    \begin{equation*}
        M \sum_{j \neq i} \sum_{k=1}^K \theta_j \cdot x_{jk}^{*}(\theta) \cdot \E[\omega_{k} | \mathbf{S} = s] - M \sum_{j \neq i} \sum_{k=1}^K \theta_j \cdot x_{jk}^{*}(\theta_{-i}) \cdot \E[\omega_{k} | \mathbf{S}_i' = s_i', \mathbf{S}_{-i} = s_{-i}].
    \end{equation*}
    We repeat the argument of Case 1.

    \medskip
    \noindent \textbf{Case 3}: The expected transfer when the other agents report truthfully is
    \begin{equation*}
        M \sum_{j \neq i} \sum_{k=1}^K \theta_j \cdot x_{jk}^{*}(\theta) \cdot \E[\omega_{k} | \mathbf{S}_i' = s_i', \mathbf{S}_{-i} = s_{-i}] - M \sum_{j \neq i} \sum_{k=1}^K \theta_j \cdot x_{jk}^{*}(\theta_{-i}) \cdot \E[\omega_{k} | \mathbf{S}_i' = s_i', \mathbf{S}_{-i} = s_{-i}].
    \end{equation*}
    Consider agent $i$ allocated slot $K$. The expected transfer of the agent is $- M \cdot \E[\omega_K | \mathbf{S}_i' = s_i', \mathbf{S}_{-i} = s_{-i}] \cdot \theta_{K+1}$, where $\theta_{K+1}$ denotes the $K$-th largest element of $\theta$. By Assumption \ref{assumption: information CTR}, there is an information structure at which the agent has a strictly profitable deviation to reporting a signal realization that minimizes the agent's payment, a contradiction.

    \medskip
    \noindent \textbf{Case 4}: The expected transfer is
    \begin{equation*}
        M \sum_{j \neq i} \sum_{k=1}^K \theta_j \cdot x_{jk}^{*}(\theta) \cdot \E[\omega_{k} | \mathbf{S} = s] - M \sum_{j \neq i} \sum_{k=1}^K \theta_j \cdot x_{jk}^{*}(\theta_{-i}) \cdot \E[\omega_{k} | \mathbf{S} = s].
    \end{equation*}
    which is equivalent to $ 
            - M \cdot \sum_{k=i}^K (\E[\omega_k | \mathbf{S} = s] - \E[\omega_{k+1} | \mathbf{S} = s]) \cdot\theta_{k+1}$
    whenever $i \leq K$ and zero otherwise, as desired.

    It remains to show the expected transfer, where expectation is taken conditional on the state, must coincide with the pivot mechanism for almost every $\omega$. In particular, the analysis above showed that the ex-post transfer must have the form
    \begin{equation*}
        \overline{t}_i(\theta, s_i', s_{-i}, \mathbf{S}_i', \mathbf{S}_{-i}, s, \mathbf{S}) = 
            - M \cdot \sum_{k=i}^K (f_k(\omega_k) - f_{k+1}(\omega_{k+1})) \cdot\theta_{k+1}
    \end{equation*}
    for $f_k: [0,1] \to \mathbb{R}$ such that $\E[f_k(\omega_k) | \mathbf{S} = s] = \E[\omega_k | \mathbf{S} = s]$ for any slot $k$ and any information structure $(s,\mathbf{S}) \in \mathcal{S} \times \Psi$. 
    
    We show $f_k(\omega_k) = \omega_k$ almost surely by induction. Consider first slot $K$ assigned to agent $K$. The net expected payoff of the agent is $M \left( \theta_K \cdot \omega_K - \theta_{K+1} \cdot f_K(\omega_K)  \right)$.
    By individual rationality, $\theta_K \geq \theta_{K+1}$ implies 
    \begin{equation*}
        \frac{\theta_K}{\theta_{K+1}} \omega_K \geq f_K(\omega_K) \quad \text{for a.e. } \, \omega_K.
    \end{equation*}
    Taking $\theta_{K+1} \to \theta_K$ implies $\omega_K \geq f_K(\omega_K)$ for a.e. $\omega_K$. It follows $\omega_K = f_K(\omega_K)$ for a.e. $\omega_K$. Now suppose $f_{l}(\omega_l) = \omega_l$ a.e. for all $k +1  \leq l \leq K$. Consider slot $k$ assigned to agent $k$. The net expected payoff of the agent is
    \begin{equation*}
        M \left( \theta_k \cdot \omega_k - (f_k(\omega_k) - \omega_{k+1}) \cdot \theta_{k+1} - \sum_{l=k+1}^K (\omega_l - \omega_{l+1}) \theta_{l+1}  \right)
    \end{equation*}
    By individual rationality and taking $\theta_l \to \theta_k$ for all $l > k$ yields
    \begin{equation*}
        \omega_k \geq f_k(\omega_k) - \omega_{k+1} + \sum_{l=k+1}^K (\omega_l - \omega_{l+1}) = f_k(\omega_k).
    \end{equation*}
    It follows $f_k(\omega_k) = \omega_k$ almost everywhere. The result follows by induction.
\end{proof}

%%%%%%%%%%%%%%%%%%%%%%%%%%%%%%%%%%%%%%%%%%%%%%%%%%%%%%%%%%%%
\subsection{Proposition \ref{prop: dynamic implementation product recs}}\label{sec: appendix proof dynamic implementation product recs}
%%%%%%%%%%%%%%%%%%%%%%%%%%%%%%%%%%%%%%%%%%%%%%%%%%%%%%%%%%%%

\begin{proof}
    Fix an arbitrary period $\tau$, history $h_{\tau} \in H_{\tau}$, advertiser $i \in N$, and type profile $\xi = (\theta, \mathbf{S}, s) \in \Xi$. Suppose the rest of the agents report $\xi_{-i}$ truthfully. The expected payoff of agent $i$ from report $\xi_i'$ at history $h_{\tau}$ is, up to the addition of a term that does not depend on the agent's report,
    \begin{equation*}
        \E \Bigg[ (\theta_i + \widehat{\omega}_{i\tau'}^{h_{\tau'}}) \sum_{\tau' = \tau}^{T} x_{i\tau'}^{*}(\xi_i', \xi_{-i}, h_{\tau'})
        +
    \sum_{j \neq i }
    \big( \theta_j + \widehat{\omega}_{j\tau'}^{h_{\tau'}} \big)
    x_{j\tau'}^{*}(\xi_i', \xi_{-i}, h_{\tau'})
         \mid \mathbf{S} = s, h_{\tau} \Bigg]
    \end{equation*}
    Using the maintained assumptions, for any $j \in N $, $\tau' \geq \tau$, and $h_{\tau'} \in H_{\tau'}$, with a slight abuse of notation,
    \begin{align*}
        \E[\widehat{\omega}_{j\tau'}^{h_{\tau'}} \cdot x_{j\tau'}^{*}(\xi_i', \xi_{-i}, h_{\tau'}) \mid \mathbf{S} = s, h_{\tau}, x^{*}_{\tau}] &= \E[\E[\widehat{\omega}_{j\tau'}^{h_{\tau'}} \cdot x_{j\tau'}^{*}(\xi_i', \xi_{-i}, h_{\tau'}) \mid \omega, h_{\tau'}, x_{\tau'}^{*}] | \mathbf{S} = s, h_{\tau}, x^{*}_{\tau}] \\
        &= \E[ \omega \cdot x_{j\tau'}^{*}(\xi_i', \xi_{-i}, h_{\tau'}) \mid \mathbf{S} = s, h_{\tau}, x^{*}_{\tau}]
    \end{align*}
    Hence, the expected payoff is given by
    \begin{equation*}
        \E \left[  \sum_{\tau' = \tau}^T \sum_{j \in N } (\theta_j + \omega_j) x_{j\tau'}^{*}(\xi_i', \xi_{-i}, h_{\tau'}) \,\middle|\, \mathbf{S} = s, h_{\tau}   \right],
    \end{equation*}
    which is maximized at $\xi_i' = \xi_i$ by optimality of the policy $x^{*}$. Since agent $i$, period $\tau$, and history $h_{\tau} \in H_{\tau}$ were arbitrary, the implementation result follows.
\end{proof}

\newpage
%%%%%%%%%%%%%%%%%%%%%%%%%%%%%%%%%%%%%%%%%%%%%%%%%%%%%%%%%%%%
\section{Online Appendix}\label{sec: appendix further results and derivations}
%%%%%%%%%%%%%%%%%%%%%%%%%%%%%%%%%%%%%%%%%%%%%%%%%%%%%%%%%%%%

%%%%%%%%%%%%%%%%%%%%%%%%%%%%%%%%%%%%%%%%%%%%%%%%%%%%%%%%%%%%
\subsection{Implementation with Message-Driven Mechanisms}\label{sec: appendix message-driven mechanisms}
%%%%%%%%%%%%%%%%%%%%%%%%%%%%%%%%%%%%%%%%%%%%%%%%%%%%%%%%%%%%

Several papers in the literature show efficient allocation rules are not implementable by message-driven mechanisms in multi-dimensional environments with interdependent values and commonly known signal spaces \citep{maskin1992auctions, dasgupta2000efficient, jehiel2001efficient, jehiel2006limits}. In particular, in a model with a finite action space, \citet{jehiel2006limits} show that in multi-dimensional environments, for generic payoff functions, only constant allocation rules are implementable among deterministic social choice functions. This result strengthens the earlier impossibility of \citet{jehiel2001efficient}, who show that when at least one agent’s signal is multi-dimensional and signals are independent across agents, the efficient allocation is not Bayesian implementable and hence, a fortiori, not posterior implementable. 

These impossibility results apply directly to our setting: even when the signal of each agent is commonly known and agents’ signal realizations are one-dimensional, the overall type is multi-dimensional due to the presence of private preference types. The additional structure that the preference component $\theta_i$ enters only agent $i$'s own payoff, while the signal component $s_i$ may affect the values of all agents, merely restricts attention to a closed subset of the overall space of payoff functions. Since the impossibility in \citet{jehiel2006limits} is established on a residual (and finitely prevalent) subset of the ambient function space, its intersection with any such closed subset is still residual (and finitely prevalent). Hence, specializing our model to satisfy the assumptions of the framework of \cite{jehiel2006limits}, for generic payoff functions consistent with our framework, every deterministic implementable allocation rule is constant on the interior of the joint type space.

We show that efficient implementation is impossible in our running examples. The first example features a continuum allocation space, a case not explicitly covered by the results of \citet{jehiel2001efficient} and \citet{jehiel2006limits}. The second example considers an environment that falls outside the generic class of \citet{jehiel2006limits} under certain conditions on the signal structure.

\begin{Ex}[Joint prediction with private biases: impossibility]\label{ex: joint prediction with private biases impossibility}
    Consider the setting of Example \ref{ex: joint prediction with private biases}. We further set $n=2$ for simplicity of exposition and assume $\sigma_i^2$ is commonly known, for each $i \in N$. Henceforth, we only use signal realizations to denote agents' information about the state for notational ease.

    \begin{Claim}\label{claim: joint prediction with private biases impossibility}
        There is no message-driven mechanism that implements the efficient allocation in posterior equilibrium.
    \end{Claim}

    \begin{proof}
        We proceed by contradiction. Suppose message-driven mechanism \( (x^{*}, t) \) implements the efficient allocation in posterior equilibrium. Then \( t \) must prevent each agent from having a profitable deviation by misreporting either her preference type or signal realization. Using standard results from the literature, we characterize transfer schemes that deter profitable deviations along each dimension of agents' types. The fact that these two classes of payment rules are distinct helps us reach the desired contradiction.

        Suppose the true profile of signal realizations is $s = (s_1,s_2)$ and is fixed to be reported truthfully by both agents.
    Consider the class of VCG transfers for agent 1:
    \begin{equation*}\label{eq: VCG}
        t_1(\theta, s; h_1) = h_1(\theta_{2}; s) + \E \left[ u_2 \big( x^{*}(\theta,s), \omega, \theta_2 \big) | s \right ] = h_1(\theta_{2}; s) - \frac{1}{4} (\theta_1 - \theta_2)^2 - \Var[\omega | s],
    \end{equation*}
    for an arbitrary function $h_1(\theta_{2}; s)$ of $\theta_{2}$. We can define the class of VCG transfers for agent 2 analogously. The VCG mechanism implements the efficient outcome in posterior equilibrium. Moreover, any transfer scheme $t$, such that $(x^{*},t)$ permits implementation in posterior equilibrium, must have this form \citep{green1977characterization,holmstrom1979groves}.\footnote{\cite{bergemann2005robust} show that with private values, as is assumed here, posterior implementation is equivalent to dominant strategy implementation.}

    Similarly, if preference types $\theta$ are fixed to be reported truthfully, and since agent $i$'s expected utility given a profile of signal realizations $s$ is supermodular in $(x, s_j)$ for each $j$, the transfer functions \( t \) such that \( (x^{*}, t) \) permits implementation in posterior equilibrium must belong to the class of generalized VCG mechanisms (Proposition 4 of \citet{bergemann2002information}) with transfers for agent 1 given by:
    \begin{equation*}
        \int_{0}^{s_1} \frac{\partial}{\partial x} \E \left[ u_2 \left( x^{*}(\theta, z_1, s_{2}), \omega, \theta_2 \right) | z_1, s_{2} \right] \frac{\partial}{\partial z_1} x^{*}(\theta, z_1, s_{2}) \, dz_1 + k_1(s_2; \theta),
    \end{equation*}
    which simplifies to
    \begin{equation*}
        t_1(\theta, s; k_1) = k_1(s_2; \theta) - (\theta_1 - \theta_2) \E[\omega | s],
    \end{equation*}
    for an arbitrary function $k_1(s_{2}; \theta)$ of $s_2$.
    
    Since the transfer function $t$ must ensure there are no profitable deviations along each dimension, it follows that there are functions $h_1$ and $k_1$ such that
    \begin{align*}
        h_1(\theta_2;s) - k_1(s_2; \theta) &= \frac{1}{4} (\theta_1 - \theta_2)^2 + \Var[\omega|s] - (\theta_1 - \theta_2) \E[\omega | s].
    \end{align*}
    Repeating the above for $s_1' \neq s_1$ and $s_2, \theta$:
    \begin{equation*}
        h_1(\theta_2;s_1', s_2) - k_1(s_2; \theta) = \frac{1}{4} (\theta_1 - \theta_2)^2 + \Var[\omega|s_1', s_2] - (\theta_1 - \theta_2) \E[\omega | s_1', s_2].
    \end{equation*}
    Taking their difference, we obtain
    \begin{equation*}
        h_1(\theta_2;s) - h_1(\theta_2;s_1', s_2) = \Var[\omega|s] - \Var[\omega|s_1', s_2] - (\theta_1 - \theta_2) \left( \E [\omega | s] - \E [\omega | s_1', s_2] \right).
    \end{equation*}
    The right-hand side varies with $ \theta_1$, while the left-hand side does not, a contradiction.
    \end{proof}
\end{Ex}

\begin{Ex}[Single-item ad auction: impossibility]\label{ex: single-item ad auction impossibility}
    Consider the setting of Example \ref{ex: ad auction}. For the simplicity of exposition, we consider a single slot. Further, we assume agents' signals are commonly known, signal realizations are one-dimensional, and there is an agent $i \in N$ whose information about the state is non-redundant: $E[\omega | \mathbf{S}_i, \mathbf{S}_{-i}] \neq  E[\omega | \mathbf{S}_{-i}]$.\footnote{Note that this environment does not belong to the generic set of \cite{jehiel2006limits} if, for example, $E[\omega | s] = \sum_{i \in N} s_i$.}

    \begin{Claim}\label{claim: single-item ad auction impossibility}
        There is no message-driven mechanism that implements the efficient allocation in posterior equilibrium.
    \end{Claim}

    \begin{proof}
    We proceed by contradiction. Suppose there is a message-driven mechanism $(x^{*}, t)$ that implements $x^{*}$. We proceed analogously to the proof of Claim \ref{claim: joint prediction with private biases impossibility}, eventually reaching a contradiction.

    Suppose the true profile of signals is $s$ and is fixed to be reported truthfully by all agents. Consider the class of VCG transfers for agent $i$:
    \begin{equation*}
        t_i(\theta, s; h_i) = h_i(\theta_{-i}; s) + \sum_{j \neq i} \theta_j x_{j}^{*}(\theta) \E[\omega | s]
    \end{equation*}
    for an arbitrary function $h_i(\theta_{-i}; s)$ of $\theta_{-i}$. Any transfer scheme $t$, such that $(x^{*}, t)$ permits implementation in posterior equilibrium, must have this form.

    Fixing $\theta$ to be reported truthfully, $t$ must belong to the class of generalized VCG mechanisms of \cite{bergemann2002information}:\footnote{We treat the allocation as discrete, treating the allocation function as deterministic by fixing a tie-breaking rule.}
    \begin{equation*}
        t_i(\theta, s; k_i) = k_i(s_{-i}; \theta)
    \end{equation*}
    for an arbitrary function $k_i(s_{-i}; \theta)$ of $s_{-i}$.

    Since the transfer function $t$ must ensure there are no profitable deviations along each dimension, it follows there are functions $h_i$ and $k_i$ such that
    \begin{equation*}
        h_i(\theta_{-i}; s) - k_i(s_{-i}; \theta) = - \sum_{j \neq i} \theta_j x_{j}^{*}(\theta) \E[\omega | s]
    \end{equation*}
    repeating the above for $s_i' \neq s_i$ such that $E[\omega | s_i', s_{-i}] \neq E[\omega | s_i, s_{-i}]$, which exists by our maintained assumption, and taking the corresponding difference, yields
    \begin{equation*}
        h_i(\theta_{-i}; s) - h_i(\theta_{-i}; s_i', s_{-i}) = \left( \E[\omega | s_i', s_{-i}] - \E[\omega | s_i, s_{-i}] \right) \sum_{j \neq i} \theta_j x_{j}^{*}(\theta).
    \end{equation*}
    The right-hand side varies with $\theta_i$ while the left-hand side does not, a contradiction.
\end{proof}
\end{Ex}

\cite{mclean2015implementation} demonstrate that if agents have \emph{zero informational size} in the terminology of \cite{mclean2002informational}, that is, the private information held by any single agent is redundant when combined with the joint information of the other agents,\footnote{\cite{mclean2015implementation} refer to this as the \emph{nonexclusive information} condition.} their generalized VCG mechanism permits implementation in posterior equilibrium.\footnote{In the generalized VCG (pivot) mechanism proposed by \cite{mclean2015implementation}, agent \(i\) pays the externality they impose on the other agents under the assumption that these agents have access to \(i\)'s information even if \(i\) were absent. In the next section, we extend the standard VCG mechanism along similar lines. However, in our pivot mechanism, we account for two types of effects: first, the cost an agent imposes on others by influencing the allocation in their favor, and second, the value they contribute to improving prediction accuracy by sharing information about the state. For the latter, the agent is compensated, as illustrated in Example \ref{ex: joint prediction with private biases: data-driven VCG}.} Moreover, \cite{mclean2015implementation} show that if each agent exerts only a small informational effect on the posterior distribution over states, their generalized VCG mechanism is approximately posterior incentive-compatible. In the limit as the informational size of all agents approaches zero, the mechanism achieves exact implementation in posterior equilibrium. Typically, however, we expect the signals of all agents to meaningfully influence predictions about the state. Our broader goal is to design mechanisms that are robust to assumptions on the underlying stochastic structure. To this end, we focus our attention on data-driven mechanisms.

%%%%%%%%%%%%%%%%%%%%%%%%%%%%%%%%%%%%%%%%%%%%%%%%%%%%%%%%%%%%
\subsection{Lottery Mechanisms}\label{appendix: surplus extraction mechanisms}
%%%%%%%%%%%%%%%%%%%%%%%%%%%%%%%%%%%%%%%%%%%%%%%%%%%%%%%%%%%%

\begin{Prop}[Implementation with lottery mechanisms]\label{prop: CM extraction}
    Suppose the state $\omega$ is fully revealed after the allocation and \eqref{eq: CM condition} holds for every agent $i \in N$. Then there exists $\phi_i: \mathcal{S} \times \Psi \times \Omega \to \mathbb{R}$ for any agent $i \in N$ such that the mechanism $(x^{*}, t)$ with transfers given by
    \begin{equation}\label{eq: transfer CM}
        t_i(\xi,\omega) = \sum_{j \neq i} v_j(x^{*}(\xi), \theta_j, s,\mathbf{S}) + \phi_i(s,\mathbf{S},\omega)
    \end{equation}
    implements $x^{*}$ in posterior equilibrium.
\end{Prop}

\begin{proof}
    Fix an arbitrary agent $i$ and suppose the rest of the agents report truthfully. Fix the true type profile to be $\xi = (\theta, s, \mathbf{S})$.

    Using the same steps as in the proof of the ``if'' direction of Theorem 2 in \cite{cremer1988full} based on Farkas' lemma, we can show there is $\Tilde{\phi}_i: \Pi_i(s_{-i}, \mathbf{S}_{-i}) \to \mathbb{R}$ such that
    \begin{equation*}
        \forall \pi_i \in \Pi_i(s_{-i}, \mathbf{S}_{-i}): \quad \E[\Tilde{\phi}_i(\pi_i, \omega) | \pi_i] = 0,
    \end{equation*}
    and
    \begin{equation*}
        \forall \pi_i, \pi_i' \in \Pi_i(s_{-i}, \mathbf{S}_{-i}) \text{ s.t. } \pi_i \neq \pi_i': \quad \E[\Tilde{\phi}_i(\pi_i, \omega) | \pi_i'] < 0.
    \end{equation*}

    For any $(s_i', \mathbf{S}_i') \in \mathcal{S}_i \times \Psi_i$, let $\pi_i(s_i', \mathbf{S}_i'; s_{-i}, \mathbf{S}_{-i})$ be the corresponding element of $\Pi_i(s_{-i}, \mathbf{S}_{-i})$. Note that for any $x \in X$ and any agent $j \in N$, whenever $\pi_i(s_i', \mathbf{S}_i'; s_{-i}, \mathbf{S}_{-i}) = \pi_i(s_i'', \mathbf{S}_i''; s_{-i}, \mathbf{S}_{-i})$, we have
    \begin{equation*}
        v_j(x, s_i', \mathbf{S}_i', s_{-i}, \mathbf{S}_{-i}) = v_j(x, s_i'', \mathbf{S}_i'', s_{-i}, \mathbf{S}_{-i}).
    \end{equation*}
    It also follows that we can choose an efficient decision rule $x^{*}$ such that
    \begin{equation*}
        x^{*}(\theta_i, s_i', \mathbf{S}_i', s_{-i}, \mathbf{S}_{-i}) = x^{*}(\theta_i, s_i'', \mathbf{S}_i'', s_{-i}, \mathbf{S}_{-i}).
    \end{equation*}
    Hence, we may write $x^{*}(\theta_i, \pi_i)$ and $v_j(x,\pi_i)$ when evaluating agent $i$'s reporting strategy, keeping the profile of the other agents fixed. That is, only the preference type and the induced posterior belief is relevant for the agent when evaluating the gross expected payoff and the efficient decision.

    Therefore, we consider the following auxiliary single-agent environment: $i$ is the single agent, the message space is $\Theta_i \times \Pi_i(s_{-i}, \mathbf{S}_{-i})$, and the mechanism is $(x^{*},t_i)$ where $x^{*}: \Theta_i \times \Pi_i(s_{-i}, \mathbf{S}_{-i}) \to X$ is the efficient decision and $t_i: \Theta_i \times \Pi_i(s_{-i}, \mathbf{S}_{-i}) \times \Omega \to \mathbb{R}$ is given by
    \begin{equation*}
        t_i(\theta_i, \pi_i, \omega) = \sum_{j \neq i} v_j(x^{*}(\theta_i, \pi_i), \theta_j, \pi_i) + \lambda \Tilde{\phi}_i(\pi_i, \omega)
    \end{equation*}
    for a constant $\lambda > 0$. Since $u_i$ is bounded, there is $\lambda^{*}$ large enough such that reporting truthfully is always optimal for the agent. 
    
    Finally, we consider the original problem. Let $\phi_i(s,\mathbf{S},\omega) = \lambda^{*} \Tilde{\phi}_i(\pi_i(s_i, \mathbf{S}_i; s_{-i}, \mathbf{S}_{-i}), \omega)$ and consider the mechanism $(x^{*}, t)$ with transfers given by \eqref{eq: transfer CM}. Then reporting truthfully is optimal for agent $i$. Indeed, only the resulting posterior is payoff-relevant, and we have shown above that agents report their posteriors truthfully in the auxiliary problem. Since agent $i$ and profile $\xi$ were arbitrary, the result follows.
\end{proof}

When agents’ types consist solely of their information about the state, full surplus extraction becomes feasible. The proof of the following result is a direct adaptation of the arguments above and is therefore omitted.

\begin{Prop}[Surplus extraction]\label{prop: CM full extraction}
    Suppose the state $\omega$ is fully revealed after the allocation and \eqref{eq: CM condition} holds for every agent $i \in N$. Further, suppose that for any agent $i$, $\Theta_i$ is a singleton and $u_i(x,\omega, \theta_i) \geq 0$ for any $x \in X$ and $\omega \in \Omega$. Then there exists $\phi_i: \mathcal{S} \times \Psi \times \Omega \to \mathbb{R}$ for any agent $i \in N$ such that the mechanism $(x^{*}, t)$ with transfers given by
    \begin{equation}\label{eq: transfer CM full}
        t_i(s,\mathbf{S},\omega) = - v_i(x^{*}(s,\mathbf{S}), \theta_i, s,\mathbf{S}) + \phi_i(s,\mathbf{S},\omega)
    \end{equation}
    implements $x^{*}$ in posterior equilibrium and for any type profile $(s,\mathbf{S})$ and agent $i \in N$ satisfies
    \begin{equation}\label{eq: surplus extraction}
        v_i(x^{*}(s, \mathbf{S}), \theta_i, s,\mathbf{S}) + \overline{t}_i(s,\mathbf{S},s,\mathbf{S}) = 0.
    \end{equation}
\end{Prop}

%%%%%%%%%%%%%%%%%%%%%%%%%%%%%%%%%%%%%%%%%%%%%%%%%%%%%%%%%%%%
\subsection{Rate of Convergence for Consistent Estimators}\label{appendix: rate of convergence}
%%%%%%%%%%%%%%%%%%%%%%%%%%%%%%%%%%%%%%%%%%%%%%%%%%%%%%%%%%%%

We build on \eqref{eq: consistency}-\eqref{eq: uniform consistency} as follows. Fix a non-negative sequence $\{q_m\}_m$. We say a sequence of estimators $\{\widehat{\omega}_m\}_m$ is \textit{consistent for $\omega$ and $x$ at rate $\{q_m\}_m$} if
\begin{equation}\label{eq: consistency rate}
    \forall \epsilon > 0: \quad \lim_{m \rightarrow \infty} \lambda(\{r \in [0,1]: q_m d_{\Omega}(\widehat{\omega}_m(\omega, r, x), \omega) > \epsilon\}) = 0.
\end{equation}
We define $\{\widehat{\omega}_m\}_m$ to be \textit{pointwise consistent at rate $\{q_m\}_m$} if it is consistent at rate $\{q_m\}_m$ for every $\omega \in \Omega$ and $x \in X$:
\begin{equation}\label{eq: pointwise consistency rate}
    \forall \epsilon > 0: \quad \lim_{m \rightarrow \infty} \lambda(\{r \in [0,1]: q_m d_{\Omega}(\widehat{\omega}_m(\omega, r, x), \omega) > \epsilon\}) = 0 \quad \forall \omega \in \Omega, x \in X.
\end{equation}
Further, $\{\widehat{\omega}_m\}_m$ is \textit{uniformly consistent at rate $\{q_m\}_m$} if the convergence is uniform:
\begin{equation}\label{eq: uniform consistency rate}
    \forall \epsilon > 0: \quad \lim_{m \rightarrow \infty} \sup_{\omega \in \Omega, x \in X} \lambda(\{r \in [0,1]: q_m d_{\Omega}(\widehat{\omega}_m(\omega, r, x), \omega) > \epsilon\}) = 0.
\end{equation}
Unlike Theorem \ref{prop: consistent estimator implementation}, since $\{q_m d_{\Omega}(\widehat{\omega}_m, \omega)\}_m$ is not guaranteed to be bounded, the following result requires explicit uniform integrability (UI) conditions. We say $\{\widehat{\omega}_m\}_m$ and $\{ q_m \}_m$ satisfy \textit{UI pointwise} if
\begin{equation}\label{eq: pointwise UI}
    \lim_{M \rightarrow \infty} \limsup_{m \to \infty} \int_{[0,1]} q_m d_{\Omega}(\widehat{\omega}_m(\omega, r, x), \omega) \mathbbm{1}_{q_m d_{\Omega}(\widehat{\omega}_m(\omega, r, x),\omega) > M} \lambda(r) = 0 \quad \forall \omega \in \Omega, x \in X.
\end{equation}
They satisfy \textit{UI uniformly} if 
\begin{equation}\label{eq: uniform UI}
    \lim_{M \rightarrow \infty} \limsup_{m \to \infty} \sup_{\omega \in \Omega, x \in X} \int_{[0,1]} q_m d_{\Omega}(\widehat{\omega}_m(\omega, r, x), \omega) \mathbbm{1}_{q_m d_{\Omega}(\widehat{\omega}_m(\omega, r, x),\omega) > M} \lambda(r) = 0.
\end{equation}
We are ready to state the formal result.

\begin{Prop}[Rate of convergence]\label{prop: rate of convergence} 
    Suppose $u_i$ is Lipschitz in $\omega$ uniformly in $x$ and $\theta_i$ for each $i \in N$. Fix a sequence of estimators $\{\widehat{\omega}_m\}_m$ and a non-negative sequence $\{q_m\}_m$  such that either of the following holds:
    \begin{enumerate}
        \item $\{\widehat{\omega}_m\}_m$ is uniformly consistent at rate $\{q_m\}_m$, and $\{\widehat{\omega}_m\}_m$ and $\{ q_m \}_m$ satisfy UI uniformly; or
        \item $\{\widehat{\omega}_m\}_m$ is pointwise consistent at rate $\{q_m\}_m$ and uniformly Lipschitz in the allocation, $\{\widehat{\omega}_m\}_m$ and $\{ q_m \}_m$ satisfy UI pointwise, posterior beliefs are Lipschitz continuous, and $\Psi$ is compact.
    \end{enumerate}
    Then there is a non-negative sequence $\{\epsilon_m\}_m$, with $q_m \epsilon_m \to 0$ as $m \to \infty$, such that every data-driven VCG mechanism for $\widehat{\omega}_m$ permits implementation in $\epsilon_m$-posterior equilibrium for every $m \in \mathbb{N}$.
\end{Prop}

As shown above, by the Lipschitz property of payoff functions, for any profile of signals $\mathbf{S} \in \Psi$ and signal realizations $s \in \mathcal{S}$, we can upper-bound $q_m \epsilon_m$ by a constant multiple of $\E[ q_m d_{\Omega}(\widehat{\omega}_m, \omega) | \mathbf{S} = s]$. Using the assumed consistency, uniform integrability, and regularity conditions, this converges to 0 uniformly.

\begin{proof}
    Fix an arbitrary $m \in \mathbb{N}$ and a data-driven VCG mechanism for $\widehat{\omega}_m$. Recall from lemmata \ref{lemma: bound on epsilon} and \ref{lemma: bound via expected error} that for each $m \in \mathbb{N}$, there is $\epsilon_m$ with
    \begin{align*}
        \epsilon_m &\leq 2 \max_{i \in N} \sum_{j \neq i} L_j \sup_{\mathbf{S} \in \Psi, s \in \mathcal{S}, x \in X}  \int_{\Omega} \int_{[0,1]} d_{\Omega}(\widehat{\omega}_m(\omega, r, x), \omega) \, d \lambda(r) \, d \pi_{\mathbf{S}}(\omega | s)\\
        &\leq 2 \max_{i \in N} \sum_{j \neq i} L_j \sup_{\omega \in \Omega, x \in X} \int_{[0,1]} d_{\Omega}(\widehat{\omega}_m(\omega, r, x), \omega) \, d \lambda(r),
    \end{align*}
    such that reporting truthfully by all agents is an $\epsilon_m$-posterior equilibrium. It follows that
    \begin{align*}
        q_m \epsilon_m &\leq 2 \max_{i \in N} \sum_{j \neq i} L_j \sup_{\mathbf{S} \in \Psi, s \in \mathcal{S}, x \in X}  \int_{\Omega} \int_{[0,1]} q_m d_{\Omega}(\widehat{\omega}_m(\omega, r, x), \omega) \, d \lambda(r) \, d \pi_{\mathbf{S}}(\omega | s)\\
        &\leq 2 \max_{i \in N} \sum_{j \neq i} L_j \sup_{\omega \in \Omega, x \in X} \int_{[0,1]} q_m d_{\Omega}(\widehat{\omega}_m(\omega, r, x), \omega) \, d \lambda(r).
    \end{align*}
    
    We prove the statement for each of the stated conditions. The claim for condition 1 follows by an argument similar to the proof of the first part of Lemma \ref{lemma: uniform convergence}. In particular, for any $M > 0$, we have
    \begin{multline*}
        \sup_{\omega \in \Omega, x \in X} \int_{[0,1]} q_m d_{\Omega}(\widehat{\omega}_m(\omega, r, x), \omega) \, d \lambda(r) = \\
        \sup_{\omega \in \Omega, x \in X} \int_{[0,1]} q_m d_{\Omega}(\widehat{\omega}_m(\omega, r, x), \omega) \mathbbm{1}_{q_m d_{\Omega}(\widehat{\omega}_m(\omega, r, x), \omega) \leq M} \, d \lambda(r) \\
        + \sup_{\omega \in \Omega, x \in X} \int_{[0,1]} q_m d_{\Omega}(\widehat{\omega}_m(\omega, r, x), \omega) \mathbbm{1}_{q_m d_{\Omega}(\widehat{\omega}_m(\omega, r, x), \omega) > M} \, d \lambda(r).
    \end{multline*}
    Defining $\mathbf{X}_m(r,\omega,x) \equiv q_m d_{\Omega}(\widehat{\omega}_m(\omega, r, x), \omega)$, note that for every $\epsilon > 0$ and $\omega \in \Omega, x \in X$,
    \begin{equation*}
        \lambda(\{r \in [0,1]: \mathbf{X}_m(\omega, r,x) > \epsilon \}) \leq \lambda(\{r \in [0,1]: q_m d_{\Omega}(\widehat{\omega}_m(\omega, r, x) > \epsilon \})
    \end{equation*}
    By \eqref{eq: uniform consistency rate}, it follows that
    \begin{equation*}
        \forall \epsilon > 0: \quad \lim_{m \rightarrow \infty} \sup_{\omega \in \Omega, x \in X} \lambda(\{r \in [0,1]: \mathbf{X}_m(\omega, r, x) > \epsilon \}) = 0.
    \end{equation*}
    The fact that
    \begin{equation*}
        \forall M > 0: \quad \lim_{m \rightarrow \infty} \sup_{\omega \in \Omega, x \in X} \int_{[0,1]} q_m d_{\Omega}(\widehat{\omega}_m(\omega, r, x), \omega) \mathbbm{1}_{q_m d_{\Omega}(\widehat{\omega}_m(\omega, r, x), \omega) \leq M} \, d \lambda(r) = 0
    \end{equation*}
    follows by an analogous argument to the first part of Lemma \ref{lemma: uniform convergence}. Moreover, by the UI condition \eqref{eq: uniform UI}, $\forall \epsilon > 0$, $\exists M > 0$ large enough such that $\forall m \geq M$,
    \begin{equation*}
        \sup_{\omega \in \Omega, x \in X} \int_{[0,1]} q_m d_{\Omega}(\widehat{\omega}_m(\omega, r, x), \omega) \mathbbm{1}_{q_m d_{\Omega}(\widehat{\omega}_m(\omega, r, x), \omega) > M} \, d \lambda(r) < \epsilon
    \end{equation*}
    Combining the two terms, the claim follows.
    
    To prove the statement for condition 2, note that, by the pointwise convergence \eqref{eq: pointwise consistency rate} and pointwise UI \eqref{eq: pointwise UI} conditions, it follows by the same steps that
    \begin{equation*}
        \lim_{m \rightarrow \infty} \int_{[0,1]} q_m d_{\Omega}(\widehat{\omega}_m(\omega, r, x), \omega) \, d \lambda(r) = 0 \quad \forall \omega \in \Omega, x \in X.
    \end{equation*}
    Hence,
    \begin{equation*}
        \lim_{m \rightarrow \infty} \int_{\Omega} \int_{[0,1]} q_m d_{\Omega}(\widehat{\omega}_m(\omega, r, x), \omega) \, d \lambda(r) \, d \pi_{\mathbf{S}}(\omega | s) = 0 \quad \forall \mathbf{S} \in \Psi, s \in \mathcal{S}, x \in X.
    \end{equation*}
    The argument showing this convergence is uniform follows analogously to the proof of the second part of Lemma \ref{lemma: uniform convergence}.
    
    Hence, in both cases, $\lim_{m \rightarrow \infty} q_m \epsilon_m = 0$, concluding the proof.
\end{proof}

%%%%%%%%%%%%%%%%%%%%%%%%%%%%%%%%%%%%%%%%%%%%%%%%%%%%%%%%%%%%
\subsection{Bayesian Interpretation of Additional Data and Alternative Specifications of Data-Driven VCG Mechanisms}\label{appendix: bayesian interpretation}
%%%%%%%%%%%%%%%%%%%%%%%%%%%%%%%%%%%%%%%%%%%%%%%%%%%%%%%%%%%%

In this section, we establish implementation results for an alternative formulation of data-driven VCG mechanisms based on a Bayesian interpretation of additional information about the payoff-relevant state. For simplicity of exposition, unlike Assumption \ref{assumption: estimator}, we assume the additional signal is not a function of the allocation. The analysis below can be extended to allow for this case under additional technical assumptions akin to those in Section \ref{sec: noisy data}.

\begin{Assumption}[Additional signal]
     There is a commonly known compact metric space $\mathcal{Y}$ endowed with the corresponding Borel $\sigma$-algebra $\cB(\mathcal{Y})$ and a commonly known measurable mapping
    \begin{equation*}
        \mathbf{Y}: \Omega \times [0,1] \rightarrow \mathcal{Y},
    \end{equation*}
    which defines a random variable $\mathbf{Y}$ on the expanded state space $(\Omega \times [0,1], \cB(\Omega) \otimes \cB([0,1]), \pi^0 \times \lambda)$. For any signal profile $\mathbf{S} \in \Psi$, the random variable $\mathbf{Y}$ is conditionally independent of $\mathbf{S}$ given the payoff-relevant state. The designer receives a realization of $\mathbf{Y}$ after the final allocation and before the final payments. This additional information is verifiable.
\end{Assumption}

We denote a generic realization of $\mathbf{Y}$ by $y \in \cY$. We denote by $\kappa_{\mathbf{Y}}: \cB(\cY) \times \Omega \rightarrow [0,1]$ the corresponding Markov kernel representing the distribution of the signal conditional on the payoff-relevant state. We represent the posterior regular conditional distribution of the payoff-relevant state conditional on the signal using the Markov kernel $\pi_{\mathbf{Y}}: \cB(\Omega) \times \cY \rightarrow [0,1]$. Finally, we denote the law of $\mathbf{Y}$ by $P_{\mathbf{Y}} = (\pi^0 \times \lambda) \circ \mathbf{Y}^{-1}$. 

Interpreting $\mathbf{Y}$ as a signal about the payoff-relevant state, there are at least two ways to define data-driven VCG mechanisms. First, we can define an estimator of the payoff-relevant state as the posterior mean $\widehat{\omega} \equiv \E[\omega | \mathbf{Y}]$. The analysis in this case follows the same steps as in Section \ref{sec: data-driven mechanisms} using the definition of data-driven VCG mechanisms in Definition \ref{def: data-driven VCG}. In particular, if the signal $\mathbf{Y}$ fully reveals the state, we obtain the ex-post case and exact implementation. A sufficient condition on the signals to obtain the result in Theorem \ref{prop: consistent estimator implementation}, along with the regularity conditions assumed in the theorem, is as follows. Consider a sequence of signals $\{\mathbf{Y}_m\}_m$ that is \textit{posterior consistent} for $\omega \in \Omega$: under the true payoff-relevant state being $\omega$, the sequence of posteriors $\{\pi_{\mathbf{Y}_m}(\cdot | y_m)\}_m$ weakly converges to the Dirac measure on $\omega$, $\delta_{\omega}$, in probability. More formally, since the Wasserstein 1-distance metrizes weak convergence in this setting, we say $\{\mathbf{Y}_m\}_m$ is \textit{posterior consistent for $\omega \in \Omega$} if 
\begin{equation}\label{eq: posterior consistency integral}
    \forall \epsilon > 0: \quad \lim_{m \rightarrow \infty}  \kappa_{\mathbf{Y}_m} \left(  \left\{ y:  W_1(\pi_{\mathbf{Y}_m}(\cdot | y), \delta_{\omega}) > \epsilon \right\} | \omega \right)  = 0.
\end{equation}
We say $\{\mathbf{Y}_m\}_m$ is \textit{pointwise posterior consistent} if \eqref{eq: posterior consistency integral} holds for every $\omega \in \Omega$:
\begin{equation}\label{eq: posterior consistency integral}
    \forall \epsilon > 0: \quad \lim_{m \rightarrow \infty}  \kappa_{\mathbf{Y}_m} \left(  \left\{ y:  W_1(\pi_{\mathbf{Y}_m}(\cdot | y), \delta_{\omega}) > \epsilon \right\} | \omega \right)  = 0 \quad \forall \omega \in \Omega.
\end{equation}
We say $\{\mathbf{Y}_m\}_m$ is \textit{uniformly posterior consistent} if \eqref{eq: posterior consistency integral} holds uniformly across $\omega \in \Omega$:
\begin{equation}\label{eq: uniform posterior consistency}
    \forall \epsilon > 0: \quad \lim_{m \rightarrow \infty} \sup_{\omega \in \Omega} \kappa_{\mathbf{Y}_m} \left(  \left\{ y:  W_1(\pi_{\mathbf{Y}_m}(\cdot | y), \delta_{\omega}) > \epsilon \right\} | \omega \right) = 0.
\end{equation}
We obtain the following result, linking pointwise posterior consistency and uniform posterior consistency to pointwise consistency and uniform consistency, respectively, of the estimator of the payoff-relevant state given by the posterior mean.

\begin{Lemma}[Consistency of the posterior mean]\label{lemma: consistency of the posterior mean}
    If $\{\mathbf{Y}_m\}_m$ is pointwise posterior consistent, then the sequence of estimators $\{\widehat{\omega}_m\}_m$ defined by
    \begin{equation*}
        \widehat{\omega}_m \equiv \E[\omega | \mathbf{Y}_m]
    \end{equation*}
    for every $m \in \N$ is pointwise consistent. Moreover, if $\{\mathbf{Y}_m\}_m$ is uniformly posterior consistent, the sequence $\{\widehat{\omega}_m\}_m$ is uniformly consistent.
\end{Lemma}

\begin{proof}
    Suppose $\{\mathbf{Y}_m\}_m$ is posterior consistent for $\omega \in \Omega$. Note that by the definition of the Wasserstein 1-distance, for any signal $\mathbf{Y}$ and $\pi_{\mathbf{Y}}(\cdot | \omega)$-a.e. $y \in \cY$,
    \begin{align*}
        d_{\Omega} \left( \int \widetilde{\omega} d \pi_{\mathbf{Y}}(\widetilde{\omega} | y), \omega \right) &= d_{\Omega} \left( \int \widetilde{\omega} d \pi_{\mathbf{Y}}(\widetilde{\omega} | y), \int \widetilde{\omega} d \delta_{\omega}(\widetilde{\omega}) \right) \\
        &\leq W_1(\pi_{\mathbf{Y}}(\cdot | y), \delta_{\omega}).
    \end{align*}
    Fix an arbitrary $\epsilon > 0$. Then, for any $m \in \N$,
    \begin{equation*}
        \lambda(\{r \in [0,1]: d_{\Omega}(\widehat{\omega}_m(\omega, r, x), \omega) > \epsilon\}) \leq \\ \kappa_{\mathbf{Y}_m} \left(  \left\{ y \in \cY:  W_1(\pi_{\mathbf{Y}_m}(\cdot | y), \delta_{\omega}) > \epsilon \right\} | \omega \right).
    \end{equation*}
    Taking the limit as $m \to \infty$ yields the first claim. Further, taking the supremum over $\omega \in \Omega$ on both sides and then taking the limit yields the second claim.
\end{proof}

The following then follows directly from Theorem \ref{prop: consistent estimator implementation}.

\begin{Cor}[Posterior consistency and the posterior mean]\label{cor: posterior consistent posterior mean implementation}
    Suppose $u_i$ is Lipschitz in $\omega$ uniformly in $x$ and $\theta_i$ for each $i \in N$. Fix a sequence of signals $\{\mathbf{Y}_m\}_m$ such that either (i) $\{\mathbf{Y}_m\}_m$ is uniformly posterior consistent or (ii) $\{\mathbf{Y}_m\}_m$ is pointwise posterior consistent, posterior beliefs are Lipschitz continuous, and $\Psi$ is compact. Then there is a non-negative sequence $\{\epsilon_m\}_m$, with $\epsilon_m \rightarrow 0$ as $m \rightarrow \infty$, such that every data-driven VCG mechanism for $\widehat{\omega}_m \equiv \E[\omega | \mathbf{Y}_m]$ permits implementation in $\epsilon_m$-posterior equilibrium for every $m \in \mathbb{N}$.
\end{Cor}

An alternative specification of the transfer rule is to take the expectation of agents' payoffs using the \textit{posterior implied by the signal $\mathbf{Y}$}. The adjusted definition of data-driven VCG mechanisms is as follows.

\begin{Def}[Bayesian data-driven VCG] \label{def: Bayesian data-driven VCG}
    A data-driven direct mechanism $(x^{*},t)$ is a \textit{Bayesian data-driven VCG mechanism} if $x^{*}$ is an efficient allocation rule and for each $i$,
    \begin{equation*}
    t_i(\xi,y) \equiv h_i(\xi_{-i},y) + \sum_{j \neq i} \E[ u_j(x^{*}(\xi),\omega,\theta_j) | y, \mathbf{Y}],
\end{equation*}
for an arbitrary integrable function $h_i$ of others' reports and a realization $y$ of the signal $\mathbf{Y}$.
\end{Def}

If $\mathbf{Y}$ fully reveals the payoff-relevant state, we again obtain exact posterior implementation following the proof of Proposition \ref{prop: ex-post}. Next, we show we also have a continuity result under this formulation, analogously to Theorem \ref{prop: consistent estimator implementation}.

\begin{Prop}[Posterior consistency in Bayesian data-driven VCG]\label{prop: posterior consistency in Bayesian data-driven VCG}
    Fix a sequence of signals $\{\mathbf{Y}_m\}_m$ such that either (i) $\{\mathbf{Y}_m\}_m$ is uniformly posterior consistent or (ii) $\{\mathbf{Y}_m\}_m$ is pointwise posterior consistent, posterior beliefs are Lipschitz continuous, and $\Psi$ is compact. Then there is a non-negative sequence $\{\epsilon_m\}_m$, with $\epsilon_m \rightarrow 0$ as $m \rightarrow \infty$, such that every Bayesian data-driven VCG mechanism for $\mathbf{Y}_m$ permits implementation in $\epsilon_m$-posterior equilibrium for every $m \in \mathbb{N}$.
\end{Prop}

The proof proceeds similarly to the proof of Theorem \ref{prop: consistent estimator implementation}. In particular, Lemma \ref{lemma: bound on epsilon} continues to be a crucial step. We replace Lemma \ref{lemma: uniform convergence} with Lemma \ref{lemma: uniform convergence under posterior consistency} stated and proved below. However, we no longer require the Lipschitz condition on payoff functions; hence, we no longer use Lemma \ref{lemma: bound via expected error}. Bayesian data-driven VCG mechanisms essentially replace agents' (expected) payoffs with their estimates. Boundedness of payoff function is sufficient to obtain the result by properties of the Wasserstein 1-distance. After proving Lemma \ref{lemma: uniform convergence under posterior consistency}, we directly proceed to proving the main statements of the proposition.

\begin{Lemma}[Uniform convergence under posterior consistency]\label{lemma: uniform convergence under posterior consistency}
    The following statements hold:
    \begin{enumerate}
        \item Fix a uniformly posterior consistent sequence of signals $\{\mathbf{Y}_m\}_m$. Then,
        \begin{equation*}
            \lim_{m \rightarrow \infty} \sup_{\omega \in \Omega} \int_{\cY} W_1(\pi_{\mathbf{Y}_m}(\cdot | y), \delta_{\omega}) d \kappa_{\mathbf{Y}_m}(y | \omega) = 0.
        \end{equation*}
        \item Fix a sequence of signals $\{\mathbf{Y}_m\}_m$ that is pointwise posterior consistent. Moreover, suppose posterior beliefs are Lipschitz continuous and $\Psi$ is compact. Then,
        \begin{equation*}
            \lim_{m \rightarrow \infty} \sup_{\mathbf{S} \in \Psi, s \in \cS} \int_{\Omega} \int_{\cY} W_1(\pi_{\mathbf{Y}_m}(\cdot | y), \delta_{\omega}) d \kappa_{\mathbf{Y}_m}(y | \omega) d \pi_{\mathbf{S}}(\omega | s) = 0.
        \end{equation*}
    \end{enumerate}
\end{Lemma}

The proof proceeds analogously to the proof of Lemma \ref{lemma: uniform convergence}.

\begin{proof}
    \medskip \noindent \textit{Part 1.} Let $\epsilon > 0$ be arbitrary. We have
    \begin{align*}
        \int_{\cY} W_1(\pi_{\mathbf{Y}_m}(\cdot | y), \delta_{\omega}) d \kappa_{\mathbf{Y}_m}(y | \omega) &= \int_{\cY} W_1(\pi_{\mathbf{Y}_m}(\cdot | y), \delta_{\omega}) \mathbbm{1}_{W_1(\pi_{\mathbf{Y}_m}(\cdot | y), \delta_{\omega}) \leq \frac{\epsilon}{2}} d \kappa_{\mathbf{Y}_m}(y | \omega) + \\
        &\quad \int_{\cY} W_1(\pi_{\mathbf{Y}_m}(\cdot | y), \delta_{\omega}) \mathbbm{1}_{W_1(\pi_{\mathbf{Y}_m}(\cdot | y), \delta_{\omega}) > \frac{\epsilon}{2}} d \kappa_{\mathbf{Y}_m}(y | \omega) \\
        &\leq \frac{\epsilon}{2} + \text{diam}(\Omega) \kappa_{\mathbf{Y}_m} \left(  \left\{ y:  W_1(\pi_{\mathbf{Y}_m}(\cdot | y), \delta_{\omega}) > \frac{\epsilon}{2} \right\} | \omega \right)
    \end{align*}
    Hence,
    \begin{equation*}
        \sup_{\omega \in \Omega} \int_{\cY} W_1(\pi_{\mathbf{Y}_m}(\cdot | y), \delta_{\omega}) d \kappa_{\mathbf{Y}_m}(y | \omega) \leq \frac{\epsilon}{2} + \text{diam}(\Omega) \sup_{\omega \in \Omega} \kappa_{\mathbf{Y}_m} \left(  \left\{ y:  W_1(\pi_{\mathbf{Y}_m}(\cdot | y), \delta_{\omega}) > \frac{\epsilon}{2} \right\} | \omega \right)
    \end{equation*}
    By the uniform posterior consistency assumption, it follows that there is $M \in \mathbb{N}$ such that, for all $m \geq M$,
    \begin{equation*}
       \sup_{\omega \in \Omega} \kappa_{\mathbf{Y}_m} \left(  \left\{ y:  W_1(\pi_{\mathbf{Y}_m}(\cdot | y), \delta_{\omega}) > \frac{\epsilon}{2} \right\} | \omega \right) < \frac{\epsilon}{2}
    \end{equation*}
    Hence, for all $m \geq M$:
    \begin{equation*}
        \sup_{\omega \in \Omega} \int_{\cY} W_1(\pi_{\mathbf{Y}_m}(\cdot | y), \delta_{\omega}) d \kappa_{\mathbf{Y}_m}(y | \omega) \leq \epsilon.
    \end{equation*}
    Since $\epsilon > 0$ was arbitrary, the result follows.

    \medskip \noindent \textit{Part 2.} Define $\Psi_m: \Psi \times \cS \rightarrow \mathbb{R}$ by
   \begin{equation*}
       \Psi_m(\mathbf{S}, s) \equiv \int_{\Omega} \int_{\cY} W_1(\pi_{\mathbf{Y}_m}(\cdot | y), \delta_{\omega}) d \kappa_{\mathbf{Y}_m}(y | \omega) d \pi_{\mathbf{S}}(\omega | s).
   \end{equation*}
    Observe that $\Psi_m \rightarrow 0$ as $m \rightarrow 0$ pointwise. We show this convergence is uniform. To this end, we show the sequence of functions $\{\Psi_m\}_m$ is uniformly equicontinuous and apply the Arzelà–Ascoli Theorem (\citet{folland1999real}, Theorem 4.44).

    Let $\epsilon > 0$ be arbitrary and fix $\delta = \frac{\epsilon}{\text{diam}(\Omega) L}$. Let $(\mathbf{S}_1, s_1)$ and $(\mathbf{S}_2, s_2)$ be arbitrary elements of $\Psi \times \mathcal{S}$ such that $d_{\Psi}(\mathbf{S}_1, \mathbf{S}_2) + d_{\mathcal{S}}(s_1, s_2) < \delta$. We have
    \begin{align*}
        |\Psi_m(\mathbf{S}_1, s_1) - \Psi_m(\mathbf{S}_2, s_2)| &= \left | \int_{\Omega} \int_{\cY} W_1(\pi_{\mathbf{Y}_m}(\cdot | y), \delta_{\omega}) d \kappa_{\mathbf{Y}_m}(y | \omega) d [ \pi_{\mathbf{S}_1}(\omega | s_1) - \pi_{\mathbf{S}_2}(\omega | s_2)] \right | \\
        &\leq \text{diam}(\Omega) \left | \int_{\Omega} d [ \pi_{\mathbf{S}_1}(\omega | s_1) - \pi_{\mathbf{S}_2}(\omega | s_2)] \right | \\
        &\leq \text{diam}(\Omega) W_1(\pi_{\mathbf{S}_1}(\cdot | s_1), \pi_{\mathbf{S}_2}(\cdot | s_2)),
    \end{align*}
    where the last inequality follows by the dual representation of $W_1$ stemming from the Kantorovich-Rubinstein Theorem (\citet{villani2021topics}, Theorem 1.14). Using the Lipschitz property, we obtain
    \begin{equation*}
        |\Psi_m(\mathbf{S}_1, s_1) - \Psi_m(\mathbf{S}_2, s_2)| \leq \text{diam}(\Omega) L (d_{\Psi}(\mathbf{S}_1, \mathbf{S}_2) + d_{\mathcal{S}}(s_1, s_2)) < \text{diam}(\Omega) L \delta = \epsilon
    \end{equation*}
    Therefore, $\{\Psi_m\}$ is uniformly equicontinuous. Moreover, the sequence is also uniformly bounded by the compactness of $\Omega$, converges pointwise to a continuous function, and the sequence of functions as well as the pointwise limit is defined on a compact domain by the compactness of $\mathcal{S}$ and $\Psi$. Hence, by the Arzelà–Ascoli Theorem, the convergence is uniform.
\end{proof}

\begin{proof}[Proof of Proposition \ref{prop: posterior consistency in Bayesian data-driven VCG}]
    By Lemma \ref{lemma: bound on epsilon}, it sufficient to show that for every agent $j \in N$
    \begin{equation*}
        \sup_{\xi = (\theta, s, \mathbf{S}) \in \Xi, x \in X} \left| \int_{\Omega} u_j(x, \omega, \theta_j) d \pi_{\mathbf{S}}(\omega | s) - \int_{\Omega} \int_{\cY} \int_{\Omega} u_j(x, \widetilde{\omega}, \theta_j) d \pi_{\mathbf{Y}}( \widetilde{\omega} | y) d \kappa_{\mathbf{Y}_m}(y | \omega) d \pi_{\mathbf{S}}(\omega | s) \right| \to 0,
    \end{equation*}
    as $m \to \infty$. To this end, note that for every $m \in \N$, $\xi = (\theta, s, \mathbf{S}) \in \Xi$, and $x \in X$, by Jensen's inequality,
    \begin{multline*}
        \left| \int_{\Omega} u_j(x, \omega, \theta_j) d \pi_{\mathbf{S}}(\omega | s) - \int_{\Omega} \int_{\cY} \int_{\Omega} u_j(x, \widetilde{\omega}, \theta_j) d \pi_{\mathbf{Y}}( \widetilde{\omega} | y) d \kappa_{\mathbf{Y}_m}(y | \omega) d \pi_{\mathbf{S}}(\omega | s) \right| \leq \\
        \int_{\Omega} \int_{\cY} \left| \int_{\Omega} u_j(x, \widetilde{\omega}, \theta_j) d [\pi_{\mathbf{Y}}( \widetilde{\omega} | y) - \delta_{\omega}(\widetilde{\omega})] \right | d \kappa_{\mathbf{Y}_m}(y | \omega) d \pi_{\mathbf{S}}(\omega | s).
    \end{multline*}
    Since we assume throughout this paper that all payoff functions are bounded, and using the Kantorovich-Rubinstein Theorem (\citet{villani2021topics}, Theorem 1.14), we can further upper-bound this by
    \begin{equation*}
        \lVert u_j \rVert_{\infty} \int_{\Omega} \int_{\cY} W_1(\pi_{\mathbf{Y}_m}(\cdot | y), \delta_{\omega}) d \kappa_{\mathbf{Y}_m}(y | \omega) d \pi_{\mathbf{S}}(\omega | s).
    \end{equation*}
    Under conditions (ii), by Lemma \ref{lemma: uniform convergence under posterior consistency},
    \begin{multline*}
        \sup_{\xi \in \Xi, x \in X} \lVert u_j \rVert_{\infty} \int_{\Omega} \int_{\cY} W_1(\pi_{\mathbf{Y}_m}(\cdot | y), \delta_{\omega}) d \kappa_{\mathbf{Y}_m}(y | \omega) d \pi_{\mathbf{S}}(\omega | s) = \\ \lVert u_j \rVert_{\infty} \sup_{\mathbf{S} \in \Psi, s \in \cS} \int_{\Omega} \int_{\cY} W_1(\pi_{\mathbf{Y}_m}(\cdot | y), \delta_{\omega}) d \kappa_{\mathbf{Y}_m}(y | \omega) d \pi_{\mathbf{S}}(\omega | s) \to 0,
    \end{multline*}
    as $m \to \infty$. Further, note that
    \begin{multline*}
        \lVert u_j \rVert_{\infty} \sup_{\mathbf{S} \in \Psi, s \in \cS} \int_{\Omega} \int_{\cY} W_1(\pi_{\mathbf{Y}_m}(\cdot | y), \delta_{\omega}) d \kappa_{\mathbf{Y}_m}(y | \omega) d \pi_{\mathbf{S}}(\omega | s) \leq \\
         \lVert u_j \rVert_{\infty} \sup_{\omega \in \Omega} \int_{\cY} W_1(\pi_{\mathbf{Y}_m}(\cdot | y), \delta_{\omega}) d \kappa_{\mathbf{Y}_m}(y | \omega).
    \end{multline*}
    The right-hand side converges to zero as $m \to \infty$ under condition (i) by Lemma \ref{lemma: uniform convergence under posterior consistency}. The result follows.
\end{proof}

%%%%%%%%%%%%%%%%%%%%%%%%%%%%%%%%%%%%%%%%%%%%%%%%%%%%%%%%%%%%
\subsection{Further Examples}
%%%%%%%%%%%%%%%%%%%%%%%%%%%%%%%%%%%%%%%%%%%%%%%%%%%%%%%%%%%%

\begin{Ex}\label{ex: rem Interdependent preferences}
    Suppose there are two agents and a single object to be allocated. The set of allocations is the probability simplex $\Delta_{1}$. The set of feasible signals $\Psi$ is a singleton. The state, preference types, and signal realizations are single-dimensional and all belong to $[0,1]$. Suppose there are no atoms in the priors on types.
    
    Suppose the payoff of agent 1 is given by
    \begin{equation*}
        u_1(x,\theta,\omega) = \theta_1 \cdot x_{1} \cdot \omega,
    \end{equation*}
    where $x_1$ is the probability agent 1 obtains the object. The payoff of agent 2 is given by
    \begin{equation*}
        u_2(x,\theta,\omega) = (- \theta_1 + \theta_2) \cdot x_{2} \cdot \omega.
    \end{equation*}
    The efficient allocation is as follows. It is efficient to allocate the object to agent 1 if $2 \theta_1 > \theta_2$. If $2 \theta_1 < \theta_2$, it is efficient to allocate the object to agent 2. In the case of a tie, any feasible allocation is efficient.

    Consider the data-driven VCG expected transfer for the ex-post case:
    \begin{equation*}
        t_1(\theta,s) = h_1(\theta_2,s_2) + \E \left[ u_2(x^{*}(\theta,s),\theta,\omega) | s \right].
    \end{equation*}
    Suppose agent 2 reports truthfully. Agent 1's expected net utility when reporting $(\theta_1',s_1')$ thus becomes
    \begin{equation*}
        h_1(\theta_2,s_2) + \left[ \theta_1 x_1^{*}(\theta_1',\theta_2) + ( - \theta_1' + \theta_2)  x_2^{*}(\theta_1',\theta_2) \right] \E \left[ \omega | s \right]. 
    \end{equation*}
    If $\theta_1 < \theta_2$ and $\E[\omega | s] > 0$, it is optimal for agent 1 to report $\theta_1' = 0$, yielding a payoff of $h_1(\theta_2,s_2) + \theta_2 \E[\omega | s]$. Therefore, reporting truthfully for all agents is not a posterior equilibrium.
\end{Ex}

%%%%%%%%%%%%%%%%%%%%%%%%%%%%%%%%%%%%%%%%%%%%%%%%%%%%%%%%%%%%
\subsection{AI Shopping Assistants: Revenue and Individual Rationality}\label{appendix: revenue AI shopping}
%%%%%%%%%%%%%%%%%%%%%%%%%%%%%%%%%%%%%%%%%%%%%%%%%%%%%%%%%%%%

Two natural additional design objectives in our AI shopping assistant application are individual rationality and a no-subsidy condition, requiring that the platform does not, in expectation, pay any agent more than her marginal contribution to the user’s surplus.

To this end, we define \emph{posterior no subsidy with initial context}:
for each advertiser $i \in N$, types $\xi \in \Xi$, estimators $\widehat{\omega}$, and initial context $c_0 \in \mathcal{C}_0$,
\begin{equation*}
    \sum_{\tau \ge 1}
    \mathbb{E}\!\left[
        t_{i\tau}(\xi,h_{\tau},\widehat{\omega}_{\tau}^{h_{\tau}})
        \,\middle|\,
        \mathbf{S}=s,\,\mathbf{C}_0=c_0
    \right]
    \le \sum_{\tau \ge 1}
    \mathbb{E}\!\left[
        x^{*}_{i\tau}(\xi,h_{\tau})\,\omega_i
        \,\middle|\,
        \mathbf{S}=s,\,\mathbf{C}_0=c_0
    \right].
\end{equation*}

We further define \emph{posterior individual rationality with initial context}: for each advertiser $i \in N$, types $\xi \in \Xi$, estimators $\widehat{\omega}$, and initial context $c_0 \in \mathcal{C}_0$,
\begin{equation*}
    \sum_{\tau \ge 1}
    \mathbb{E}\!\left[
        x^{*}_{i\tau}(\xi,h_{\tau})\,\theta_i
        +
        t_{i\tau}(\xi,h_{\tau},\widehat{\omega}_{\tau}^{h_{\tau}})
        \,\middle|\,
        \mathbf{S}=s,\,\mathbf{C}_0=c_0
    \right]
    \ge 0.
\end{equation*}

We also define the following adjustments to the data-driven dynamic team mechanism resembling the pivot mechanism. Suppose $\Psi$ and $\mathcal{C}$ are both compact. Let $x^{+, -i}$ be the optimal policy for the following adjusted social problem excluding agent $i$:
\begin{equation*}
 \max_{\mathbf{S}_i \in \Psi_i,\; s_i \in \mathcal{S}_i} \max_{\{x_{\tau}: H_{\tau} \to X\}_{\tau=1}^T} 
    \E\!\left[ \sum_{\tau = 1}^T \sum_{j \neq i }
    \big(\theta_j + \omega_{j\tau}\big)
        x_{j\tau}(h_{\tau})
        \,\middle|\,
        \mathbf{S}=s,\,\mathbf{C}_0=c_0
    \right].
\end{equation*}
Further, let $x^{-, -i}$ be the optimal policy for the following:
\begin{equation*}
    \min_{\mathbf{S}_i \in \Psi_i,\; s_i \in \mathcal{S}_i} \max_{\{x_{\tau}: H_{\tau} \to X\}_{\tau=1}^T} 
    \E\!\left[ \sum_{\tau = 1}^T \sum_{j \neq i }
    \big(\theta_j + \omega_{j\tau}\big)
        x_{j\tau}(h_{\tau})
        \,\middle|\,
        \mathbf{S}=s,\,\mathbf{C}_0=c_0
    \right].
\end{equation*}
Then consider ``Groves'' payments in the data-driven dynamic team mechanism that satisfy
\begin{equation*}
    \sum_{\tau = 1}^T h_{i\tau}^{+}(\xi_{-i},h_{\tau},\widehat{\omega}^{*})
    = -
    \max_{\mathbf{S}_i \in \Psi_i,\; s_i \in \mathcal{S}_i} \E\!\left[ \sum_{\tau = 1}^T \sum_{j \neq i }
    \big(\theta_j + \omega_{j}\big)
        x_{j\tau}^{+, -i}(\xi_{-i}, h_{\tau})
        \,\middle|\,
        \mathbf{S}=s,\,\mathbf{C}_0=c_0
    \right],
\end{equation*}
and
\begin{equation*}
    \sum_{\tau = 1}^T h_{i\tau}^{-}(\xi_{-i},h_{\tau},\widehat{\omega}^{*})
    = -
    \min_{\mathbf{S}_i \in \Psi_i,\; s_i \in \mathcal{S}_i} \E\!\left[ \sum_{\tau = 1}^T \sum_{j \neq i }
    \big(\theta_j + \omega_{j}\big)
        x_{j\tau}^{-, -i}(\xi_{-i}, h_{\tau})
        \,\middle|\,
        \mathbf{S}=s,\,\mathbf{C}_0=c_0
    \right].
\end{equation*}

We obtain the following result.

\begin{Prop}[Subsidies and participation]
    The data-driven dynamic team mechanism with Groves adjustment $h^{+}$ satisfies posterior no subsidy with initial context. With the Groves adjustment $h^{-}$, it satisfies posterior individual rationality with initial context.
\end{Prop}

\begin{proof}
    Fix an arbitrary agent $i$, types $\xi \in \Xi$ and initial context $c_0 \in \mathcal{C}_0$. Under the Groves adjustment $h^{+}$, the expected transfer satisfies
    \begin{multline*}
        \sum_{\tau =1}^T \E \left[ \omega_i x_{i \tau}^{*}(\xi, h_{\tau}) + \sum_{j \neq i } (\theta_j + \omega_{j}) x_{j\tau}^{*}(\xi, h_{\tau}) \,\middle|\,
        \mathbf{S}=s,\,\mathbf{C}_0=c_0
    \right] \\
        -
    \max_{\mathbf{S}_i \in \Psi_i,\; s_i \in \mathcal{S}_i} \E\!\left[ \sum_{\tau = 1}^T \sum_{j \neq i }
    \big(\theta_j + \omega_{j}\big)
        x_{j\tau}^{+, -i}(\xi_{-i}, h_{\tau})
        \,\middle|\,
        \mathbf{S}=s,\,\mathbf{C}_0=c_0
    \right].
    \end{multline*}
    By the construction of $x^{*}$ and $x^{+,-i}$,
    \begin{multline*}
         \E \left[ \sum_{\tau =1}^T \sum_{j \neq i } (\theta_j + \omega_{j}) x_{j\tau}^{*}(\xi, h_{\tau}) \,\middle|\,
        \mathbf{S}=s,\,\mathbf{C}_0=c_0
    \right] \\
        \leq
    \max_{\mathbf{S}_i \in \Psi_i,\; s_i \in \mathcal{S}_i} \E\!\left[ \sum_{\tau = 1}^T \sum_{j \neq i }
    \big(\theta_j + \omega_{j}\big)
        x_{j\tau}^{+, -i}(\xi_{-i}, h_{\tau})
        \,\middle|\,
        \mathbf{S}=s,\,\mathbf{C}_0=c_0
    \right].
    \end{multline*}
    The claim regarding no subsidies follows.

    Also note that, by the construction of $x^{*}$ and $x^{-,-i}$,
    \begin{multline*}
         \E \left[ \sum_{\tau =1}^T \sum_{j \in N } (\theta_j + \omega_{j}) x_{j\tau}^{*}(\xi, h_{\tau}) \,\middle|\,
        \mathbf{S}=s,\,\mathbf{C}_0=c_0
    \right] \\
        \geq
    \min_{\mathbf{S}_i \in \Psi_i,\; s_i \in \mathcal{S}_i} \E\!\left[ \sum_{\tau = 1}^T \sum_{j \neq i }
    \big(\theta_j + \omega_{j}\big)
        x_{j\tau}^{-, -i}(\xi_{-i}, h_{\tau})
        \,\middle|\,
        \mathbf{S}=s,\,\mathbf{C}_0=c_0
    \right].
    \end{multline*}
    The claim regarding individual rationality now also follows.
\end{proof}

It may be infeasible to satisfy the no-subsidy and individual rationality conditions simultaneously. Nevertheless, when the expected value of $\omega_i$ is sufficiently large, agent~$i$ receives enough expected surplus under $h^{+}$ to satisfy her participation constraint. Similarly, when the expectation is non-negative and the agent’s information is not socially valuable, for instance, when the initial context together with other agents’ information suffices for decision making, the transfer scheme also satisfies the participation constraint. In this case, the adjusted social problems introduced above coincide.

\end{document}